\documentclass[twocolumns,tradictabstract,a4paper]{aa}
\usepackage{graphicx}
\usepackage{txfonts}

\newcommand{\helixp}{\textsc{HeLIx$^+$}}
\newcommand{\ve}[1]{{\rm\bf {#1}}}
\newcommand{\df}{{\rm d}}

\def\ms{~m s$^{-1}$}
\def\deg{^{\circ}}

\begin{document}

\title{Comparison of inversion codes for polarized line formation in 
MHD simulations. I. Milne-Eddington codes}

%

\author{J.M.~Borrero\inst{1} \and B.W.~Lites\inst{2} \and A.~Lagg\inst{3} \and R.~Rezaei\inst{1} \and M.~Rempel\inst{2}}
\institute{{Kiepenheuer Institut f\"ur Sonnenphysik, Sch\"oneckstr. 6, 79104, Freiburg, Germany.}
\email{borrero@kis.uni-freiburg.de, rrezaei@kis.uni-freiburg.de} \and {High Altitude Observatory (NCAR), 3090 Center Green Dr., Boulder, CO~80301, USA.}
\email{lites@hao.ucar.edu, rempel@hao.ucar.edu} \and {Max Planck Institute for Solar System Research, 
Justus-von-Liebig-Weg 3, 37077, G\"ottingen, Germany.} \email{lagg@mps.mpg.de}}
\date{Recieved / Accepted}

\abstract{Milne-Eddington (M-E) inversion codes for the radiative transfer equation are the most widely used tools to infer the magnetic field from observations of the polarization 
signals in photospheric and chromospheric spectral lines. Unfortunately, a comprehensive comparison between the different M-E codes available to the solar physics community is still 
missing, and so is a physical interpretation of their inferences. In this contribution we offer a comparison between three of those codes (VFISV, ASP/HAO, and \helixp). These codes 
are used to invert synthetic Stokes profiles that were previously obtained from realistic non-grey three-dimensional magnetohydrodynamical (3D MHD) simulations. The results of the inversion are compared with each other and with 
those from the MHD simulations. In the first case, the M-E codes retrieve values for the magnetic field strength, inclination and line-of-sight velocity that agree with each other within 
$\sigma_B \leq 35$ (Gauss), $\sigma_\gamma \leq 1.2\deg$, and $\sigma_{\rm v} \leq 10$\ms, respectively. Additionally, M-E inversion codes agree with the numerical simulations, when compared at a 
fixed optical depth, within $\sigma_B \leq 130$ (Gauss), $\sigma_\gamma \leq 5\deg$, and $\sigma_{\rm v} \leq 320$\ms. Finally, we show that employing generalized response functions to determine 
the height at which M-E codes measure physical parameters is more meaningful than comparing at a fixed geometrical height or optical depth. In this case the differences between M-E inferences
and the 3D MHD simulations decrease to $\sigma_B \leq 90$ (Gauss), $\sigma_\gamma \leq 3\deg$, and $\sigma_{\rm v} \leq 90$\ms.}
\authorrunning{Borrero et al.}
\titlerunning{Comparison ME inversion codes and MHD simulations}
\maketitle

\section{Introduction and motivation}
\label{section:intro}

Inversion codes for the radiative transfer equation are, arguably, the best tool available
to infer the physical properties of the solar atmosphere. Even though these inversion codes have been used successfully 
in multiple investigations \citep[see reviews by][]{hector2001,jc2003review,luis2006review,basilio2007review}, a large 
portion of the solar physics community still feel that they do not adequately address the questions of convergence and uniqueness.
This has led many researchers to rely on simpler methods 
in their investigations: center-of-gravity or bisector analysis to determine the line-of-sight component 
of the velocity \citep{rolf2004,morten2013}, separation between $I+V$ and $I-V$ to determine the line-of-sight 
component of the magnetic field \citep{liu2012,couvidat2012}, separation of $\sigma$ components in Stokes-$I$ 
to determine the total magnetic field strength \citep{horst1993,penn2006}, weak-field approximation to determine 
the magnetic field vector \citep{jefferies1991,nazaret2005}, and so forth.\\

Nowadays, inversion codes for the radiative transfer equation are used much more often.
Indeed, many data pipelines for space-borne and ground-based instruments routinely apply inversion codes 
for the radiative transfer equation to analyze the data. That is the case of Hinode/SP \citep{lites2007csac}
\footnote{Inversions with the MERLIN inversion code of Hinode/SP data are readily available at the CSAC webpage: 
\url{http://www.csac.hao.ucar.edu/csac/archive.jsp}}, SDO/HMI \citep{borrero2011}\footnote{Inversions with the 
VFISV inversion code of HMI data are available through the JSOC webpage: \url{http://jsoc.stanford.edu/HMI/Vector_products.html}},
SOLIS/VSM \citep{keller2003}\footnote{Inversions with the VFISV inversion code of SOLIS/VSM data are available at:
\url{http://solis.nso.edu/0/solis_data4.html}}. Future space-missions, such as Solar Orbiter, also plan on including 
inversion codes on their data-processing pipelines \citep{castillo2006,david2007}. The most widely used inversion codes are the so-called 
Milne-Eddington codes (hereafter M-E codes). Although M-E codes operate under rather restrictive assumptions about 
the thermodynamics of the solar plasma \citep{auer1977,landolfi1982}, they are often regarded as being able 
to retrieve reliable values for the magnetic and kinematic properties of the solar photosphere, and even 
chromoshere \citep{lagg2007}. However, the interpretation of M-E inferences is not straightforward, as the Milne-Eddington
solution for the radiative transfer equation assumes that the magnetic field and velocity are constant with height through
the solar atmosphere, which we know is not the case.\\

The purposes of this paper are twofold. On the one hand, we will address concerns about the convergence and uniqueness 
of the physical parameters retrieved by M-E inversion codes. On the other hand, we will investigate the meaning
of M-E inferences in the presence of atmospheres where the magnetic field and velocity vary with height. To this end,
we solve the radiative transfer equation using physical parameters derived from realistic three-dimensional non-grey
magnetohydrodynamic simulations (Section~\ref{section:description_mhd}), and produce synthetic Stokes profiles 
(Section~\ref{section:description_synthesis}) for two widely used magnetically sensitive neutral iron lines. The Stokes 
profiles are then inverted using three different inversion codes that operate under the Milne-Eddington approximation, 
but employ different optimization algorithms (Section~\ref{section:description_ic}). To study whether the three inversion 
codes converge to the same solution, their results are compared to the others' (Section~\ref{section:comparison_ic_ic}).
In addition, we investigate the meaning of M-E inferences by comparing the results from the three inversion codes to
the original values from the three-dimensional magneto-hydrodynamical (3D MHD) simulations (Section~\ref{section:comparison_ic_mhd}) in two different ways: a) assuming
that the information provided by the spectral lines comes from a single optical depth in the solar Photosphere, and b) employing
response functions to determine exactly the layers that contribute to the formation of the selected spectral lines. Finally,
we provide averaged response functions in different solar structures (granulation, sunspots, etc) and for different physical parameters,
(Section~\ref{section:hof}) in order to offer a quantitative explanation as to which layers the selected neutral iron lines are 
sensitive.\\

\section{3D non-grey MHD simulations}
\label{section:description_mhd}

\begin{figure*}
\begin{center}
\includegraphics[width=19cm]{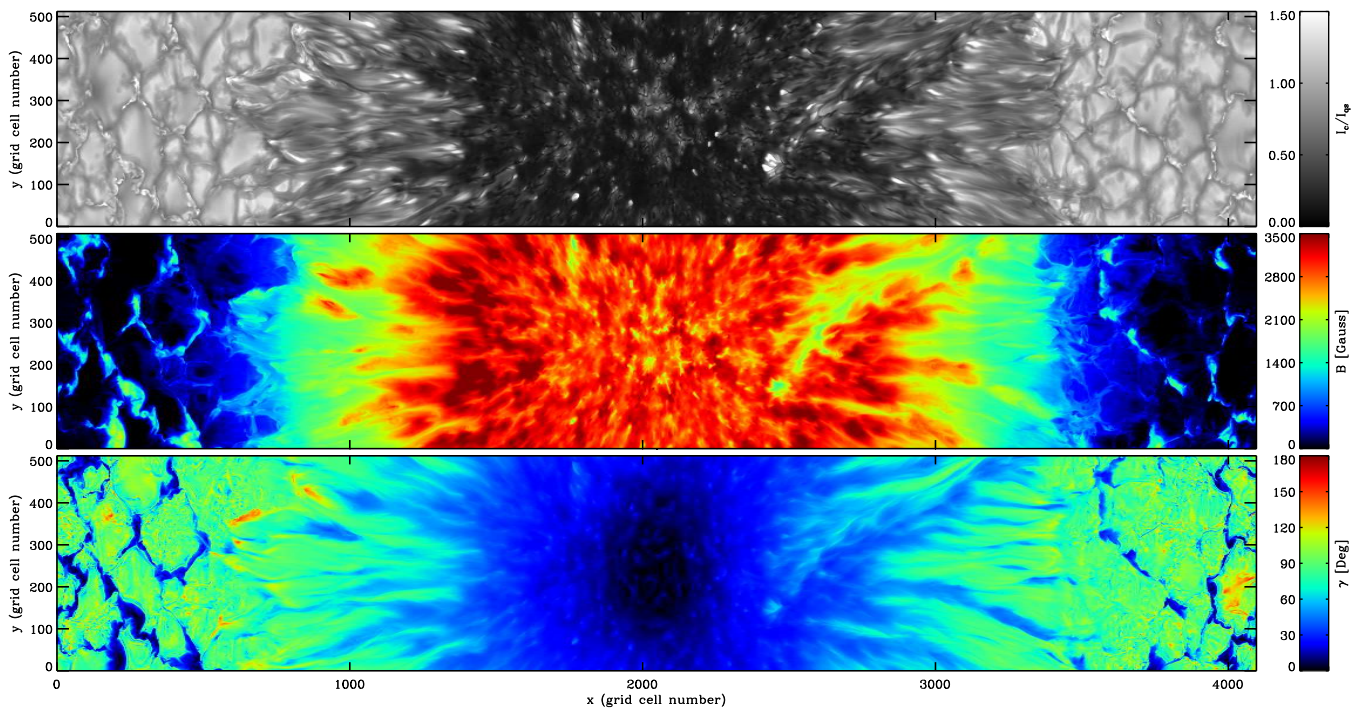}
\caption{Overview of the magnetohydrodynamic simulations employed in this work. The upper panel shows a map of continuum intensity at 
$\lambda_c=630$ nm, $I_{\rm c}(x,y)$, normalized to the average continuum intensity on the quiet Sun ($I_{\rm qs}$).
The middle and bottom panels show maps of magnetic field strength $B(x,y)$ and inclination $\gamma(x,y)$, respectively, at an 
optical depth $\tau_c = 10^{-2}$. The conversion between $z$ and $\tau_c$ is described in Section~\ref{section:description_synthesis}.
The maps in this figure have a horizontal extension of 4096$\times$512 grid points or 49.152$\times$6.144 Mm. The figure has been 
compressed along the $X$-axis so as to fit the entire box on the same panel.}
\label{figure:simul_highlight}
\end{center}
\end{figure*}

Our investigation is based on a non-grey sunspot simulation following the setup described in \citet{matthias2012}. 
These are sunspot models in a domain of the size $49.152\times 49.152\times 6.144\,\mbox{Mm}^3$
that were computed using grey radiative transfer and different grid resolutions. To obtain it, we restarted a non-grey simulation from
the model with $16\times 16\times 12\,\mbox{km}^3$ resolution in \citet{matthias2012} and evolved it for an additional 15 minutes 
with non-grey radiative transfer at a higher resolution of $12\times 12\times 8$~km. At this resolution the domain has a size
of $4096\times 4096\times 768$ grid points. Figure~\ref{figure:simul_highlight} displays a $4096\times 512$ subsection of the domain.
The maps correspond to three physical parameters (continuum intensity, magnetic field strength $B$, and inclination of the magnetic
field with respect to the observer's line-of-sight $\gamma$) at a fixed optical depth. The horizontal slice contains regions that are 
representative of umbra, penumbra and a $\sim 200$~G plage region surrounding the spot.

\section{Synthesis of Stokes profiles}
\label{section:description_synthesis}

The three dimensional magneto-hydrodynamic simulations provide data cubes for the temperature $T(\ve{r})$, gas pressure $P_g(\ve{r})$, density 
$\rho(\ve{r})$, and velocity and magnetic field vectors $\ve{v}(\ve{r})$ and $\ve{B}(\ve{r})$. Here $\ve{r}$ refers to the position in
Cartesian coordinates: $\ve{r}=(x,y,z)$. With this information, and assuming that the observer looks down into the solar
atmosphere along the $z$-direction\footnote{Thus, the simulation box is considered to be located at disk center: $\Theta=0\deg$, with $\Theta$
being the heliocentric angle.}, it is possible to solve the radiative transfer equation for polarized light \citep{jc2003book} along the 
vertical $z$-direction for every ray path with fixed $(x,y)$ values,\\

\begin{equation}
\frac{\partial \ve{I}(z,\lambda)}{\partial z} = \hat{\mathcal{K}}(z,\lambda)[\ve{S}(z,\lambda) - \ve{I}(z,\lambda)] \;,
\label{equation:rtez}
\end{equation}

\noindent where $\ve{I} = (I,Q,U,V)^{\dagger}$ ($^\dagger$ denotes transposition) is the Stokes vector, $\hat{\mathcal{K}}$ is the absorption matrix, 
and $\ve{S}$ the source function. The four components of the Stokes vector $I_j$ ($j=1,..,4$) are commonly referred to as {\it Stokes parameters}, 
and their wavelength dependence, $I_j(\lambda)$, are referred to as {\it Stokes profiles}. To solve the radiative transfer equation 
we employ the synthesis module of the SIR code \citep[Stokes Inversion based 
on Response functions;][]{basilio1992}. The SIR code assumes Local Thermodynamic Equilibrium to compute the population of the atomic 
levels, and therefore the source function depends only on the local temperature $\ve{S} = (B[T(z),\lambda],0,0,0)^{\dagger}$, with 
$B[T(z),\lambda]$ being Planck's function for a given temperature and wavelength. The numerical integration of the radiative transfer equation is done in 
the optical depth scale $\tau_c$, which is related to the $z$-coordinate through

\begin{equation} 
\df \tau_c = -\rho(z) \kappa(z,\lambda_c) \df z \;,
\label{equation:ztau}
\end{equation}

\noindent where $\kappa(z,\lambda_c)=\kappa_c(z)$ is the absorption coefficient per unit mass in a wavelength where there are no spectral lines (i.e., continuum). In 
the $\tau_c$-scale, Eq.~\ref{equation:rtez} is written as\\

\begin{equation}
\frac{\partial \ve{I}(\tau_c,\lambda)}{\partial \tau_c} = \hat{\mathcal{K}}^{'}(\tau_c,\lambda)[\ve{I}(\tau_c,\lambda)-\ve{S}(\tau_c,\lambda)] \;,
\label{equation:rtet}
\end{equation}

\noindent where $\hat{\mathcal{K}}^{'}=\hat{\mathcal{K}}/(\rho \kappa_c)$.  In order to go from Eq.~\ref{equation:rtez} to Eq.~\ref{equation:rtet} we need to solve 
Eq.~\ref{equation:ztau} along each ray path. To do so, we need to know $\rho(z)$ and $\kappa_c(z)$. The former is readily available through the MHD simulations. The continuum 
opacity $\kappa_c(z)$, however, depends on several thermodynamic parameters: temperature $T(z)$, gas pressure $P_g(z)$, and electron 
pressure $P_e(z)$. Again, the first two are provided by the MHD simulations, but the last must be computed by other means.  To this end we employ 
an iterative technique, described in \citet{mihalas1970book}, that solves the Saha ionization equation for 83 elements plus
contributions from H$^{-}$, H$^{+}$, and H$_2^{+}$. Once $P_g(z)$, $P_e(z)$ and $T(z)$ are known, we employ 
SIR's opacity package \citep[based on][]{wittmann1974} to determine $\kappa_c(z)$. These calculations include contributions from H, He, H$^{-}$, He$^{-}$, H$_2^{-}$, 
H$_2^{+}$, C, Mg, and Na, as well as Thomson scattering by free electrons and Rayleigh scattering by H, He and H$_2$.\\

In addition, a boundary condition is needed for the solution of Eq.~\ref{equation:ztau}. In our case we consider that $\tau_c=10^{-6}$ on
the uppermost layer of the simulation box $z_{\rm max}=2048$ km. The conversion from the $z$-scale to the $\tau_c$-scale is only affected by 
the choice of our boundary condition only close to the uppermost layer, but it has no effect in the region where photospheric spectral lines
are formed, $\tau_c \in [1,10^{-4}]$.\\

Although the vertical grid size of $\Delta z = 8$ km can be considered as a very good resolution from the point of view of
MHD simulations, it does not necessarily guarantee that, once we convert to the optical depth scale, the step size in this new scale, $\Delta(\log\tau_c)$, 
is small enough to properly integrate the radiative transfer equation using SIR's synthesis module \citep[Hermitian algorithm by][]{luis1998}. In particular,
layers where the MHD simulations show large temperature changes within a few grid points\footnote{This happens commonly in the upper layers
of the simulation domain, were shocks tend to produce very steep temperature variations.} are prone to produce overshooting effects in the Hermitian
algorithm. Although overshooting can be avoided by implementing better-behaved integration schemes \citep{jaime2013}, in our case, we have 
opted for a spline reinterpolation of the stratification in the physical parameters, after the $z-\tau_c$ conversion, into a 
finer grid with $\Delta(\log\tau_c)=0.01$. This ensures that the Hermitian algorithm performs adequately.\\

For the investigations in this paper we have synthesized two magnetically-sensitive spectral lines commonly used in spectropolarimetry. The
properties of these spectral lines are summarized in Table~\ref{table:lines}. The spectral synthesis has been carried out
with a wavelength sampling of 10 m{\AA} from $-500$ m{\AA} to $+500$ m{\AA} around the central laboratory wavelength ($\lambda_0$)
of each spectral line. The continuum between both lines has been determined considering how the wings of each line blend into that of the
other line. Owing to the variations with optical depth ($\tau_c$) in the physical quantities present in the MHD simulations,
the Stokes profiles synthesized with the SIR code are, in general, asymmetric \citep{landolfi1996}.\\

\begin{table*}
\begin{center}
\caption{Spectral lines synthesized in this work. Central wavelengths $\lambda_0$, excitation potential of the lower level $\chi_{\rm low}$,
and electronic configurations of the lower and upper levels are taken from \protect\citet{nave1994}. The oscillator strength of the first line
has been taken from \protect\citet{bard1991}. The oscillator strength of the second spectral line has been determined from the value of the first one
and employing a measured ratio of 2.8 between the gf factors of both lines ({\it private communication} from Brian C. Fawcett from the 
Rutherford Appleton Laboratory). The $\alpha$ and $\sigma$ parameters are the temperature
exponent and cross section values (in units of the square of the Bohr radius $a_0^2$), respectively, needed for the calculation of the line-broadening 
by collisions with neutral atoms under the ABO theory. In particular, the numbers provided have been obtained by interpolation on the tables 
provided by \protect\citet{anstee1995}.\label{table:lines}}
\begin{tabular}{|ccccccccc|}
\hline
Element & Ion & $\lambda_0$ & $\log(gf)$ & $\chi_{\rm low}$ & $\alpha$ & $\sigma/(a_0^2)$ & Upper & Lower \\
        &     & [{\AA}]   &            & $[eV]$         &          &                 &        &       \\
\hline
Fe      & I & 6301.5012 & -0.718 & 3.654 & 0.243 & 840 & $^5$P$_2$ & $^5$D$_0$ \\
Fe      & I & 6302.4936 & -1.165 & 3.686 & 0.241 & 856 & $^5$P$_1$ & $^5$D$_0$ \\
\hline
\end{tabular}
\end{center}
\end{table*}

Finally, it is important to mention that, although the original simulation box contains 4096$\times$4096 grid points on the $XY$-plane, for the tests presented in this paper we have 
only made use of a 4096$\times$16 slice. This slice corresponds to $x \in [1,4096]$ and $y \in [248,264]$ in Fig.~\ref{figure:simul_highlight}. As already mentioned in 
Section~\ref{section:description_mhd}, along the $x$-direction this slice contains granulation, penumbra and umbra. Because the granulation is close to the sunspot 
it cannot be fully regarded as quiet Sun. The total number of points included in the analysis, $2^{16}$, is large enough so as to allow for statistical comparisons 
(see Sections~\ref{section:comparison_ic_ic},~\ref{section:comparison_fixed_tau}, and ~\ref{section:comparison_rf}).\\

\section{Milne-Eddington inversion codes}
\label{section:description_ic}

Unlike the SIR code that takes into account the full dependence of the physical parameters on optical depth $\tau_c$ (see Sect.~\ref{section:description_synthesis}), 
Milne-Eddington (M-E) codes solve the radiative transfer equation under the Milne-Eddington approximation \citep{auer1977,landolfi1982}. This approximation assumes
that many physical parameters relevant to the formation of spectral lines in solar (or stellar) atmospheres are constant or, in other words, independent of 
$\tau_c$. These parameters are: $\eta_0$ (ratio between the absorption coefficient at the line-core and continuum), $a$ (damping), $\Delta\lambda_D$ (Doppler
width of the spectral line), $\ve{B}$ (magnetic field vector, usually expressed in spherical coordinates $B$, $\gamma$, $\phi$), and $\rm v_{\rm los}$ (line-of-sight 
component of the velocity). In addition to this, the M-E approximation considers that the Source function varies linearly with optical depth: 
$\ve{S}(\tau_c) = (S_0 + \tau_c S_1,0,0,0)$ (see Eq.~\ref{equation:rtet}). Owing to these assumptions, the solution to the polarized radiative transfer equation
can be obtained analytically \citep{unno1956,rachkovski1962}, thereby greatly improving the speed of the forward solution: $\ve{I}=f(\ve{M})$, where $f(\ve{M})$ 
\citep[see for instance][Eq.~1 or Eqs. 9.44-9.45, respectively]{landolfi1982,jc2003book} is an analytic function of\\ 

\begin{equation}
\ve{M}=[\eta_0, a, \Delta\lambda_D, S_0, S_1, B, \gamma, \phi, {\rm v}_{\rm los}] \;.
\label{equation:m}
\end{equation}

The physical parameters $\ve{M}$ enter the solution of the radiative transfer equation (Eq.~\ref{equation:rtet}) in a straightforward fashion 
\citep[see][Eqs.~7.44-7.45]{jc2003book}, and therefore no further attempt is made to derive any of them from the underlying microphysics. Indeed, 
one can surmise that the thermodynamic parameters $[\eta_0, a, \Delta\lambda_D, S_0, S_1]$ could be derived from the temperature 
$T$, density $\rho$, electron pressure $P_e$, and gas pressure $P_g$. For instance, it could be considered that $\eta_0 \propto \rho$, $\Delta\lambda_D \propto \sqrt{T}$ 
\citep[see][Eqs.~6.42 and 7.40]{jc2003book}, etc. However, in doing so, we would have to solve the Saha and Boltzmann equations numerically, and iterate to obtain the electron
pressure from a given temperature and gas pressure (see Sect.~\ref{section:description_synthesis}), and so on. These numerical computations would defeat the original purpose
of having an analytic solution to the radiative transfer equation (Eq.~\ref{equation:rtet}). Fortunately, this does not apply to the magnetic and kinematic parameters, 
and therefore results for $[B, \gamma, \phi, {\rm v}_{\rm los}]$ can be readily interpreted.\\

The applicability M-E inversion codes is usually limited to the inversion of the observed Stokes vector in a single spectral line. Alternatively, 
if more lines are included in the analysis, one must take care that those spectral lines are close in wavelength and sample similar layers on the solar atmosphere. 
Inverting spectral lines that are far away in wavelength or formed in very different layers implies that a new set $\ve{M}$ (Eq.~\ref{equation:m}) is needed
for each line, which would multiply the number of free parameters during the inversion. This happens because each line is formed in layers
characterized by very different physical conditions. A pair of lines that are close in wavelength are Fe I 6301.50 
and 6302.49 {\AA} (see Table~\ref{table:lines}). Although they do not sample exactly the same photospheric layers \citep{mariam2006} they are 
close enough so as to allow for M-E inversions using the same set of $\ve{M}$ \citep{david2010fepair}. Indeed, these two lines have become the lines of 
choice by many spectropolarimeters both on ground \citep{baur1981,polis1,polis2,ali2007} and space-borne \citep{ichimoto2008} instruments. Other spectral 
lines, such as the line pair Fe I 5247.06 and 5250.22 {\AA}, are formed even closer \citep{stenflo1973,hector2008}. However, owing to the large
temperature dependence of the latter pair, we have chosen the former one for our work.\\

Because the physical parameters in $\ve{M}$ (Eq.~\ref{equation:m}) are considered independent of the optical depth $\tau_c$, the M-E solution for the radiative transfer equation
is incapable of reproducing asymmetric Stokes profiles. This implies that, in general, the Milne Eddington inversion of the Stokes profiles synthesized 
from realistic MHD simulations (see Section~\ref{section:description_synthesis}) will not be able to fully reproduce those profiles. This is a well-known
limitation of M-E inversion codes. However, because of their speed and simplicity (e.g., analytic solution), M-E codes have become the most 
widely used codes to study the solar magnetic field. Therefore, the question is whether, in spite of the drawbacks listed above, inversion codes based on the Milne-Eddington
approximation are capable of reliably inferring the magnetic and kinematic properties of the solar atmosphere. Answering this question is the main purpose of this paper.
We will address it in two different ways. In Section~\ref{section:comparison_ic_ic} we will study whether different M-E inversion codes agree
on the magnetic and kinematic properties of the atmosphere after inverting the profiles synthesized in Section~\ref{section:description_synthesis}. 
Then, in Section~\ref{section:comparison_ic_mhd} we will investigate how those results compare with the original magnetic and kinematic parameters of the MHD simulations. 
In the following we describe the inversion codes that will be tested in this paper.\\

\subsection{Very Fast Inversion of the Stokes Vector inversion code: VFISV}
\label{section:vfisv}

The Very Fast Inversion of the Stokes Vector (VFISV) is a Milne-Eddington inversion code developed to analyze data from the Helioseismic and Magnetic
Imager instrument onboard the Solar Dynamics Observatory satellite \citep{scherrer2012}. Because of the unique characteristics of this instrument,
VFISV is designed to invert one spectral line at a time. In addition, it assumes that the Zeeman pattern can be described under a normal Zeeman triplet 
($J=1 \rightarrow J=0$ transition). For these reasons, in the inversions carried out in this work, VFISV will only consider the Fe I 6302.4936 {\AA} spectral line
(see Table~\ref{table:lines}). The VFISV inversion strategy uses analytic derivatives of the Stokes vector with respect to the free parameters, and employs
a combination of the Levenberg-Marquardt method \citep{press1986} and Singular Value Decomposition to fit the observed profiles and retrieve the physical 
parameters, $\ve{M}$ (Eq.~\ref{equation:m}), of the solar atmosphere. The initial guess for $B$, $\gamma$, $\phi$, and ${\rm v}_{\rm los}$ is obtained from 
an Artificial Neural Network that has been specifically trained, using back-propagation \citep{bishop1995}, for the aforementioned spectral line. In addition 
to this, VFISV is able to re-initialize the inversion process using random values of $\ve{M}$ if the inversion does not converge after a predetermined number 
of iterations. A detailed description of the code can be found in \citet{borrero2011}.\\

\subsection{Advanced Stokes Polarimeter inversion code: ASP/HAO}
\label{section:asphao}

The Advanced Stokes Polarimeter \citep[ASP; ][]{elmore1992} inversion code is a M-E inversion that evolved from early
attempts \citep{auer1977} to invert observed Stokes profiles measured with the HAO Stokes I and Stokes II instruments. 
In the first modification to the code of \citet{auer1977}, \citet{skumanich1987} relaxed a number of approximations that 
were leading to non-convergent behaviour, and demonstrated the important role of scattered/stray light by fitting only the 
Stokes polarization profiles $Q, U, V$. The first systematic inversions of Stokes profiles measured across sunspots 
resulted from the further refinement of the ASP/HAO code \citep{lites1990}. This modification included the innovation of 
fitting the Stokes $I$ profile with an added variable fraction of a stray/scattered light profile pre-determined from 
quiet Sun profiles. Not only was the fractional admixture of stray light determined, but the code also allows the stray 
light profile to shift in wavelength. Once data were available from the ASP instrument, the code had been modified to 
invert simultaneously two or more lines of the same multiplet \citep{lites1993}. During the period of frequent usage of 
the ASP instrument (1993-2006), the ASP/HAO code was applied routinely to many data sets, and continued to see refinement. 
In particular, several means of initialization of the variables were developed, culminating in usage of a five-parameter 
genetic algorithm \citep{charbonneau1995} as the default initialization. Furthermore, the code was extended to allow two 
magnetic components plus the non-magnetic stray/scattered light component \citep[see, for example,][]{lites2002}.\\
 
The ASP/HAO code also uses a Levenberg-Marquardt method for iterative fitting of the profiles, modified for accelerated
convergence \citep{lites1990}. The system of linear equations is solved using stepwise regression \citep{jennrich1977} rather  
than singular value decomposition because the latter was found to be unstable in some cases. The magnetic splitting
pattern for transitions is calculated in generality under the L-S coupling assumption. The genetic algorithm is used to 
set initial values of $B$, $\gamma$, $\Delta\lambda_D$, $\eta_0$, and ${\rm v}_{\rm los}$. Other simpler algorithms are 
used to initialize the remainder of the variables. Even with only these five parameters, the genetic solution requires 
computational time comparable to that of the Levenberg-Marquardt procedure. For the least-squares fitting, derivatives 
are computed from analytical expressions. Although there is a pre-established maximum number of iteration, the code usually
converges in ten or fewer \citep{lites1990}.\\

\subsection{Helium Line Information eXtractor: \helixp{}}
\label{section:helix}

The M-E code \helixp{} \cite[]{lagg2004,lagg2009} was developed to analyze the spectral region around the He~\textsc{i} 10830~\AA{} infrared triplet. This triplet often 
occurs in multi-lobed profiles, indicative for the complex velocity and magnetic field morphology present in an usually optically thin layer in the upper chromosphere, 
the formation region of this triplet \cite[]{xu2012,sasso2011,lagg2007,solanki2003}. Several photospheric lines and telluric blends in the spectral vicinity are 
interfering with this triplet. \helixp{} is optimized to treat multiple He~\textsc{i} components, the photospheric and telluric lines simultaneously and thereby to obtain 
the atmospheric conditions in the photosphere and the chromosphere with a single inversion. The minimization algorithm is based on the genetic algorithm PIKAIA 
\cite[]{charbonneau1995}, well suited for finding the global minimum in the large parameter space resulting from this complex conditions, independent of the selection 
of the initial guess values. \helixp{} can also take into account the Paschen-Back effect \cite[]{sasso2006} and the Hanle effect \cite[see HAZEL code by][]{asensioramos2008} 
in the He~\textsc{i} infrared triplet. However, for the application in this paper \helixp{} only considers the Zeeman of the spectral lines
in Table~\ref{table:lines} under the assumption of L-S coupling. The flexible wavelength weighting scheme available in \helixp{} was selected to 
match the scheme of VFISV and ASP/HAO.\\

Before proceeding with the comparison of the inversion codes described above, there are a few things that must be taken care of. The first one is to
make sure that the codes are compatible as far as their synthesis modules are concerned. In other words, that for the same set of $\ve{M}$ parameters 
(Eq.~\ref{equation:m}), the three codes yield the same Stokes vector $\ve{I}=f(\ve{M})$. This exercise revealed that, while ASP/HAO (Sect.~\ref{section:asphao}) 
and \helixp (Sect.~\ref{section:helix}) agreed almost perfectly, the VFISV code (Sect.~\ref{section:vfisv}) yielded results that differ from the others' 
at the $10^{-3}$ level. After tracking down the source of discrepancies, an error was found in VFISV when computing the imaginary part of the Voigt-Faraday 
function needed for the calculation of the magneto-optical effects (i.e., anti-symmetric part of the absorption matrix). It turned out that VFISV was 
needlessly multiplying this function by a factor of two. After correcting this bug, all three codes agreed at the $10^{-6}$ level\footnote{This bug was found 
during our first ISSI meeting in Bern in January 2010. It was subsequently reported to the HMI team in Stanford and High Altitude Observatory, who 
corrected the bug on their version of VFISV before any HMI data had been analyzed. This is mentioned in \citet{rebeca2014}.}.\\

The next step was to agree on the inversion set-up. In this work we are only interested in the effect of employing different minimization algorithms, 
thus we will not consider instrumental effects such as limited spectral/spatial sampling, photon noise, etc. These questions have already been addressed 
elsewhere \citep[see for instance][]{david2010imax,borrero2007,borrero2011}. This implies that physical parameters such as macroturbulent velocity 
$v_{\rm mac}$ and magnetic filling factor $\alpha_{\rm mag}$ (commonly used in inversions) were not considered, and we only accounted for the physical 
parameters contained in \ve{M} (Eq.~\ref{equation:m}).\\

Finally, since we have not considered the effect of photon noise we give the same weights to all four Stokes parameters during the inversion: $w_i=w_q=w_u=w_v=1$. 
These weights appear in the $\chi^2$-merit function that is being minimized during the inversion process\\

\begin{equation}
\chi^2 = \frac{1}{4N-L} \sum\limits_{i=1}^{4}\sum\limits_{j=1}^{N} \left\{I_i^{\rm obs}(\lambda_j)-I_i^{\rm me}(\lambda_j,\ve{M})\right\}^2 w_i^2 \;,
\label{equation:chisq}
\end{equation}

\noindent where the index $i=1,...,4$ refers to the four components of the Stokes vector ($I_1=I, I_2=Q$, etc) and the index $j=1,...,N$ runs for all
wavelength positions ($N$ is the total number of wavelength positions). $I_i^{\rm obs}$ and $I_i^{\rm me}$ refer to the observed and Milne-Eddington Stokes
profiles, respectively, with the latter being a function of the physical parameters in $\ve{M}$ (Eq.~\ref{equation:m}). $L$ in Eq.~\ref{equation:chisq} 
refers to the total number of free parameters in $\ve{M}$. Strictly speaking, \helixp{} does not minimize $\chi^2$ but instead the genetic algorithm 
maximizes the so-called fitness function, which has been set-up to coincide with the inverse of $\chi^2$ (see Sect.~\ref{section:helix}).\\

\section{Comparison between inversion codes}
\label{section:comparison_ic_ic}

The analytic function $f$ that solves the radiative transfer equation $\ve{I}=f(\ve{M})$ (Eq.~\ref{equation:rtet}) is non-linear and represents, even under 
the simplifications of the Milne-Eddington approximation, a transcendental equation. For this reason, the inverse problem, that is, obtaining the physical 
parameters of the solar atmosphere from observations of the Stokes vector, cannot be analytically attained. Instead one must resort to fitting algorithms 
(e.g., merit function minimization; see Eq.~\ref{equation:chisq}). This usually involves the use of non-linear iterative techniques that can sometimes fall into local-minima, or present uniqueness 
problems in the solution. To address this problem and to study to what accuracy the physical parameters of the solar atmosphere can be inferred, we will
perform in this section a comparison of the solutions obtained with the three different M-E inversion codes described in Section~\ref{section:description_ic}.
These codes employ rather different optimization algorithms, and therefore they are very suitable for our purpose. The comparison will be made by 
inverting the Stokes profiles synthesized (Sect.~\ref{section:description_synthesis}) from three-dimensional MHD simulations (Sect.~\ref{section:description_mhd}).\\

\begin{figure*}
\begin{center}
\begin{tabular}{ccc}
\includegraphics[width=5.5cm]{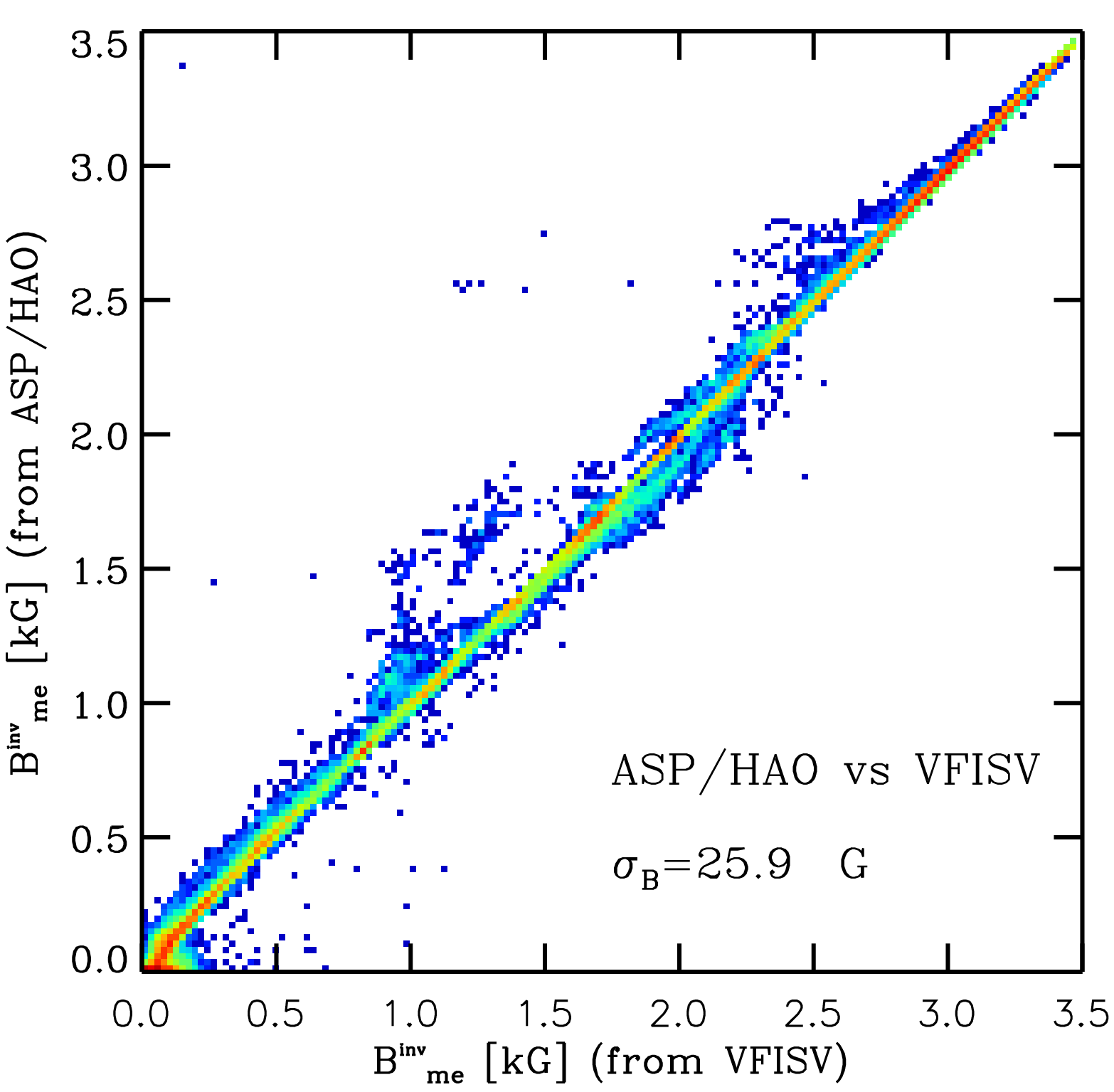} &
\includegraphics[width=5.5cm]{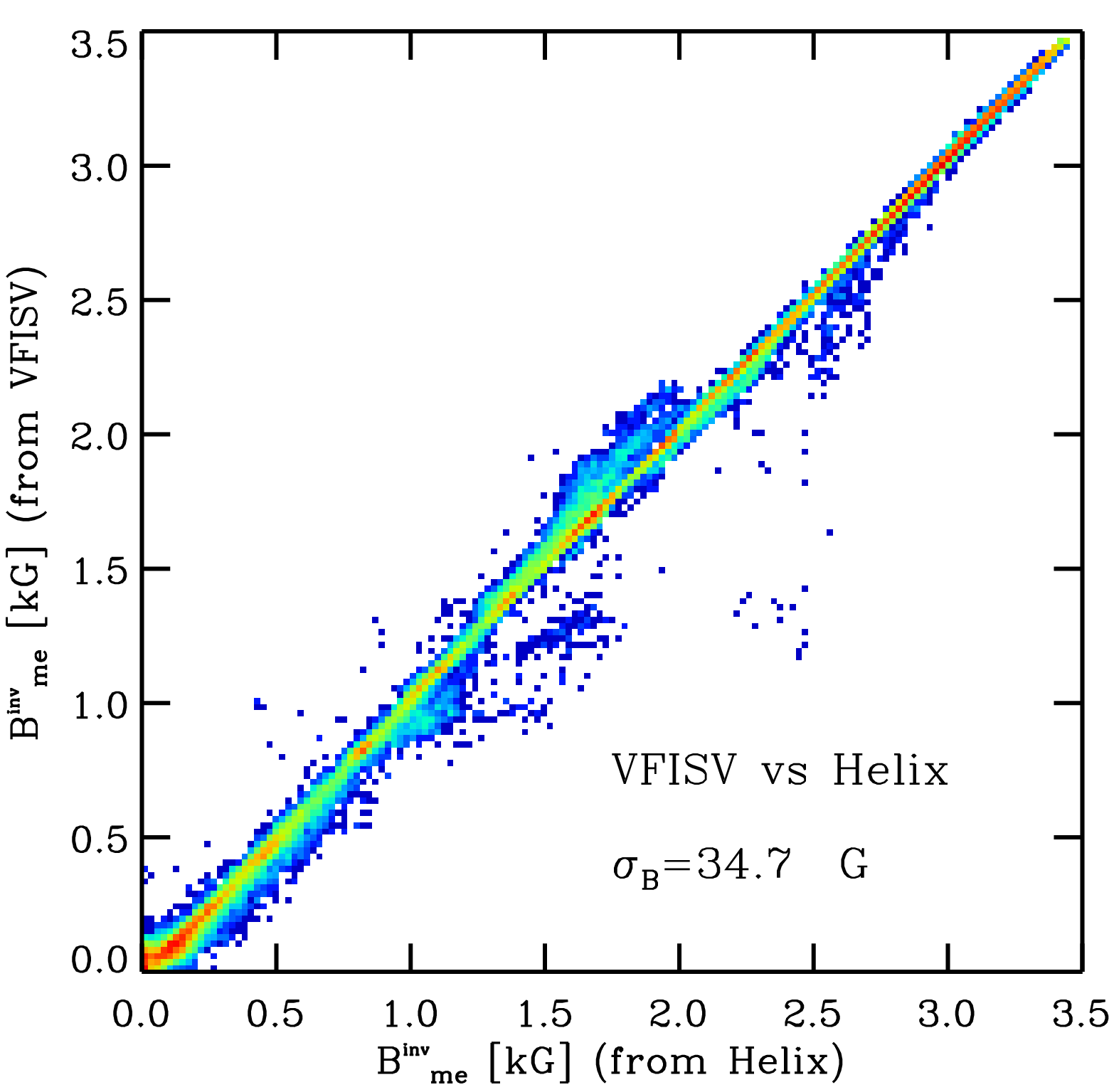} &
\includegraphics[width=5.5cm]{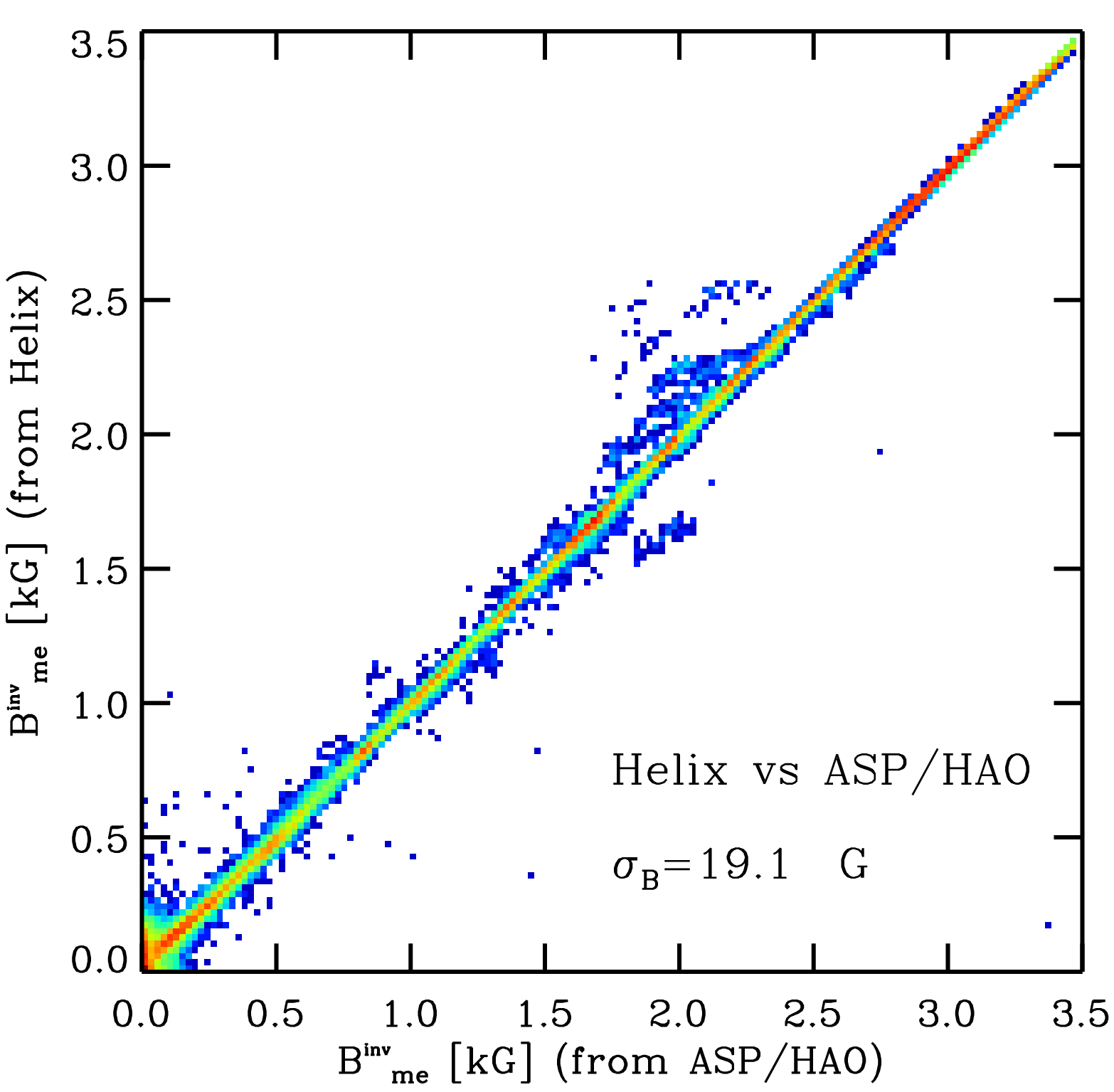} \\
\includegraphics[width=5.5cm]{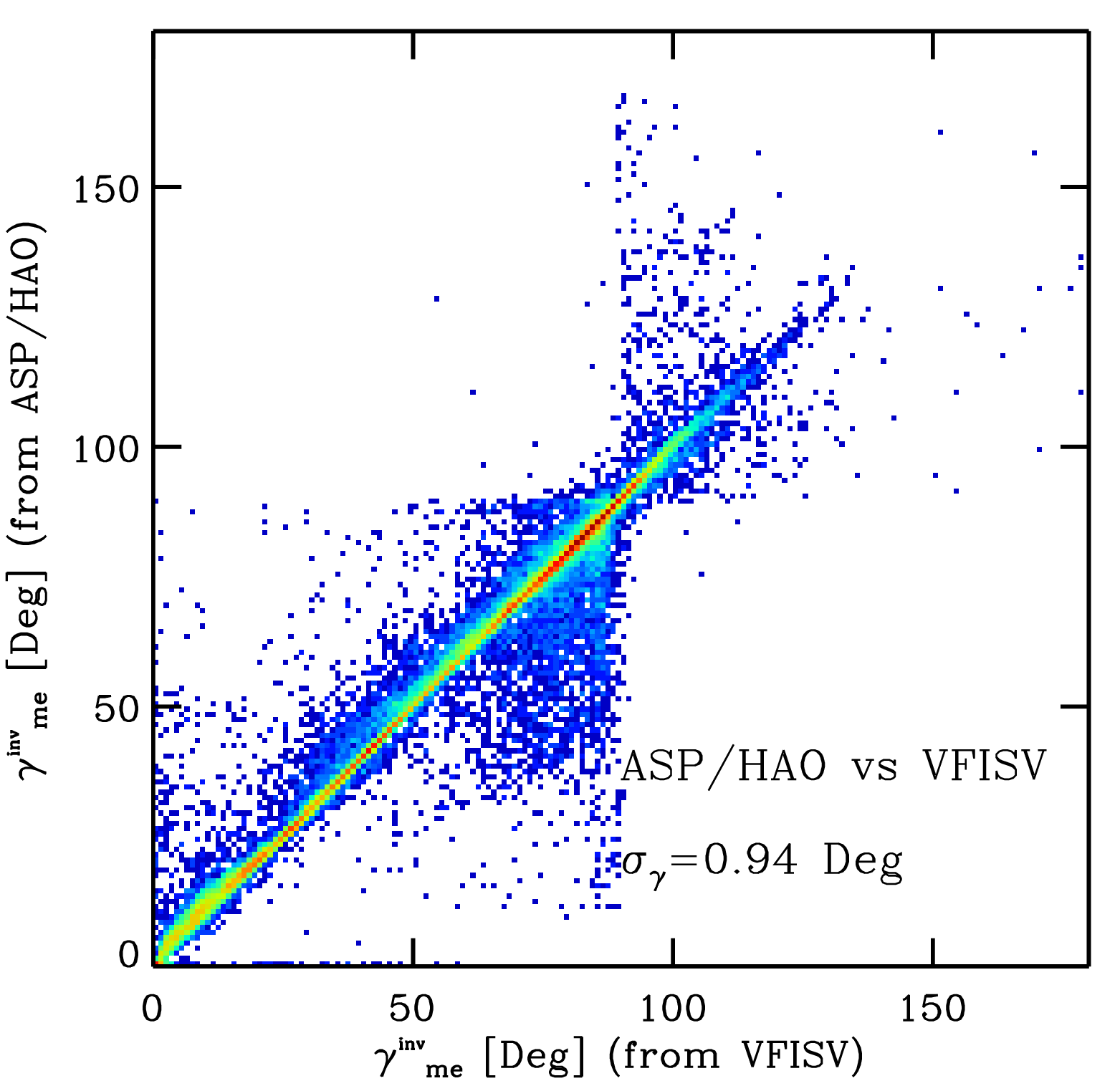} &
\includegraphics[width=5.5cm]{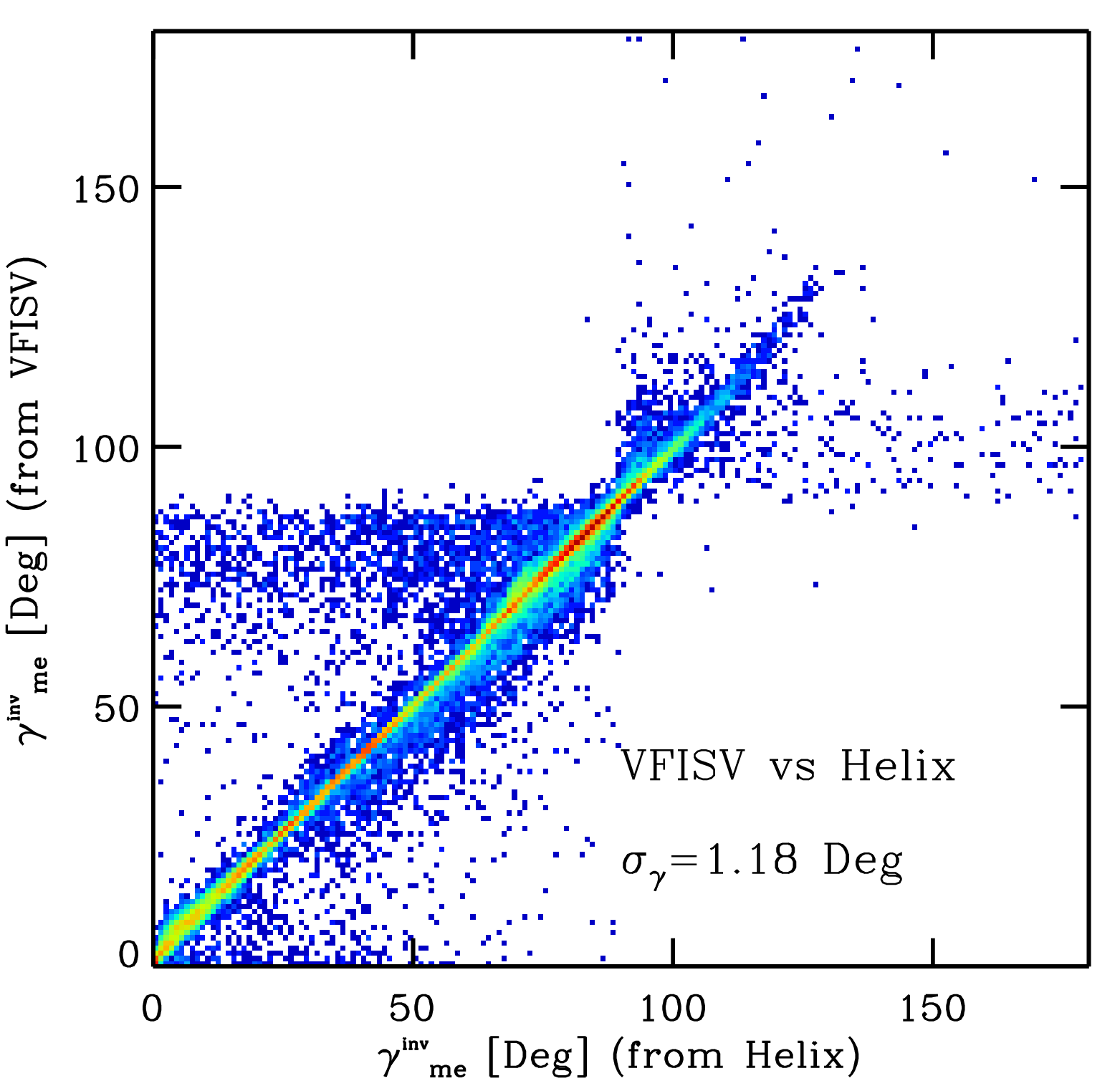} &
\includegraphics[width=5.5cm]{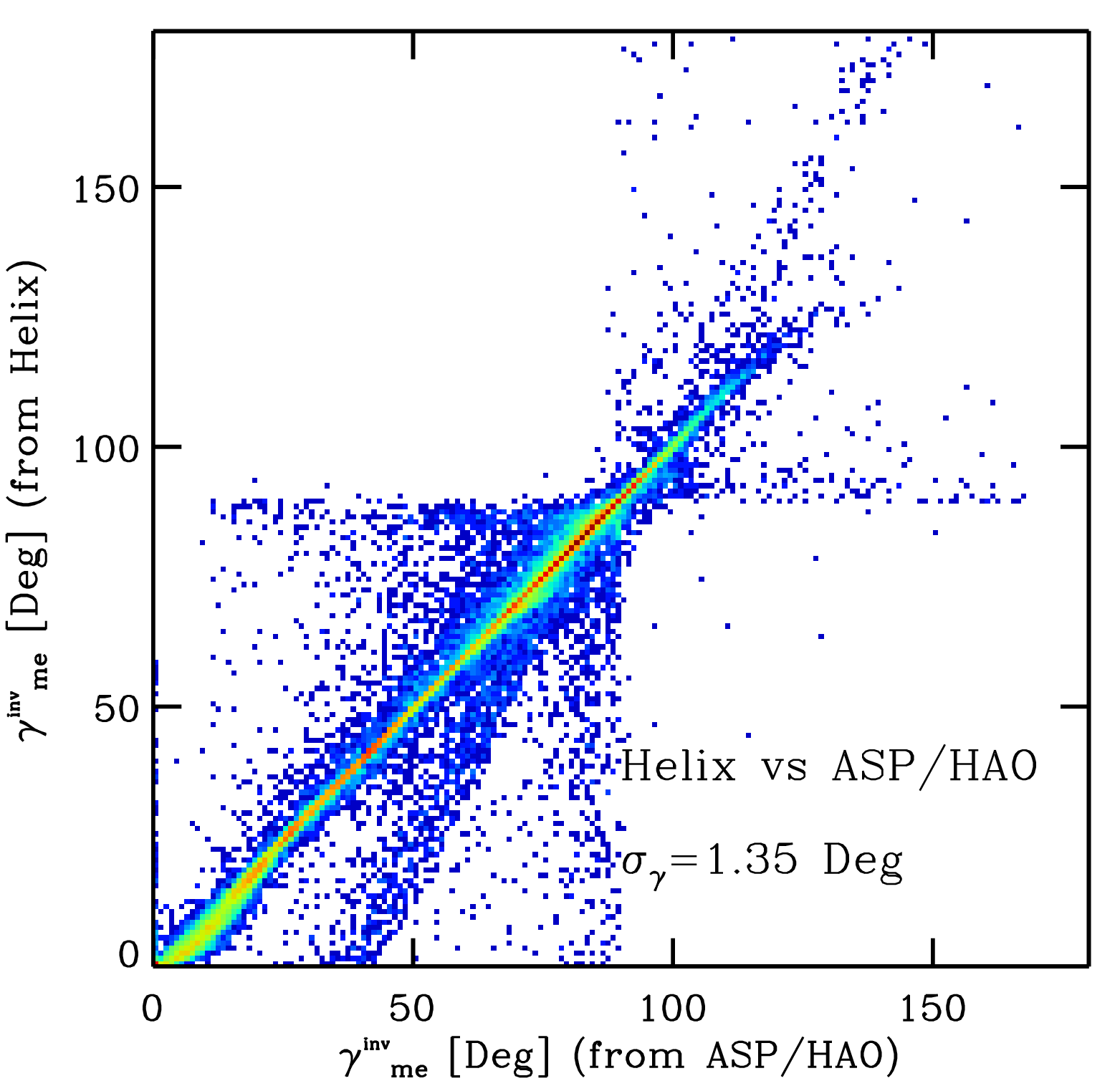} \\
\includegraphics[width=5.5cm]{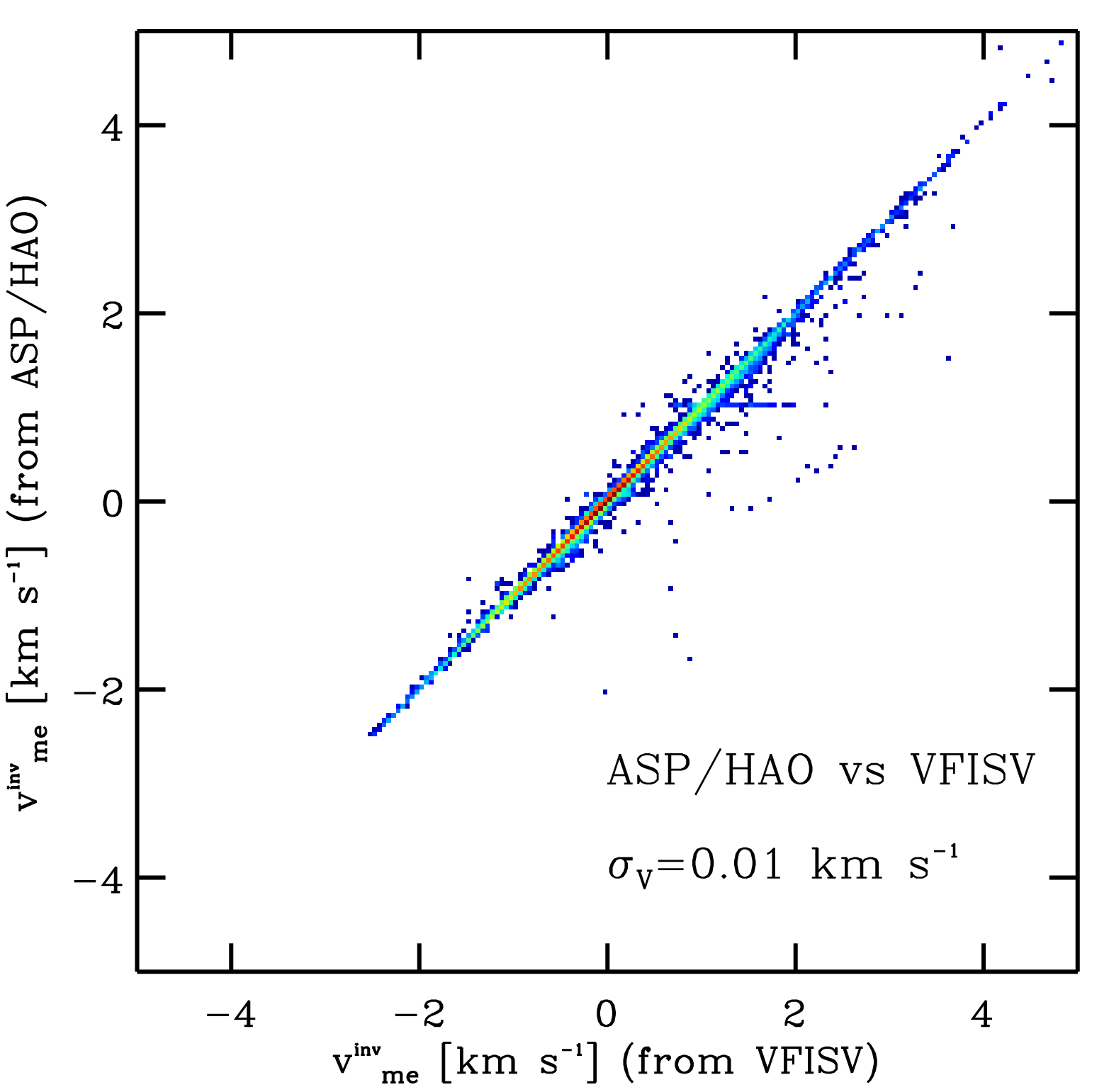} &
\includegraphics[width=5.5cm]{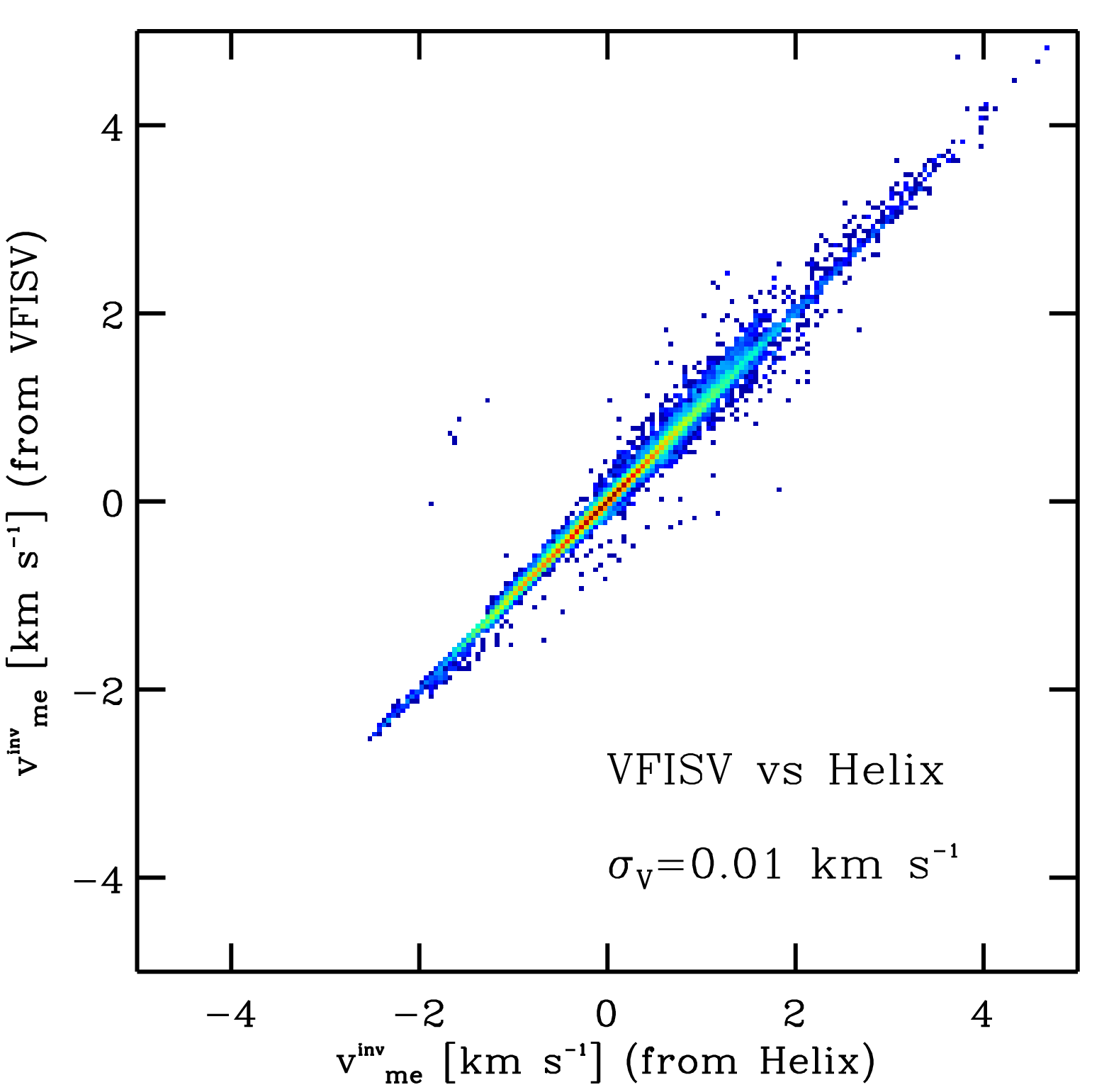} &
\includegraphics[width=5.5cm]{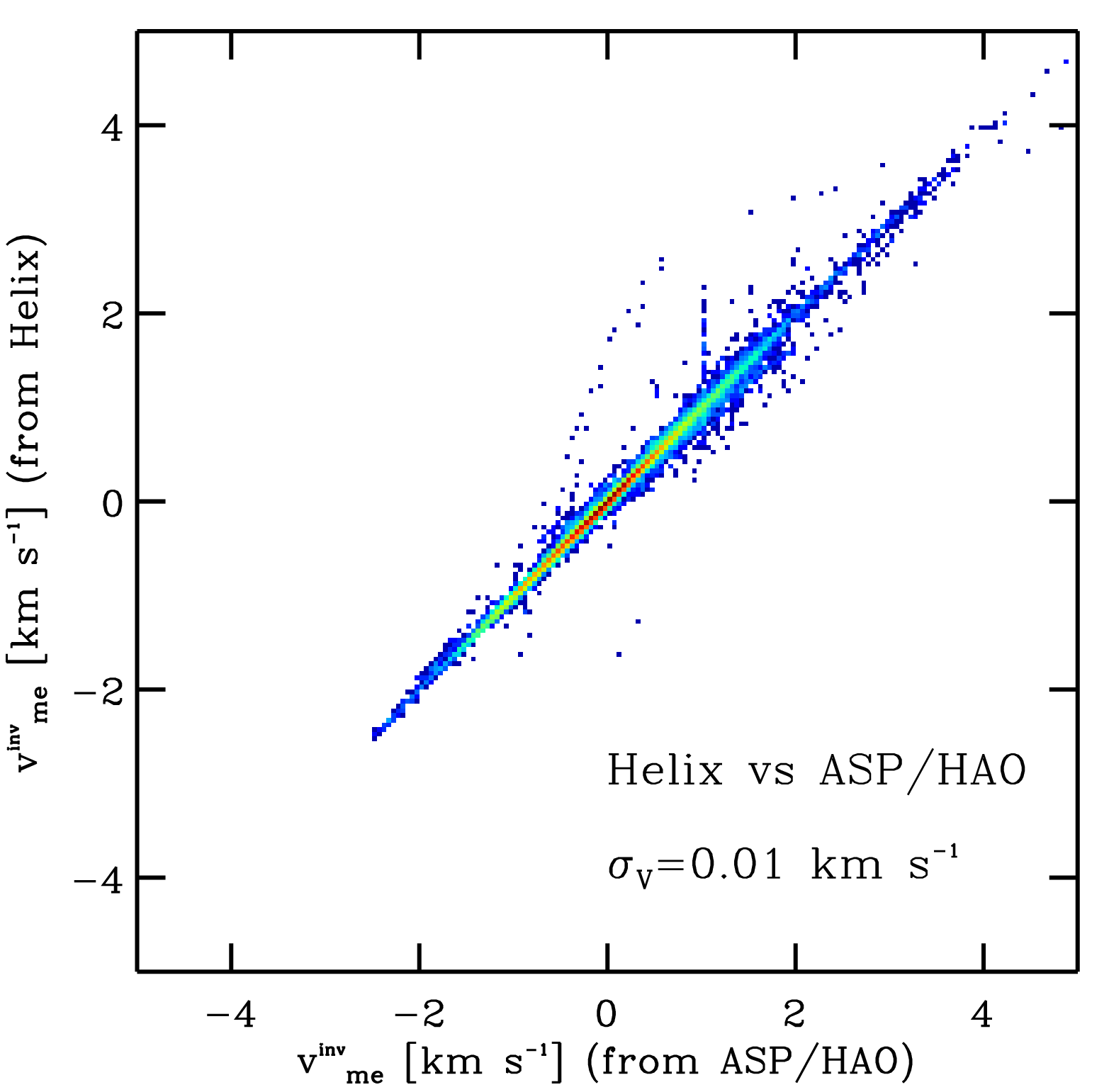} \\
\end{tabular}
\caption{Comparison between the physical parameters obtained through M-E inversion codes: magnetic field strength $B$ (top row), inclination of
the magnetic field vector with respect to the observer's line-of-sight $\gamma$ (middle row), and line-of-sight velocity ${\rm v}_{\rm los}$ (bottom row). The
comparison between the ASP/HAO code and VFISV is displayed in the left column, while VFISV-\helixp and \helixp-ASP/HAO are compared on the middle and right columns,
respectively. The color scale indicates a logarithmic density plot, where red regions contain about 3000 points from Fig.~\ref{figure:simul_highlight}.}
\label{figure:ic_ic}
\end{center}
\end{figure*}

The results from the inversion of $2^{16}$ points (see Sect.~\ref{section:description_synthesis}) using VFISV, \helixp, and the ASP/HAO inversion 
are presented in Figure~\ref{figure:ic_ic} for the magnetic field strength $B$ (top row), for the inclination 
of the magnetic vector with respect to the observer's line-of-sight $\gamma$ (middle row), and finally, for the 
line-of-sight component of the velocity ${\rm v}_{\rm los}$ (bottom row). The left column in this figure compares the ASP/HAO code with VFISV,
the middle column compares VFISV with \helixp, and the right column compares \helixp with the ASP/HAO code. All panels in Fig.~\ref{figure:ic_ic} are 
logarithmic density plots, with red and blue colors indicating regions of high ($\approx 10^3$) and low ($\approx 1$) density, respectively. In addition, each panel provides 
the standard deviation $\sigma$ of the difference between the results of different codes for $B$, $\gamma$, and ${\rm v}_{\rm los}$. The best agreement in the 
magnetic field strength $B$ is found between the ASP/HAO and the \helixp code, while for the inclination $\gamma$, VFISV and ASP/HAO present the most similar results. 
In the case of the line-of-sight velocity ${\rm v}_{\rm los}$ all codes agree equally well. It is important to mention here that ASP/HAO presented a few hundred 
non-convergent points that have not been considered in this comparison. These points appear on the center of granules (Fig.~\ref{figure:simul_highlight}). Here 
the ASP code yielded by design when no solution was reached zero magnetic field, $B=0$, and an inclination that was either $\gamma=0\deg$ or $\gamma=180\deg$. 
The suspected cause of this condition is a failure of the genetic initialization algorithm to provide a good initial guess.\\

In general, the three codes agree with each other at the level of $\sigma_B \leq 35$ (Gauss), $\sigma_\gamma \leq 1.2\deg$, and $\sigma_{\rm v} \leq 10$\ms. 
These values are comparable or even better than those reported in \citet{borrero2007,david2010fixedtau,fleck2011,couvidat2012} where similar comparisons to ours
were carried out but employing simpler methods, such as, center of gravity, weak-field approximation, MDI-like algorithms, etc. A direct comparison with those works
cannot be done, however, because some of them add photon noise, perform spectral degradation, or analyze ideal (e.g., symmetric) profiles, etc. At any rate, our 
results make us confident that, despite comments to the contrary from those who are not users of inversion codes, lack of convergence does not seem 
to play a major role in the inferences done through M-E inversion codes. The question of uniqueness will be addressed in Section~\ref{section:comparison_ic_mhd}, 
where we will compare the physical parameters derived from the inversion to those from the numerical simuations used in the synthesis (see Sect.~\ref{section:description_synthesis}).\\

It is worth noting that, according to Figure~\ref{figure:ic_ic}, the discrepancies between the inversion codes increase
in those points where $B \approx 1.0-2.0$ kG and $\gamma \geq 75\deg$. Stokes profiles in this region belong to the penumbra (see Fig.~\ref{figure:simul_highlight}),
where the physical quantities undergo rapid changes with $\tau_c$ both in observations \citep{jorge1992ncp} and simulations \citep{borrero2010}.
It is therefore not surprising that M-E codes, that assume that the physical quantities are constant with $\tau_c$ (see Sect.~\ref{section:description_ic}), present larger 
differences here. This effect is not seen in the line-of-sight component of the velocity ${\rm v}_{\rm los}$ (Fig.~\ref{figure:ic_ic}; bottom panels) because we assume that the 
observer looks at the simulation box as if the sunspot was at disk center (see Sect.~\ref{section:description_synthesis}), and therefore the velocities in the
penumbra, which are mainly horizontal in nature (i.e., contained in the $XY$-plane), contribute little to ${\rm v}_{\rm los}$.\\

On the other hand, close to the region where $B \geq 2.5$ kG and $\gamma \leq 15\deg$ results seem to be particularly consistent between the three 
codes. Points in this region are located in the umbra (see Fig.~\ref{figure:simul_highlight}). Unlike the penumbra, the umbra is characterized by smooth
variations of the physical quantities with $\tau_c$, thus resulting in a better agreement between the M-E inversion codes.\\

\section{Comparison between inversion codes and 3D MHD simulations}
\label{section:comparison_ic_mhd}

\subsection{At a fixed optical depth $\tau_c$.}
\label{section:comparison_fixed_tau}

As explained in Section~\ref{section:description_ic}, M-E inversion codes are usually restricted to single-line inversions or to line-pairs
that sample very similar optical depths. Interestingly, even if a single spectral line (or several spectral lines formed very close to each other) 
is inverted, one must always take into account that the concept of \emph{height of formation of a spectral line} is a fuzzy one \citep{jc1996,jorge1996}. 
In photospheric spectral lines (e.g., absorption lines) the continuum and line-wing in Stokes $I$ generally sample deeper layers ($\tau_c \approx 1$) than the line 
core ($\tau_c \approx 10^{-2}-10^{-4}$), but there is always a broad overlap region as a single wavelength cannot be ascribed to any particular optical depth. 
In addition, the aforementioned relationship only holds for Stokes $I$, and it can be very different for Stokes $Q$, $U$, and $V$. Furthermore, the photospheric layers sampled 
by spectral lines strongly depend on the physical parameter measured \citep{basilio1994}. For instance, we do not sense the same layer when measuring
the magnetic field or velocity at $\Delta\lambda=0$ (i.e., zero-crossing) in Stokes $V$. All these effects make it difficult to compare the results obtained
through the application of M-E inversion codes (Sect.~\ref{section:description_ic}) to Stokes profiles synthesized from MHD simulations (Sect.~\ref{section:description_synthesis}).
While the former retrieve a single value the magnetic and kinematic parameters ($B$, $\gamma$, $\phi$, ${\rm v}_{\rm los}$) at each point on the $XY$-plane 
(Fig.~\ref{figure:simul_highlight}), the latter provide the full dependence on optical depth ($B(\tau_c)$, $\gamma(\tau_c)$, $\phi(\tau_c)$, 
${\rm v}_{\rm los}(\tau_c)$). The problem therefore boils down to devising a strategy to compare the results from M-E inversion with those from 3D MHD simulations. Several methods 
have been used in the past.\\

One of these methods considers that the value of the physical parameters inferred from the M-E inversion must correspond to a certain height (either on an optical
depth $\tau_c$, or on a geometrical height $z$) in the atmosphere. To determine that height, one computes the $\tau_c$-dependence of the standard deviation of 
the difference between the MHD model parameters and the constant M-E parameters. The optical depth, $\tau_c^{*}$, where the standard deviation is smallest is then considered to be the
height where the spectral line(s) provide the most accurate information. A very similar approach is to calculate the correlation coefficient between the physical 
parameters from 3D MHD simulations at different heights and those from M-E inversions, and take $\tau_c^{*}$ as the height with the highest correlation \citep{kucera1998,fleck2011}. 
These two approaches have the disadvantage that, because $\tau_c^{*}$ is obtained statistically from an ensemble of very different atmospheres, whatever value is obtained 
for $\tau_c^{*}$, it does not consider that the layers where the spectral line(s) provide most information depends on the physical parameters of the atmosphere 
itself \citep{jc1996,jorge1996}.\\

A slightly different approach is to determine, for each point on the simulation, the height at which the results from the M-E inversion and
3D MHD simulations coincide. This approach has the advantage that it allows us to study changes in the layer where the spectral line(s) provides most
information \citep[see][Fig.~6]{david2010fixedtau}. On the other hand, a shortcoming of this method is that it assumes that the error is zero,
as it takes the height at which M-E inversions and numerical simulations give the same result.\\

\begin{figure*}
\begin{center}
\begin{tabular}{ccc}
\includegraphics[width=5.5cm]{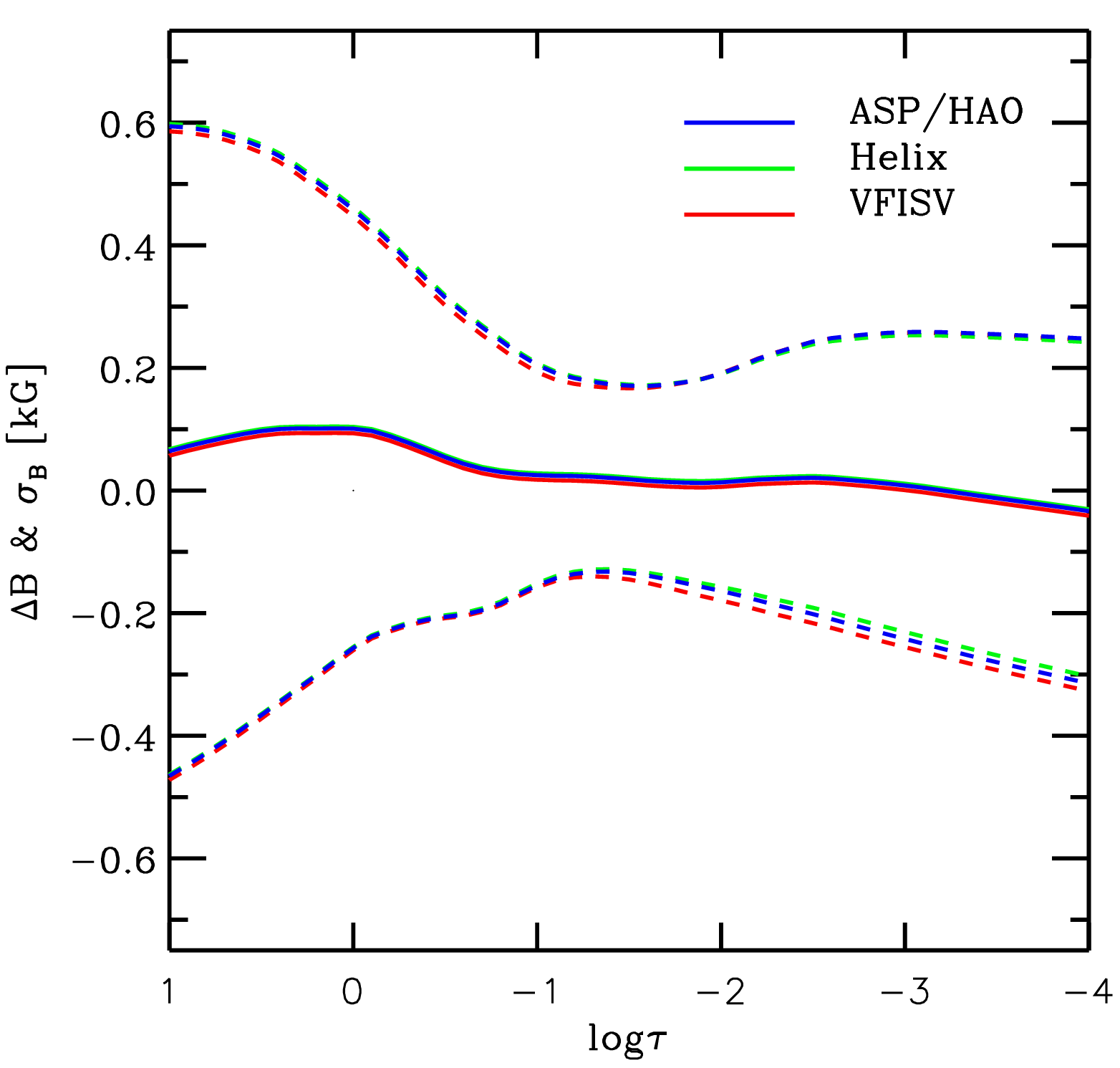} &
\includegraphics[width=5.5cm]{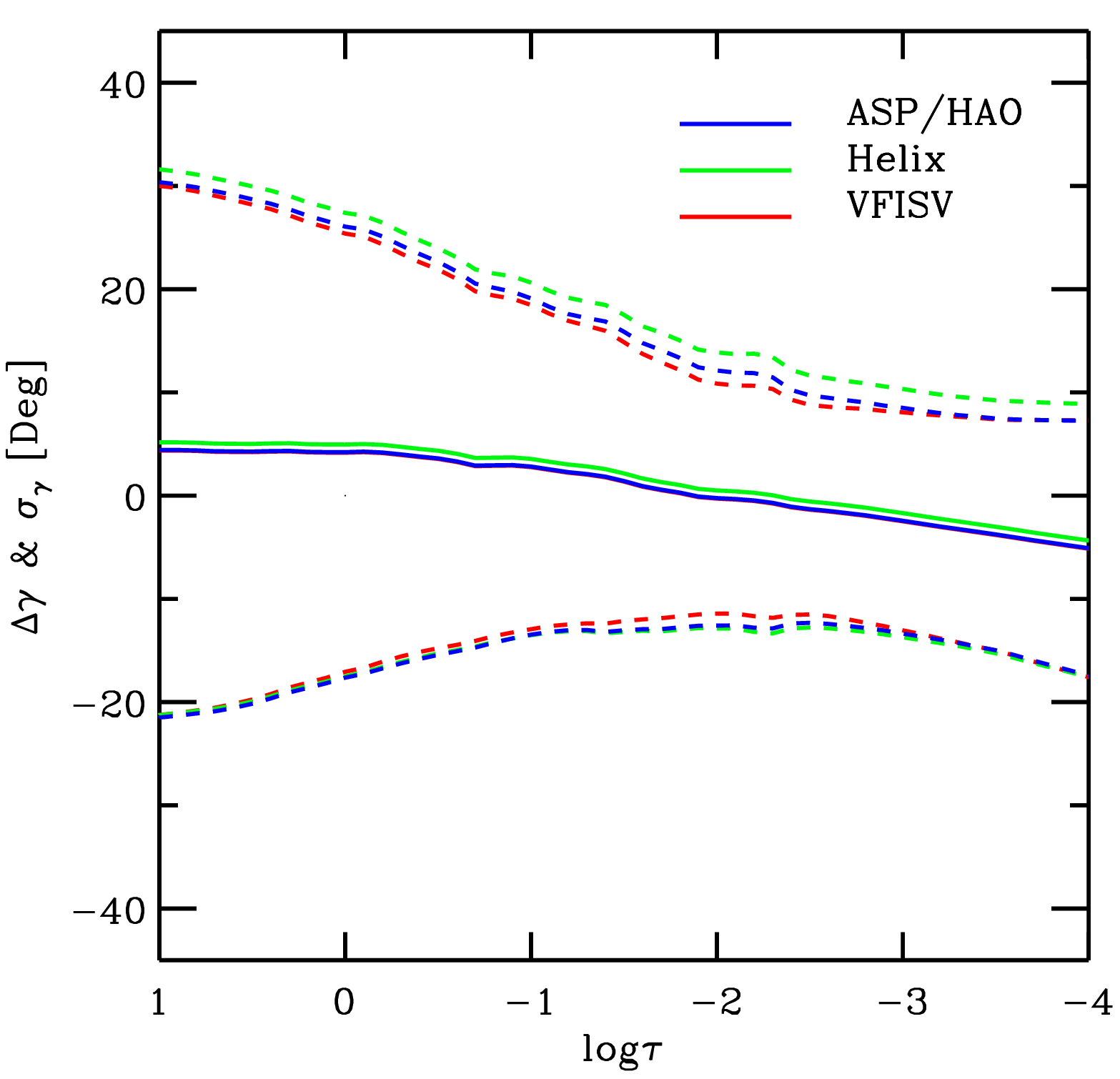} &
\includegraphics[width=5.5cm]{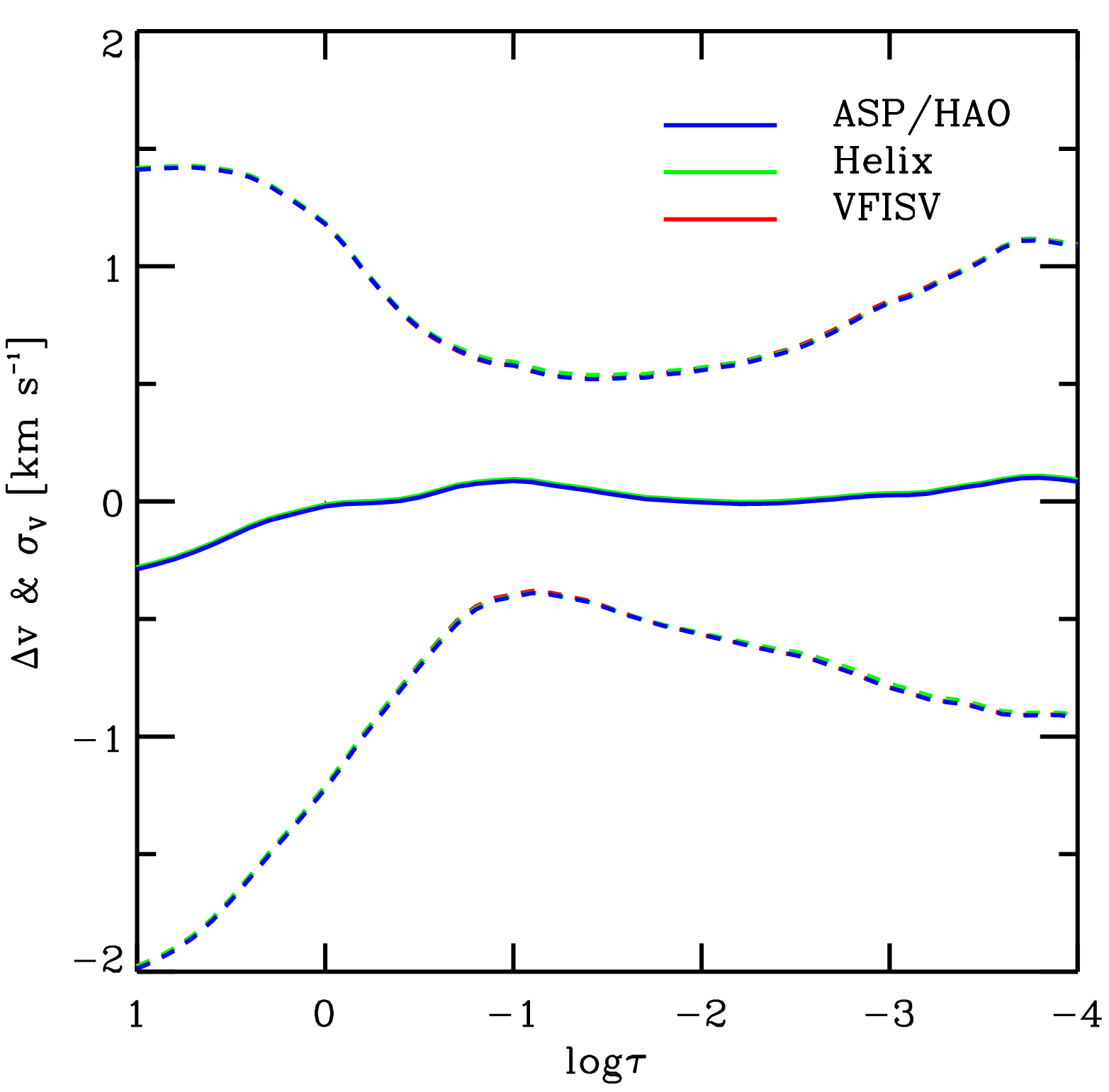} \\
\end{tabular}
\caption{Mean differences (solid) and standard deviation (dashed) as a function of the optical depth $\tau_c$ between the original 
(MHD simulations) and inferred (M-E inversion) physical parameters: magnetic field strength $B$ (left), inclination of the magnetic 
field with respect to the line-of-sight $\gamma$ (middle), and line-of-sight component of the velocity ${\rm v}_{\rm los}$ (left). Red, blue, 
and green colors correspond to the VFISV (Sect.~\ref{section:vfisv}), ASP/HAO (Sect.~\ref{section:asphao}), and \helixp (Sect.~\ref{section:helix}) 
inversion codes, respectively.}
\label{figure:sigma_tau}
\end{center}
\end{figure*}

In this work we are interested in investigating to which accuracy M-E inversion codes can infer the magnetic and kinematic properties of
the solar atmosphere, and therefore we will follow the first approach described above, that is, we determine a standard deviation
between the M-E results and the MHD simulations at each optical depth ($\tau_c$) and for each physical parameter we are interested in: $B$, $\gamma$,
and ${\rm v}_{\rm los}$. The result of this process is presented in Figure~\ref{figure:sigma_tau}. In this figure, the solid-color lines represent
the mean of the differences between the M-E inversions and the MHD simulations: $\Delta B$ (left column), $\Delta\gamma$ (middle column),
and $\Delta v$ (right row). These curves allow us to see systematic differences between the inferred (through M-E inversions) and original
(from MHD simulations) parameters. The dashed-color lines indicate the standard deviation between the two aforementioned values. From these 
plots we have determined that the optical depth, $\tau_c^{*}$, where the differences between the inferred vales and the original ones are smallest, are: 
$\log\tau_c^{*} \approx -1.4$ for $B$, $\log\tau_c^{*} \approx -1.7$ for $\gamma$, and $\log\tau_c^{*} \approx -1.0$ for ${\rm v}_{\rm los}$. The fact that  $\tau_c^{*}$
is different for each physical parameter strongly supports the idea that a spectral line provides information about different photospheric
layers depending on the physical parameter measured \citep{basilio1994}.\\

\begin{figure*}
\begin{center}
\begin{tabular}{ccc}
\includegraphics[width=5.5cm]{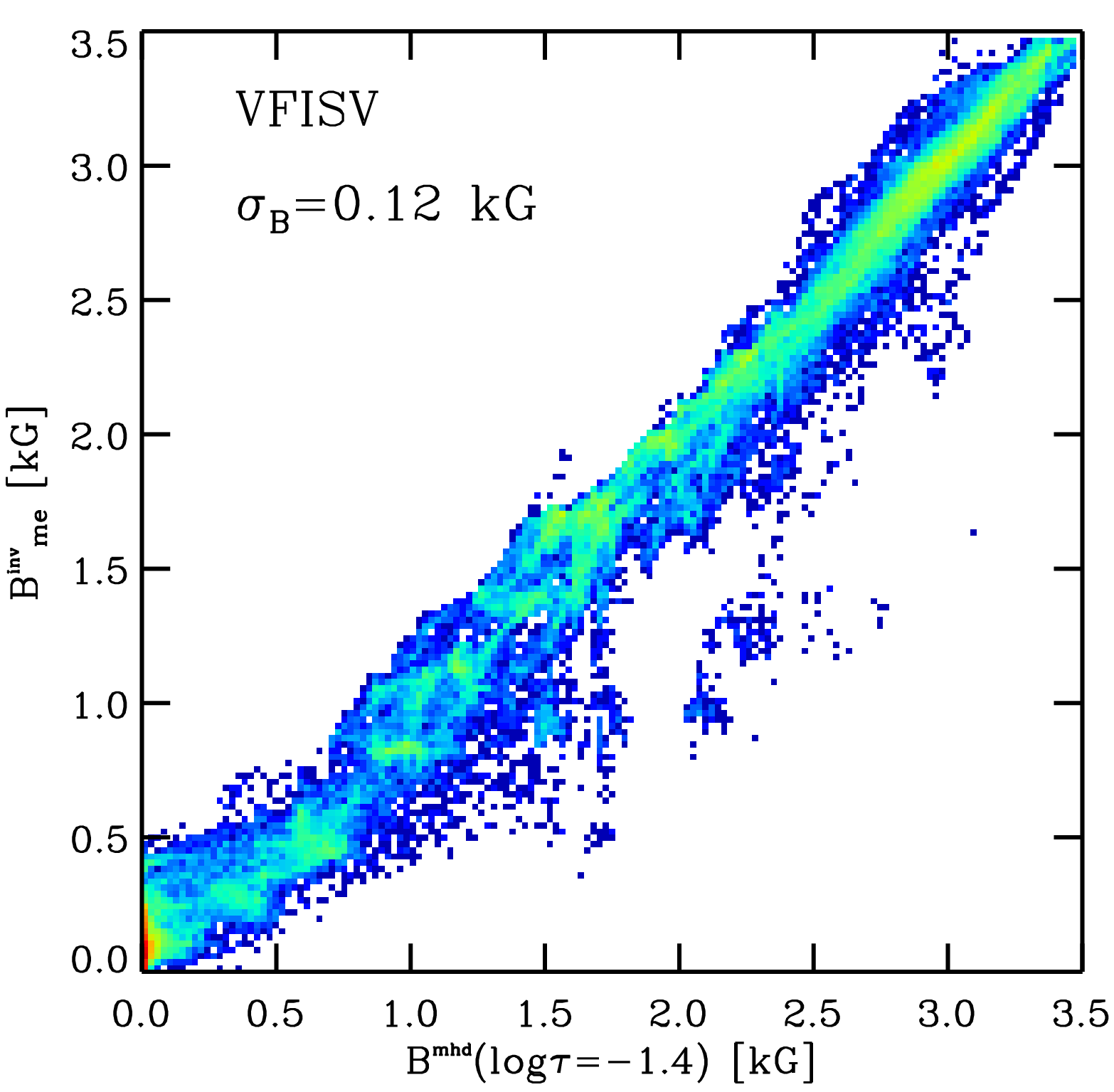} &
\includegraphics[width=5.5cm]{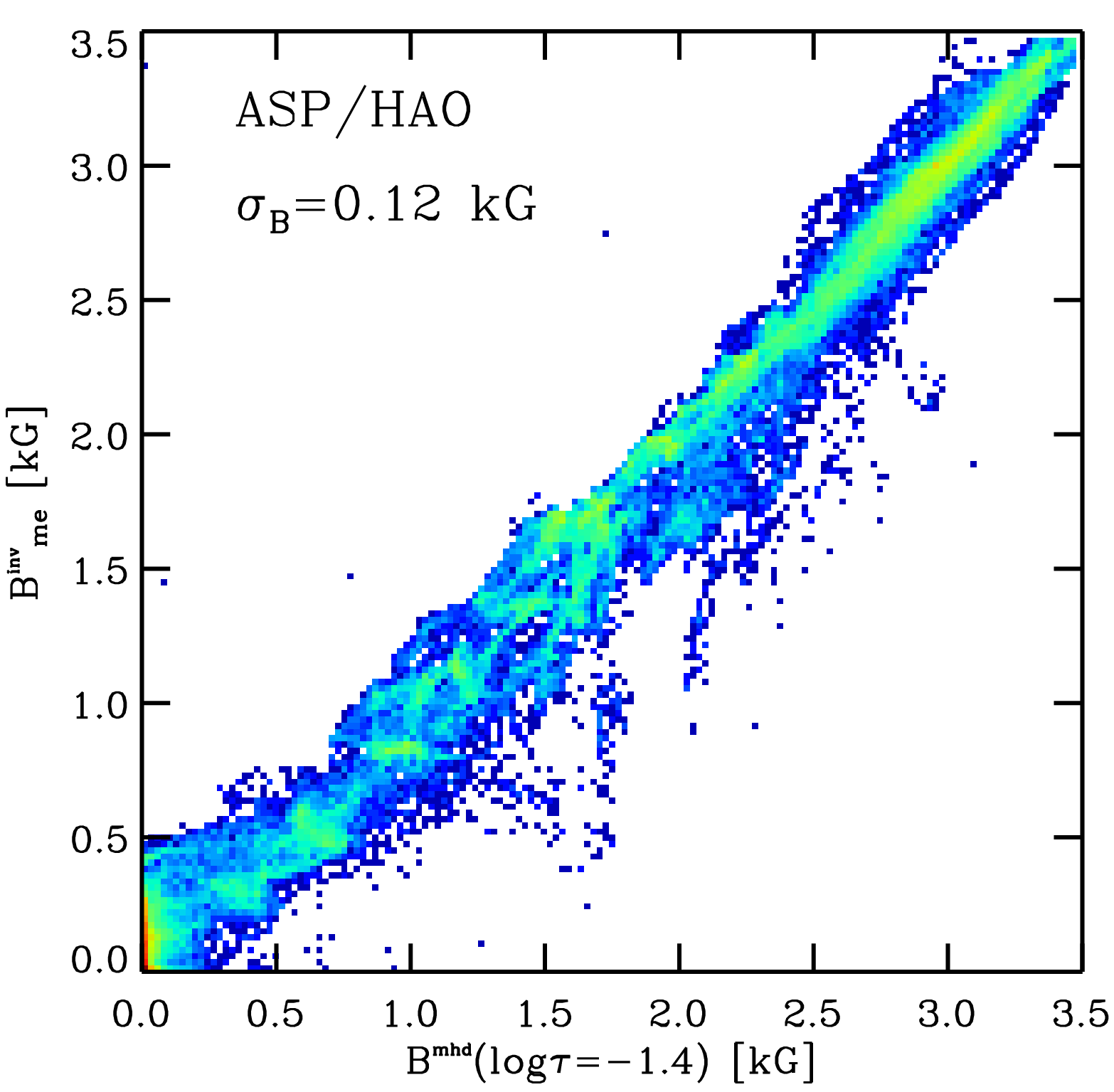} &
\includegraphics[width=5.5cm]{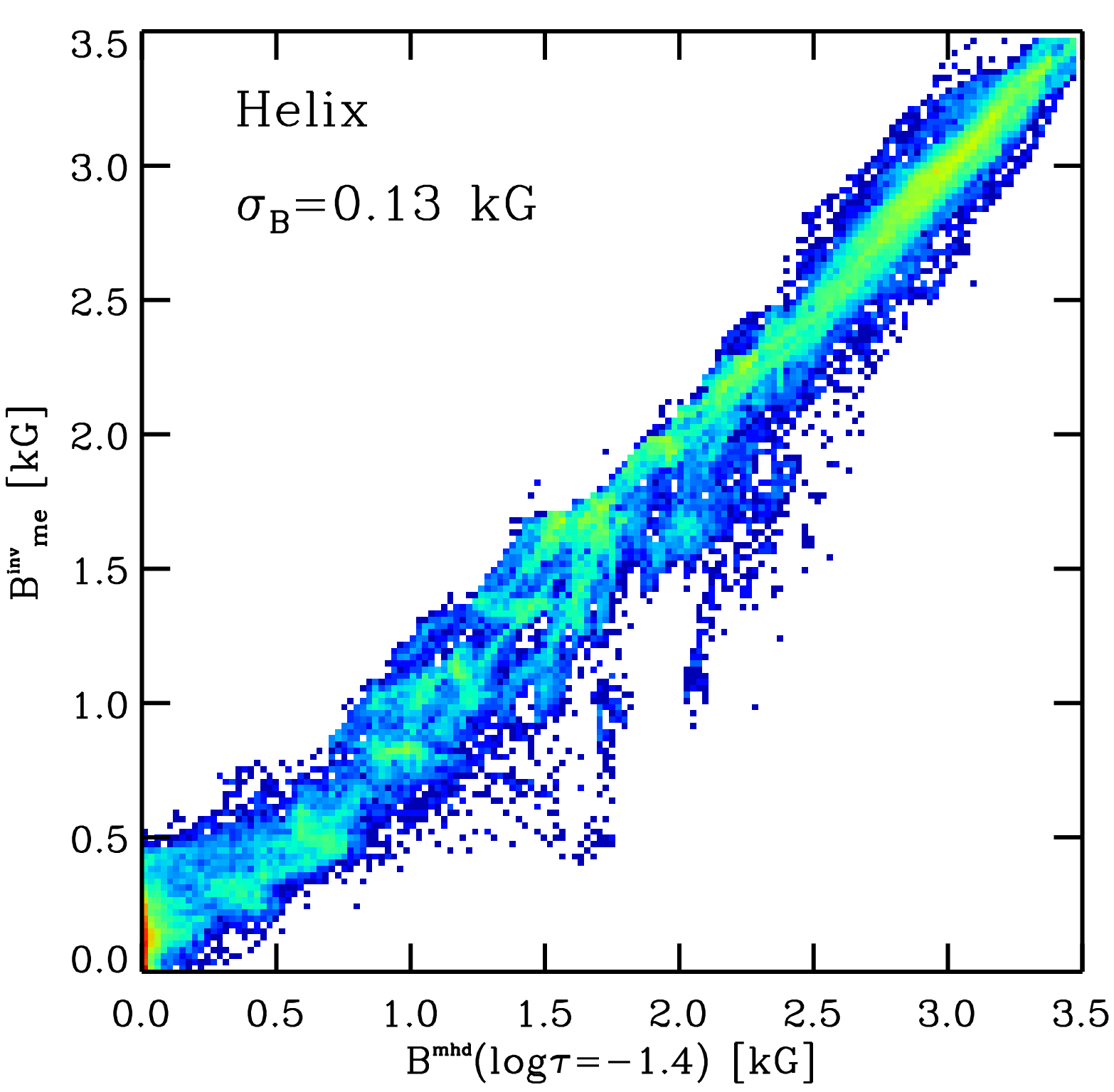} \\
\includegraphics[width=5.5cm]{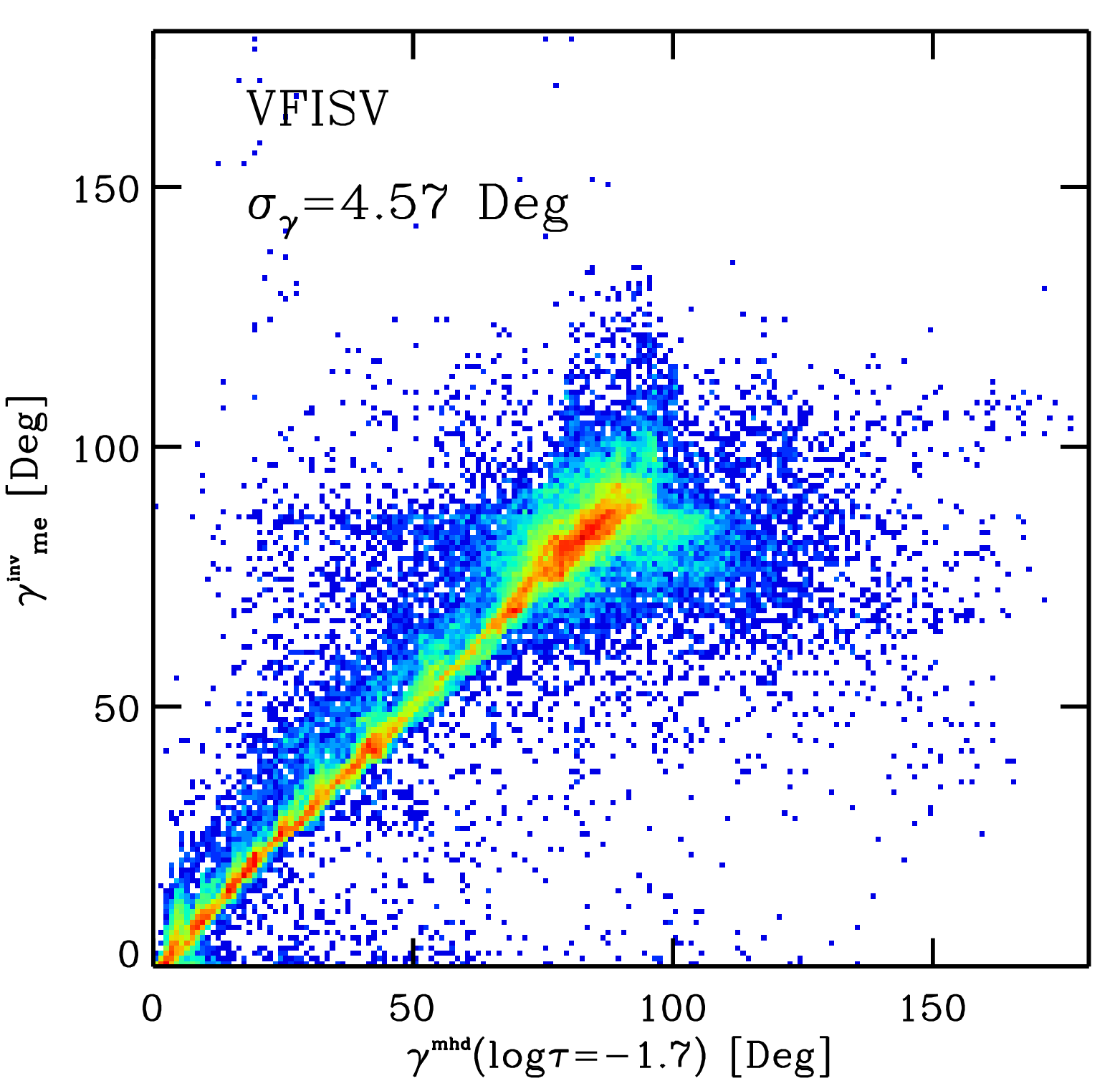} &
\includegraphics[width=5.5cm]{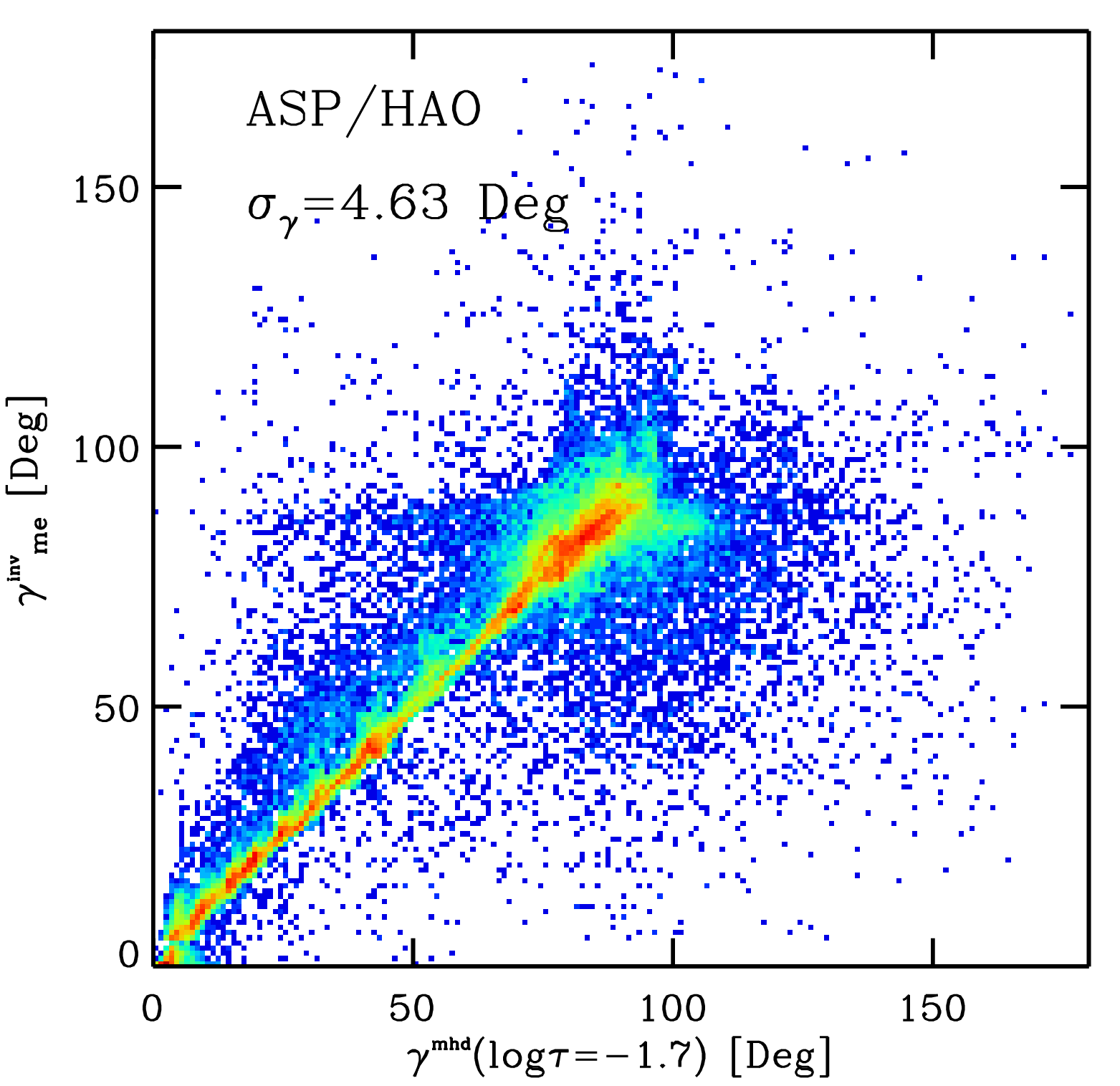} &
\includegraphics[width=5.5cm]{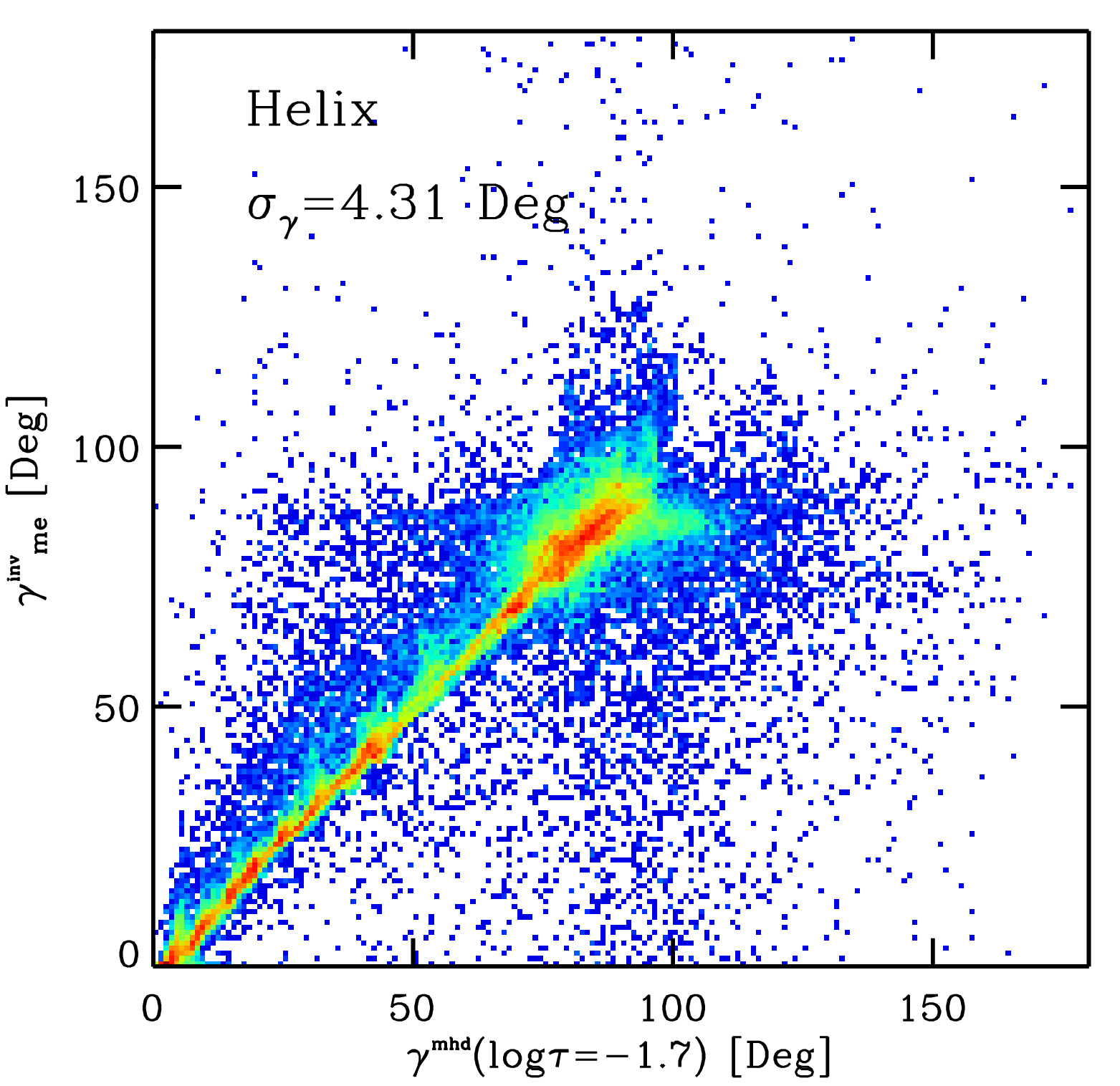} \\
\includegraphics[width=5.5cm]{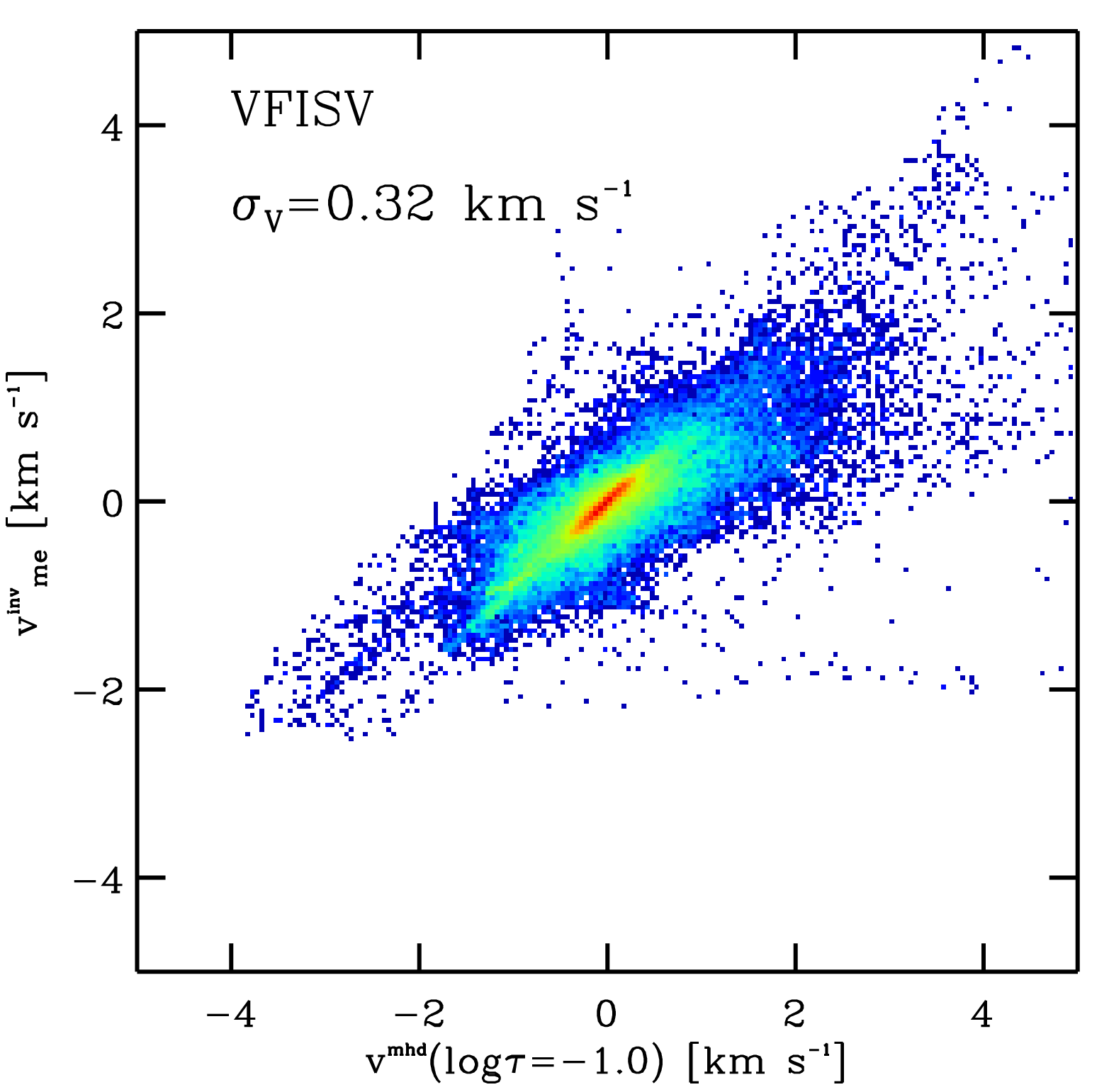} &
\includegraphics[width=5.5cm]{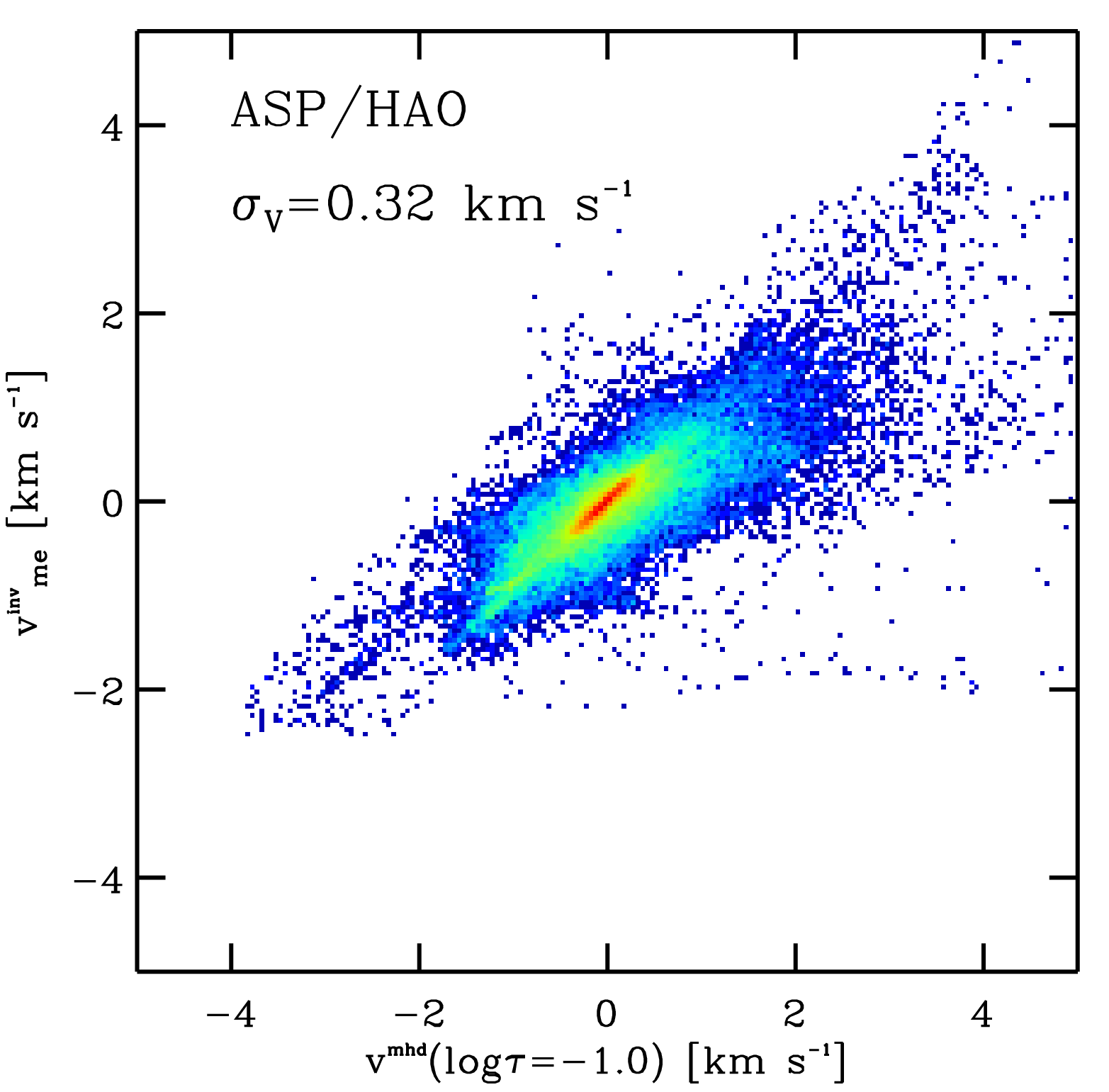} &
\includegraphics[width=5.5cm]{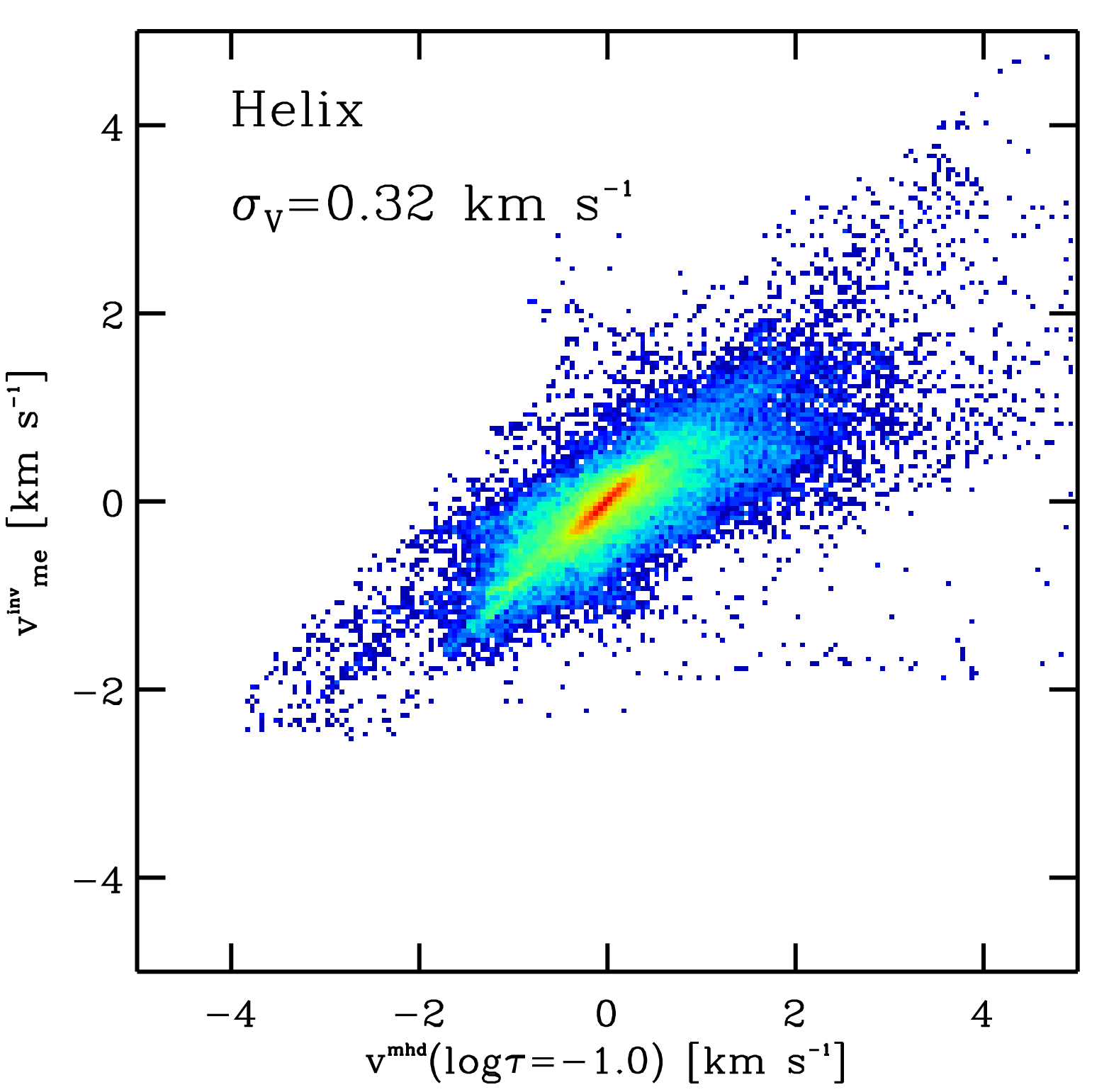} \\
\end{tabular}
\caption{Scatter-density plots of the physical parameters inferred through M-E inversion codes (vertical axis) and the original values
from the 3D MHD numerical simulations (horizontal axis). The values from the simulations are taken at $\log\tau_c^{*} = -1.4, -1.7, -1.0$
for $B$ (top rows), $\gamma$ (middle rows), and ${\rm v}_{\rm los}$ (bottom rows), respectively. Left column corresponds to VFISV, middle one to
ASP/HAO, and right one to \helixp. Standard deviations $\sigma$ are also indicated for each physical parameter and for each inversion code.}
\label{figure:ic_mhd_fixed}
\end{center}
\end{figure*}

Figure~\ref{figure:ic_mhd_fixed} presents similar scatter-density plots to those in Fig.~\ref{figure:ic_ic} but comparing the results
obtained through the M-E inversions with those from the 3D MHD numerical simulations at the heights of maximum information, $\tau_c^{*}$, 
that have just been determined. From this figure we obtained the following errors: $\sigma_B < 130$ G (magnetic field strength), $\sigma_\gamma < 5\deg$ 
(inclination of the magnetic field vector with respect to the observer's line of sight), $\sigma_{\rm v} < 320$\ms (line-of-sight component of the velocity). 
As previously observed and explained (Section~\ref{section:comparison_ic_ic}; Fig.~\ref{figure:ic_ic}) the errors increase in the penumbra, but decrease in the umbra.\\

\subsection{Using Response Functions}
\label{section:comparison_rf}

As mentioned in Section~\ref{section:comparison_fixed_tau}, it is not possible to assign a single optical depth to the measurement
of a particular physical parameter, even if the measurement is done at a single wavelength. The proper way to compare the $\tau_c$-independent
results from M-E inversions with the $\tau_c$-dependent values from MHD simulations, is to compare (at every point in the $XY$-plane in 
Fig.~\ref{figure:simul_highlight}) the former with an average of the latter. We define this average as\\

\begin{equation}
\widetilde{\rm X}^{\rm mhd} = \int\limits_{0}^{\infty} w_{\rm x}(\tau_c) {\rm X}^{\rm mhd}(\tau_c) \df \tau_c \;,
\label{equation:xmhd_rf}
\end{equation}

\noindent where $w_{\rm x}(\tau_c)$ is a weighting function that conveys the information as to which atmospheric layers the spectral line(s) is
sensitive to. We emphasize that $w_{\rm x}(\tau_c)$ is different for each physical parameter ${\rm X}$ (e.g., ${\rm X}=B$, ${\rm X}=\gamma$, and 
${\rm X}={\rm v}_{\rm los}$), and for each $(x,y)$ point on the simulation. In this sense, $w_{\rm x}(\tau_c)$ can be understood as a sensitivity 
kernel, akin to those employed in Helioseismology \citep[see][and references therein]{birch2000}. In this work, we consider the following weighting 
function,\\

\begin{equation}
w_{\rm x}(\tau_c) = \frac{\sum\limits_{j=1}^{4}\int\limits \Big\| \frac{\partial I_j(\lambda)}{\partial {\rm X}(\tau_c)} \Big \| \df \lambda}
{\int\limits_{0}^{\infty} \left\{\sum\limits_{j=1}^{4}\int\limits \Big\| \frac{\partial I_j(\lambda)}{\partial {\rm X}(\tau_c)} \Big \| \df \lambda\right\}\df\tau_c} \;,
\label{equation:weight_function}
\end{equation}

\noindent where index $j$ runs for all four components of the Stokes vector (see Sect.~\ref{section:description_synthesis}). The partial derivatives
of $I_j$ (Stokes profiles) with respect to ${\rm X}$ are the so-called {\it Response Functions} \citep{beckers1975rf,caccin1977rf,egidio1977rf}, and can be
interpreted as the changes in the $j$-component of the Stokes vector $\ve{I}(\lambda)$ when a small perturbation is added at some optical depth $\tau_c$.
The weighting functions for \helixp and ASP/HAO codes are obtained through a wavelength integral that includes both spectral lines in Table~\ref{table:lines}.
However, for VFISV, the wavelength integral includes only the second spectral line in this table (see Sect.~\ref{section:vfisv}). We note that the denominator
in Eq.~\ref{equation:weight_function} ensures that $w_{\rm x}(\tau_c)$ is normalized to unity.\\

The absolute value is taken inside the integral in Eq.~\ref{equation:weight_function} because the response function
can be positive at some wavelengths but negative at others. If we perform a straight integral summation, the contribution at different wavelengths might cancel 
out. This situation is not desirable because a negative response is a response after all, and still indicates that the spectral line is sensitive at that 
wavelength for a particular ${\rm X}(\tau_c)$ perturbation.\\

To perform the comparison between M-E inversion codes and MHD simulations according to Eqs.~\ref{equation:xmhd_rf} and ~\ref{equation:weight_function},
we need to calculate the response functions of the Stokes vector with respect to $B$, $\gamma$, and ${\rm v}_{\rm los}$ as a function of wavelength, $\lambda$,
and optical depth, $\tau_c$, for the spectral lines in Table~\ref{table:lines}. This was done, for every point on the $XY$-plane in Fig.~\ref{figure:simul_highlight},
employing the SIR inversion code \citep[][]{basilio1992}.\\ 

\begin{figure*}
\begin{center}
\begin{tabular}{ccc}
\includegraphics[width=5.5cm]{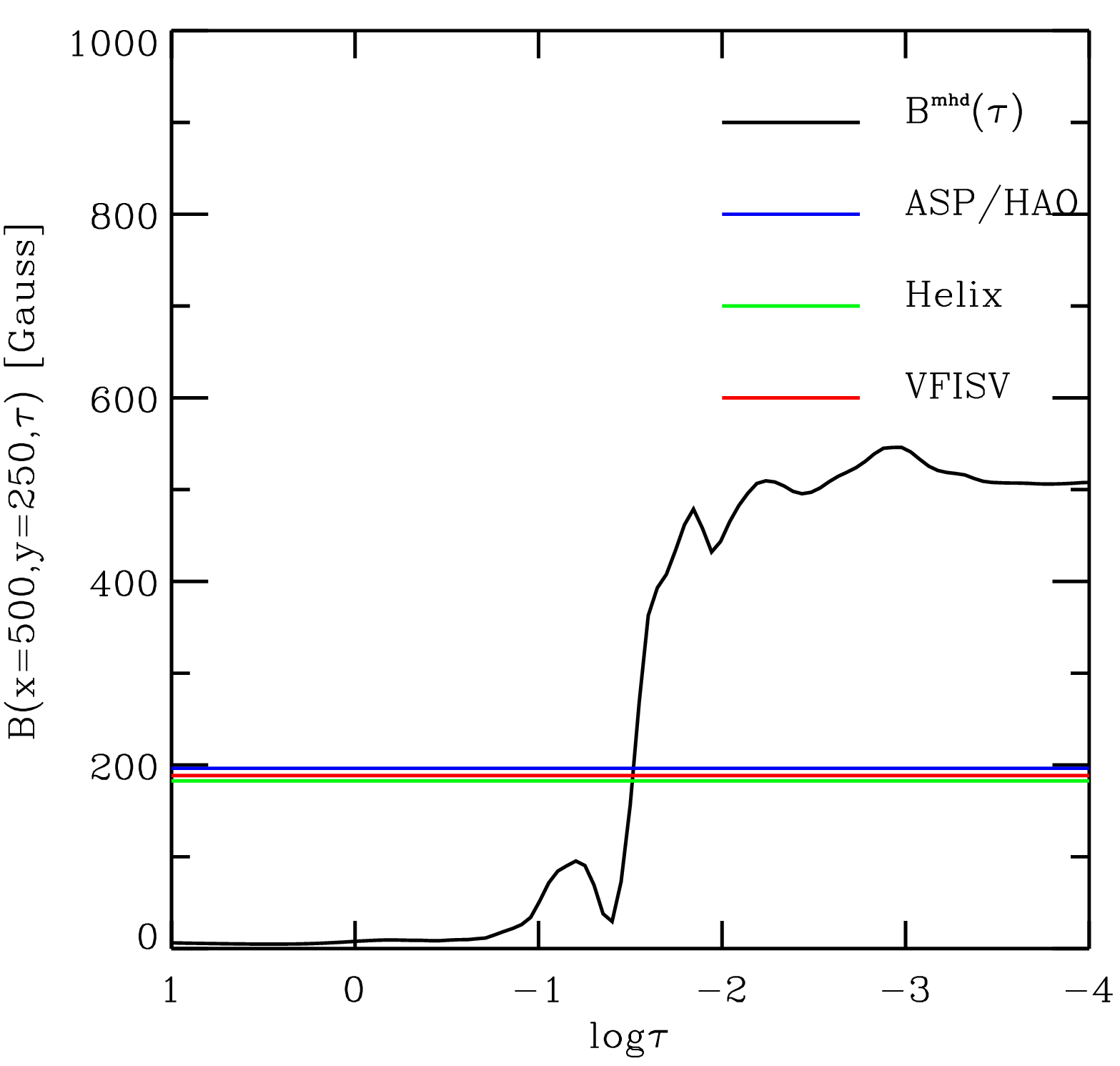} &
\includegraphics[width=5.5cm]{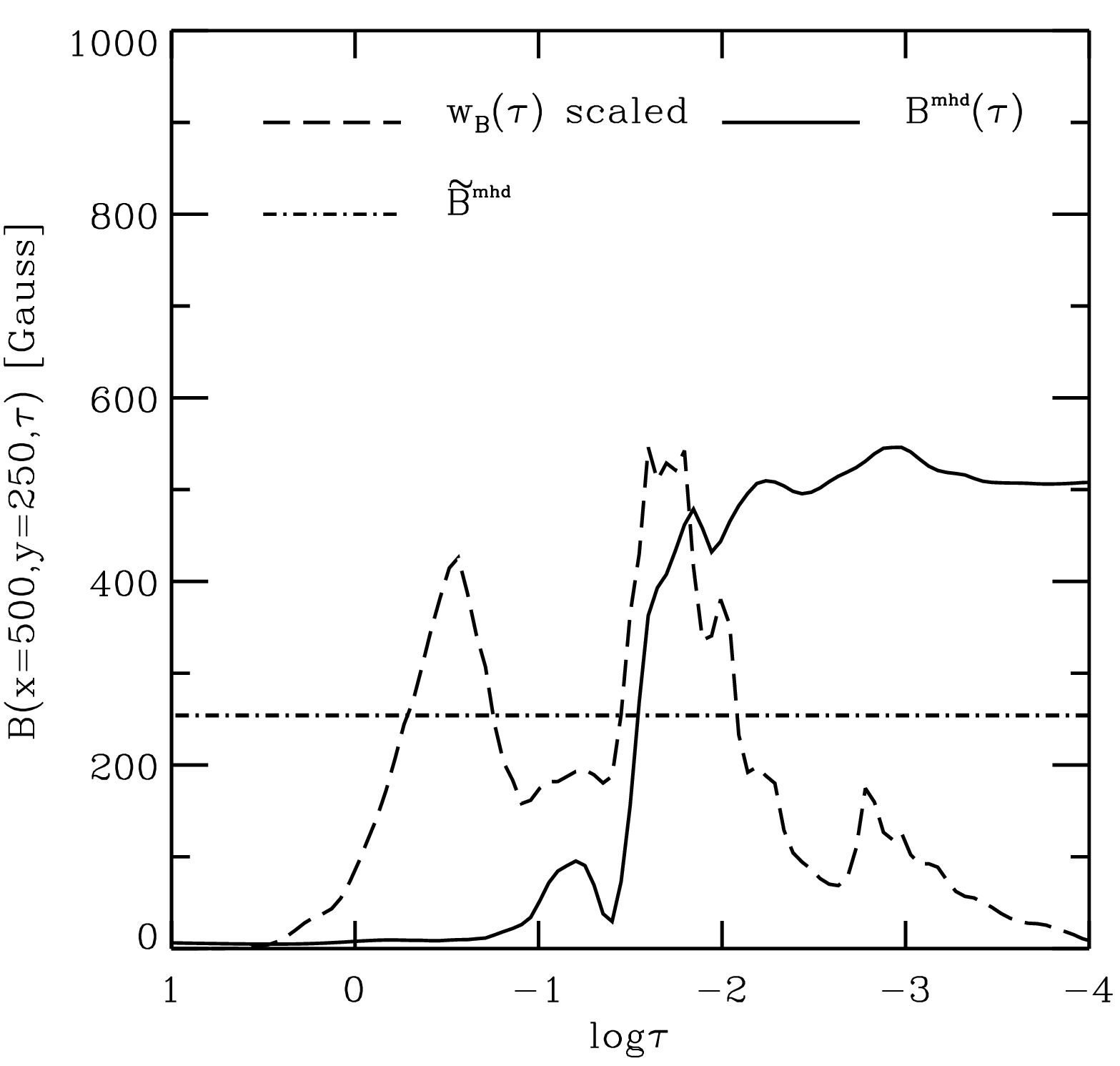} &
\includegraphics[width=5.5cm]{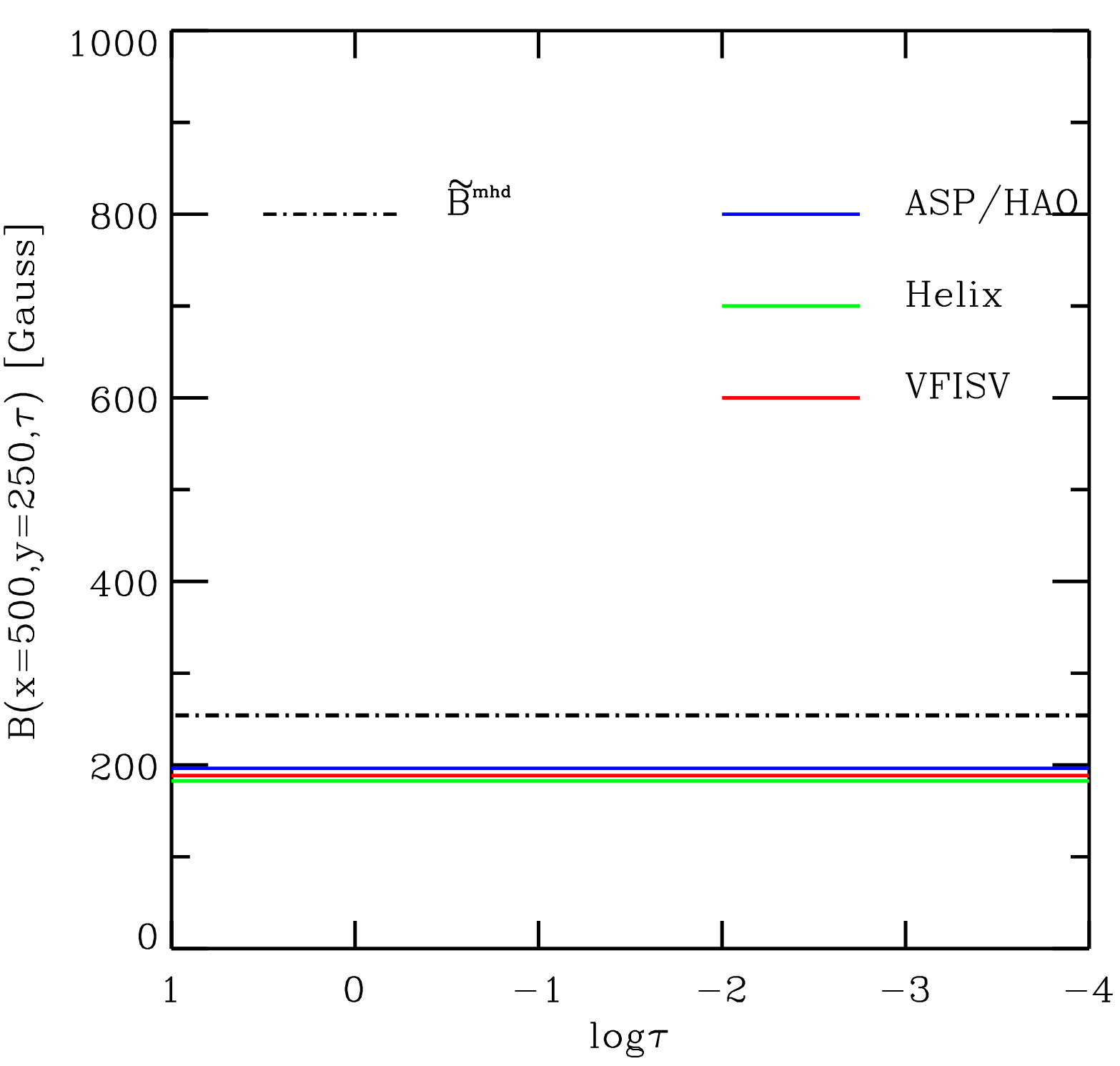}\\
\end{tabular}
\caption{{\it Left panel:} original stratification of the magnetic field strength as a function of optical depth from the 3D MHD simulation $B^{\rm mhd}(\tau_c)$
(solid-black line) for a particular point of the simulation domain $(x=500,y=250)$ (see Fig.~\ref{figure:simul_highlight}). The color lines represent the results
from the Milne-Eddington inversion of the Stokes profiles synthesized from the MHD simulation. {\it Middle panel:} here the solid-black is the same as on
the left panel, $B^{\rm mhd}(\tau_c)$; the dashed-black line represents the weighting function $w_B(\tau_c)$ employed to calculated the weighted average 
$\widetilde{B}^{\rm mhd}$ (dashed-dotted black line) according to Eq.~\ref{equation:xmhd_rf}. {\it Right panel:} comparison between the weighted average value
of the magnetic field $\widetilde{B}^{\rm mhd}$ (dashed-dotted black line) from the numerical simulations with the magnetic field obtained from the M-E inversions
(solid-color lines).}
\label{figure:weight_function}
\end{center}
\end{figure*}

Figure~\ref{figure:weight_function} illustrates an example of the process we have just described. On the leftmost panel we plot (solid-black line) 
the dependence of the magnetic field strength on the optical depth given by the MHD simulations, $B^{\rm mhd}(\tau_c)$, for point $(x=500,y=250)$ in 
Fig.~\ref{figure:simul_highlight}. We also plot (solid-color lines) the results from the inversion with the three different M-E inversion 
codes described in Sect.~\ref{section:description_ic}. Due to the assumptions of the Milne-Eddington model, the color lines are constant 
with $\tau_c$. The question we posed earlier in this section
was about comparing the results from the M-E inversions with $B^{\rm mhd}(\tau_c)$. In Sect.~\ref{section:comparison_fixed_tau} we compared with $B^{\rm mhd}(\log\tau_c^{*}=-1.4)$.
However, in this section we employ the weighting function, $w_{B}(\tau_c)$, showcased in the middle panel of Fig.~\ref{figure:weight_function} (dashed-black line).
For simplicity, in this example we consider that $w_{B}(\tau_c)$ is the same for the three M-E inversion codes tested, even though we know they are not because VFISV includes
only one spectral line in the wavelength integral in Eq.~\ref{equation:weight_function}. With it, a weighted average of $B^{\rm mhd}(\tau_c)$ is then calculated 
according to Eq.~\ref{equation:xmhd_rf}, resulting in $\widetilde{B}^{\rm mhd}$ (dashed-dotted line in middle panel). This is the actual value that is then 
compared to the results from the Milne-Eddington inversions (right panel in Fig.~\ref{figure:weight_function}). The process must then be repeated for 
$\gamma$, and ${\rm v}_{\rm los}$ and for all $2^{16}$ points in the simulation.
For each case, a new weighting function must be calculated because the sensitivity of the spectral line depends on both the physical parameter measured
and the properties of the atmosphere where it is measured.\\

\begin{figure*}
\begin{center}
\begin{tabular}{ccc}
\includegraphics[width=5.5cm]{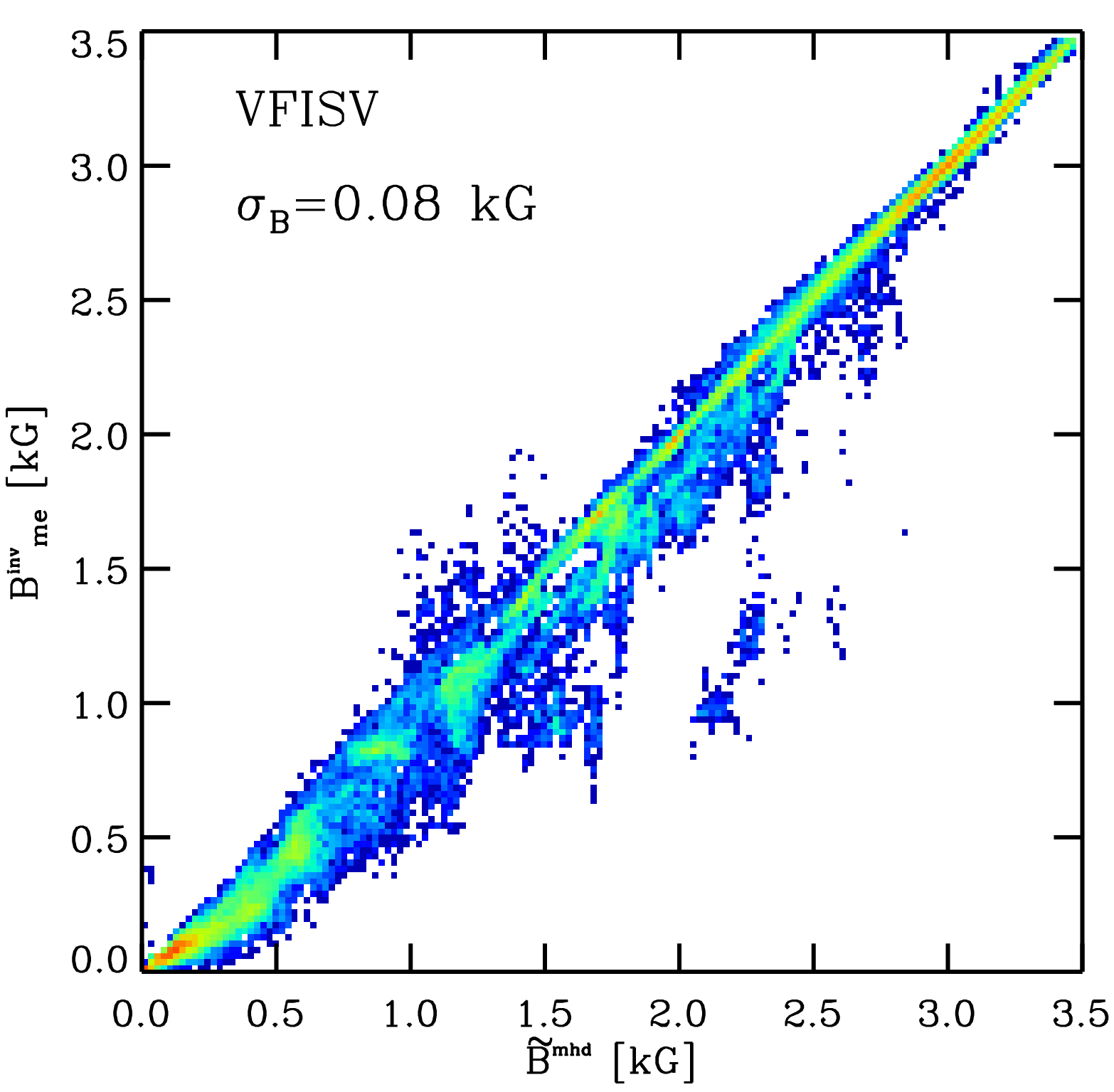} &
\includegraphics[width=5.5cm]{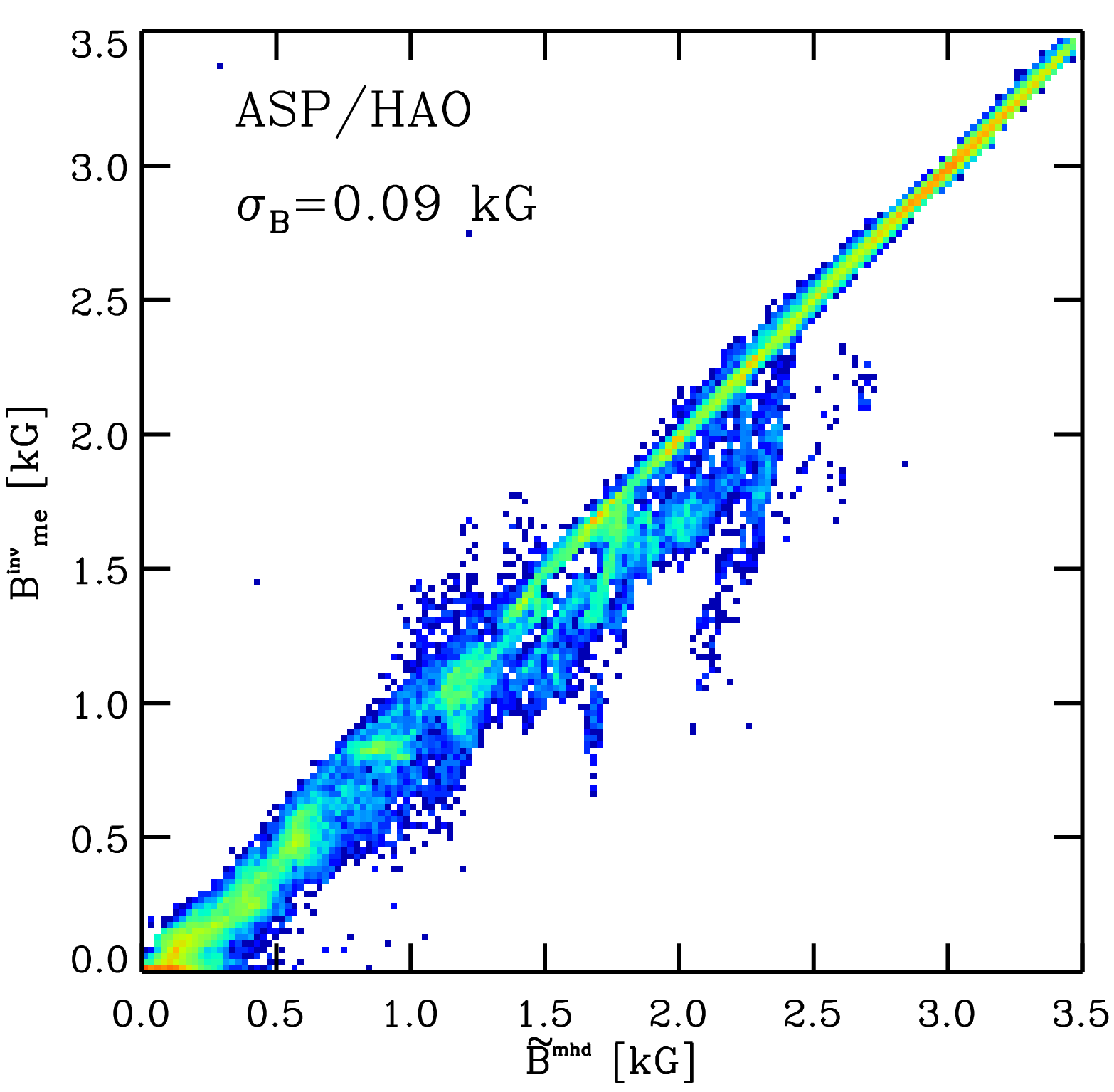} &
\includegraphics[width=5.5cm]{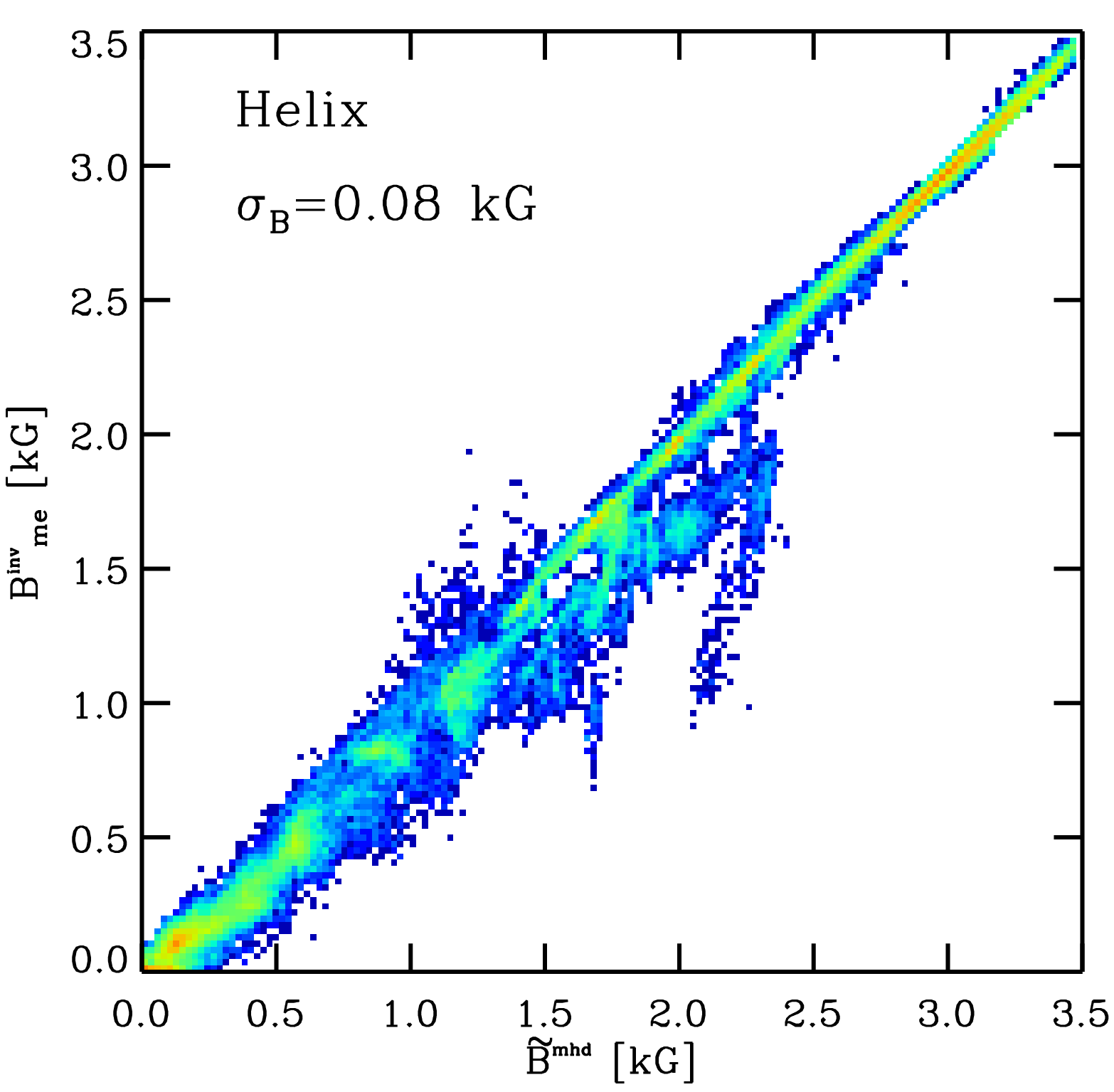} \\
\includegraphics[width=5.5cm]{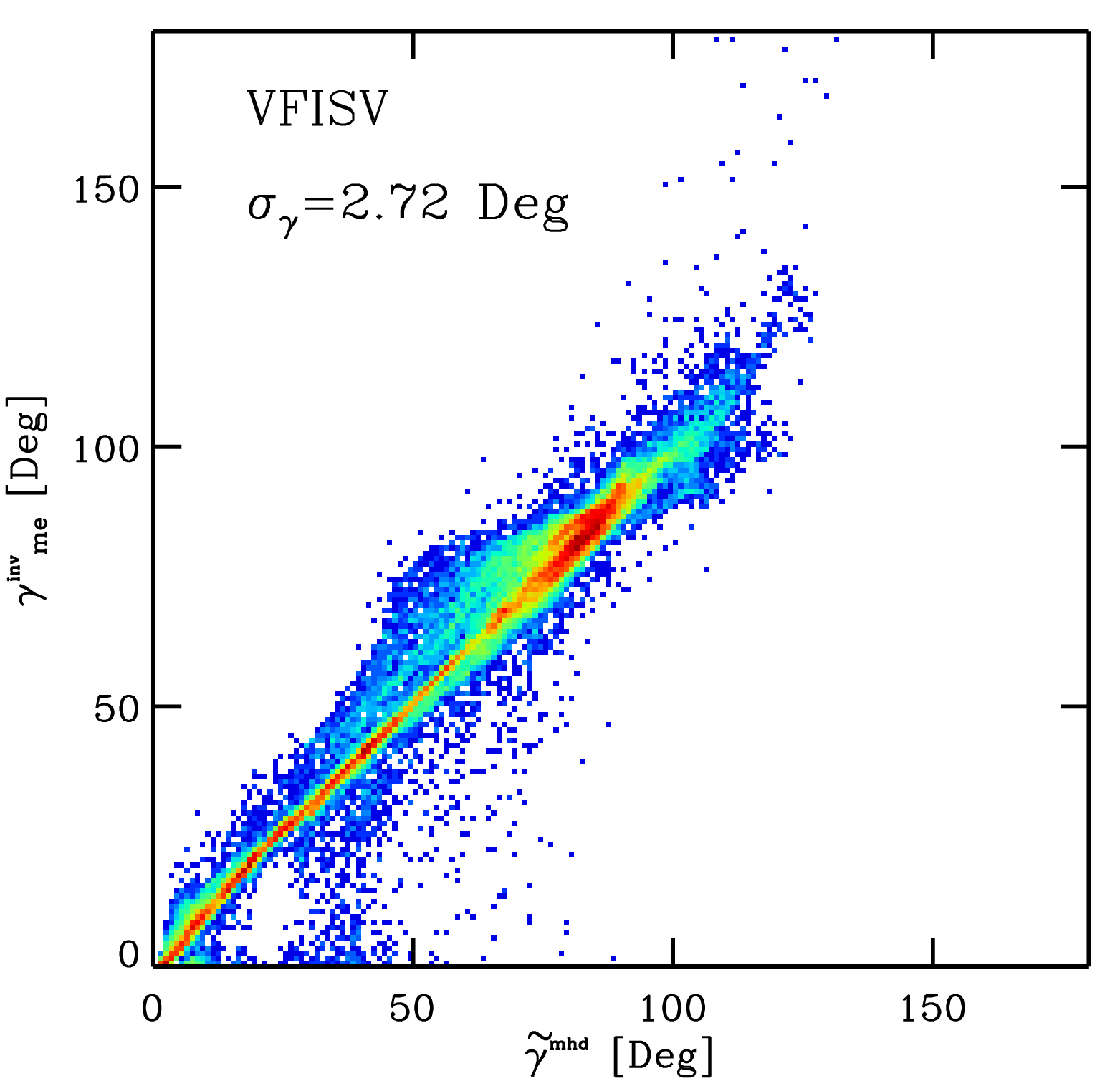} &
\includegraphics[width=5.5cm]{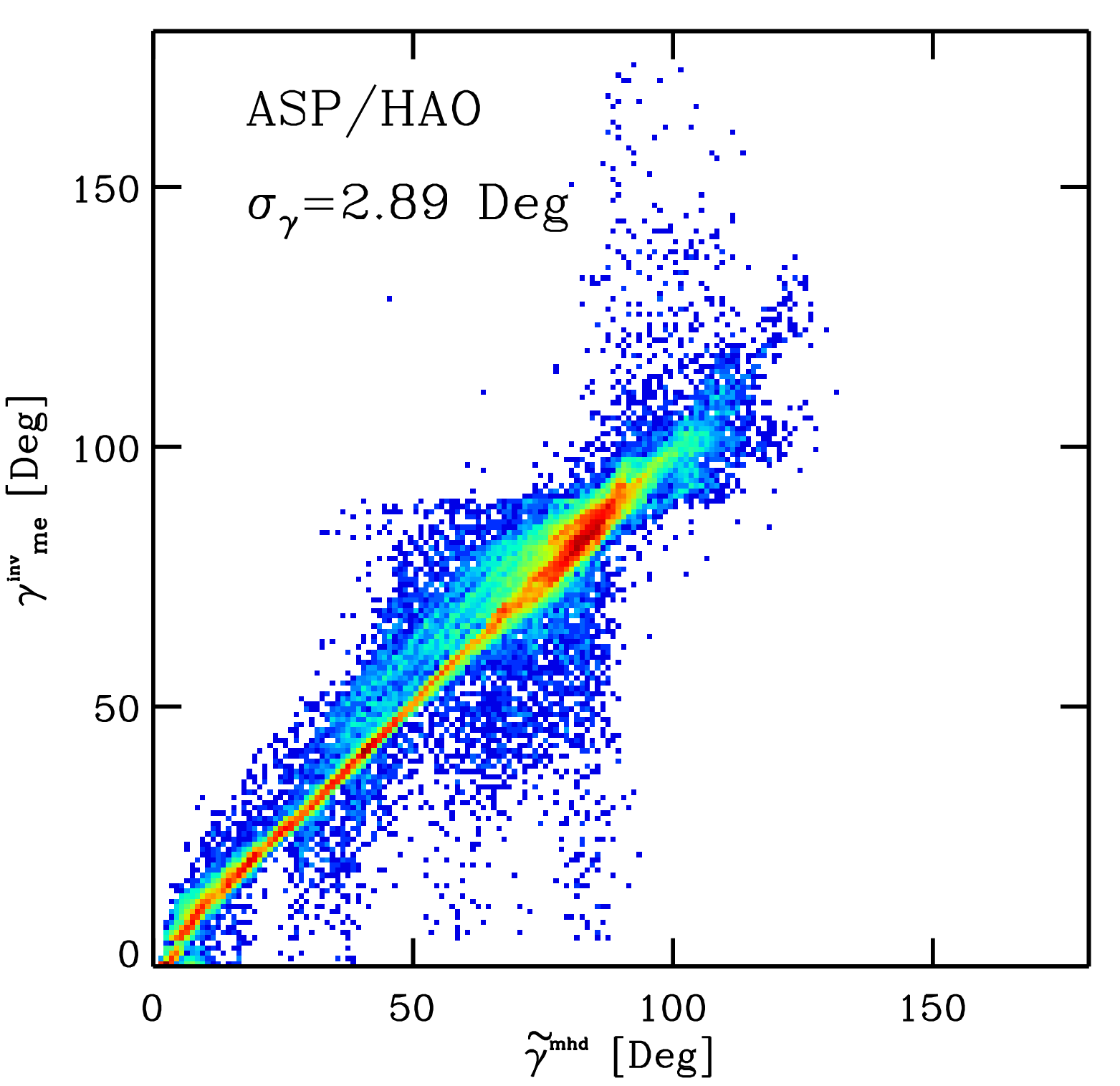} &
\includegraphics[width=5.5cm]{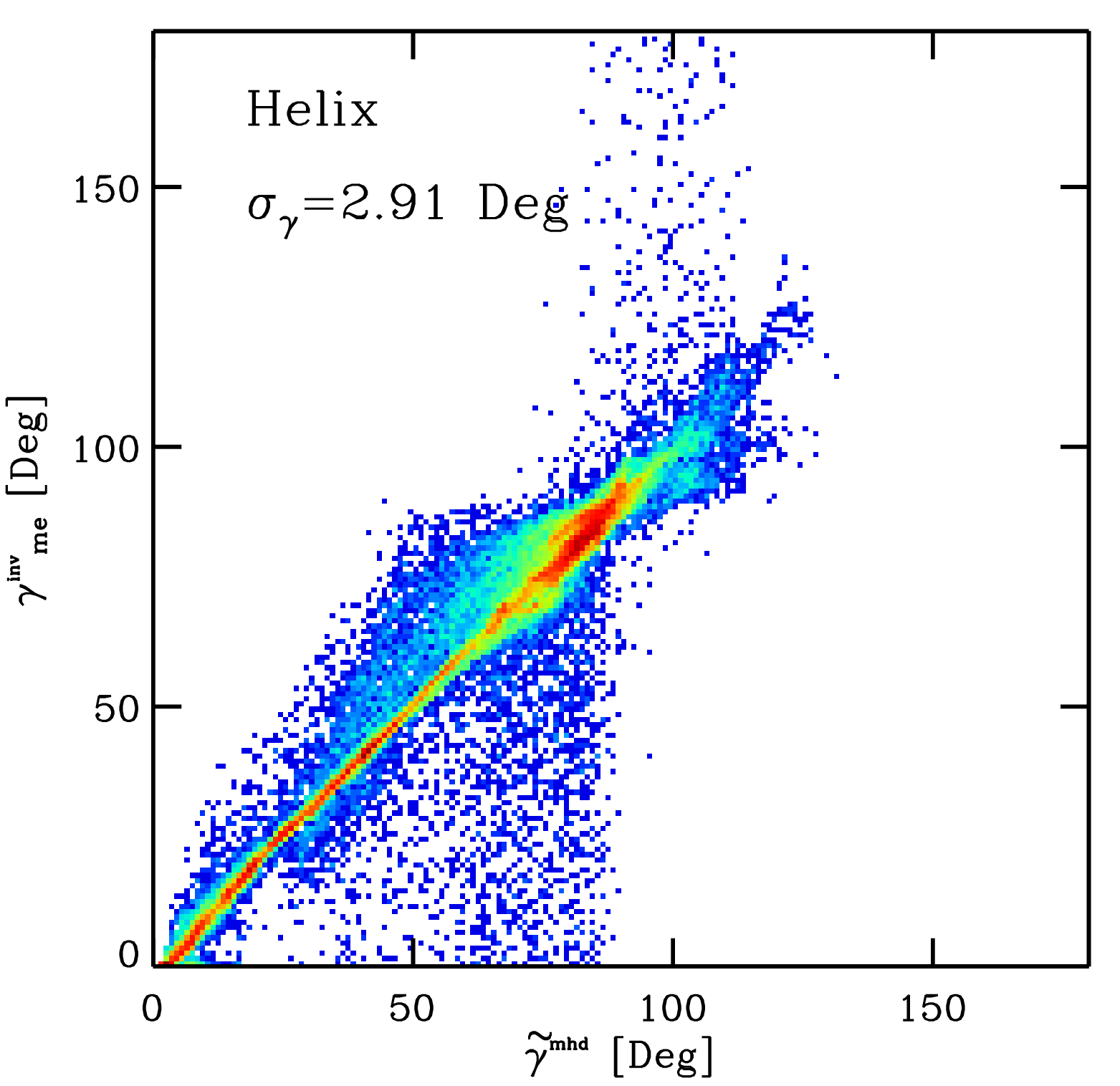} \\
\includegraphics[width=5.5cm]{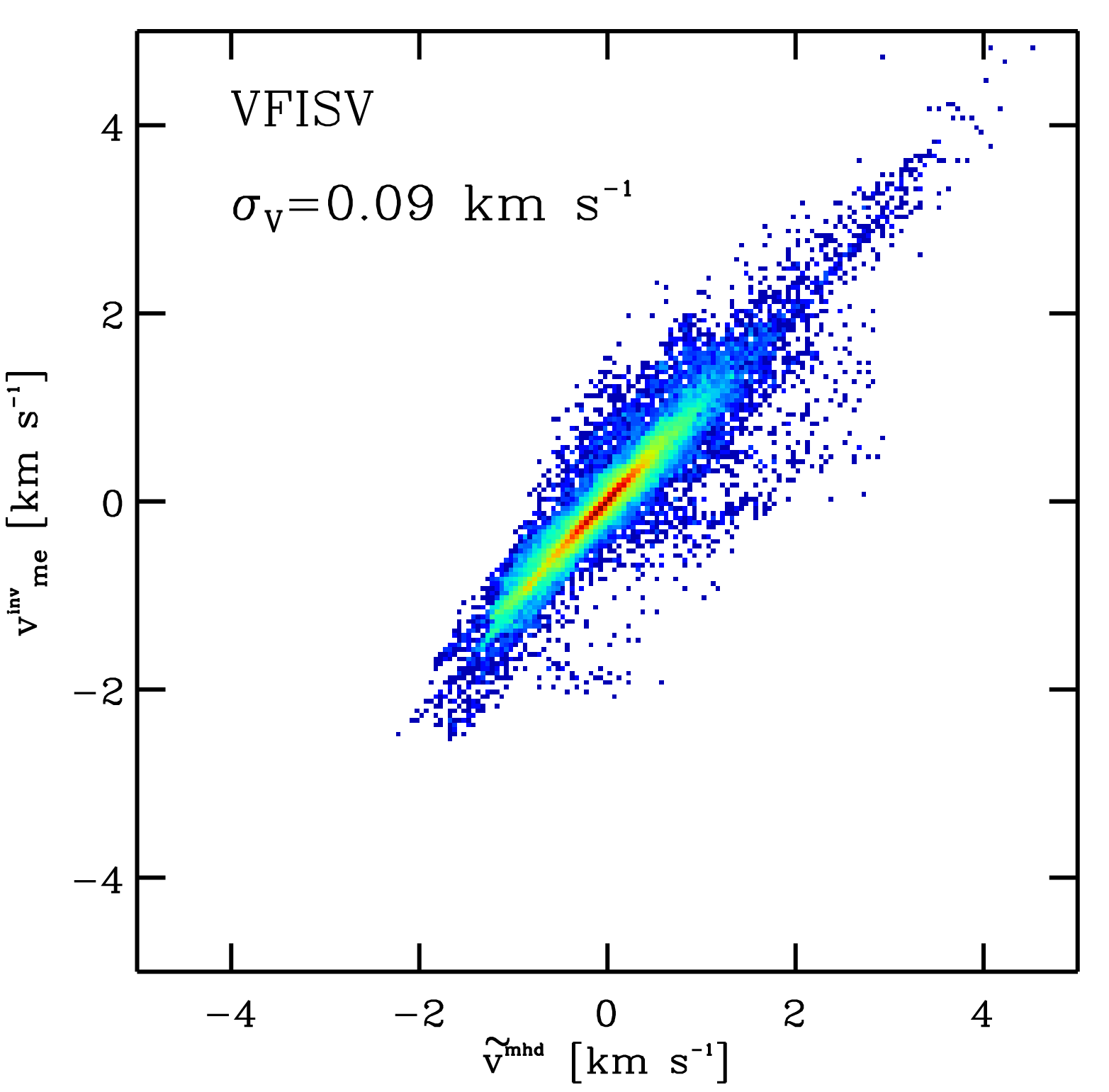} &
\includegraphics[width=5.5cm]{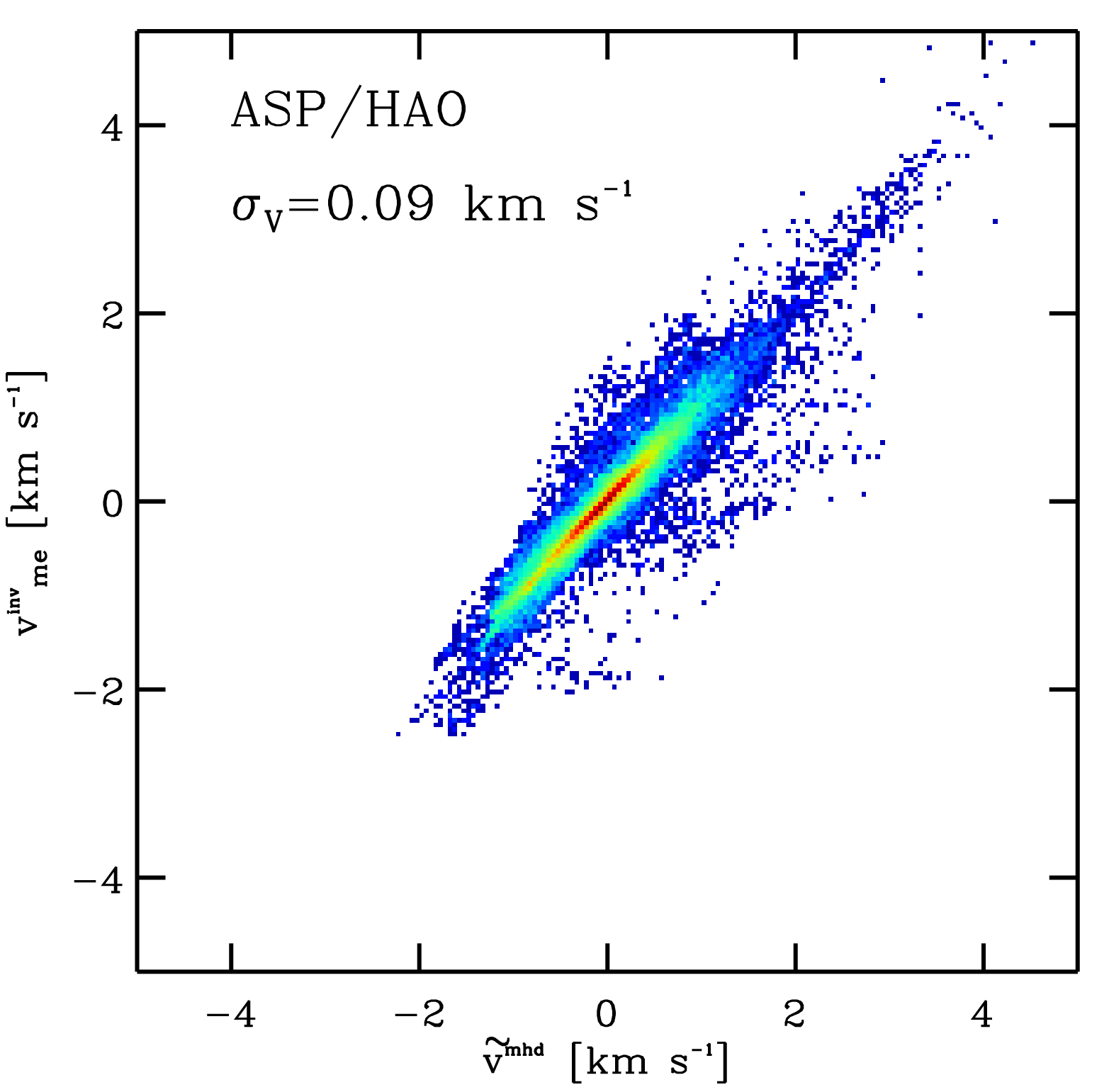} &
\includegraphics[width=5.5cm]{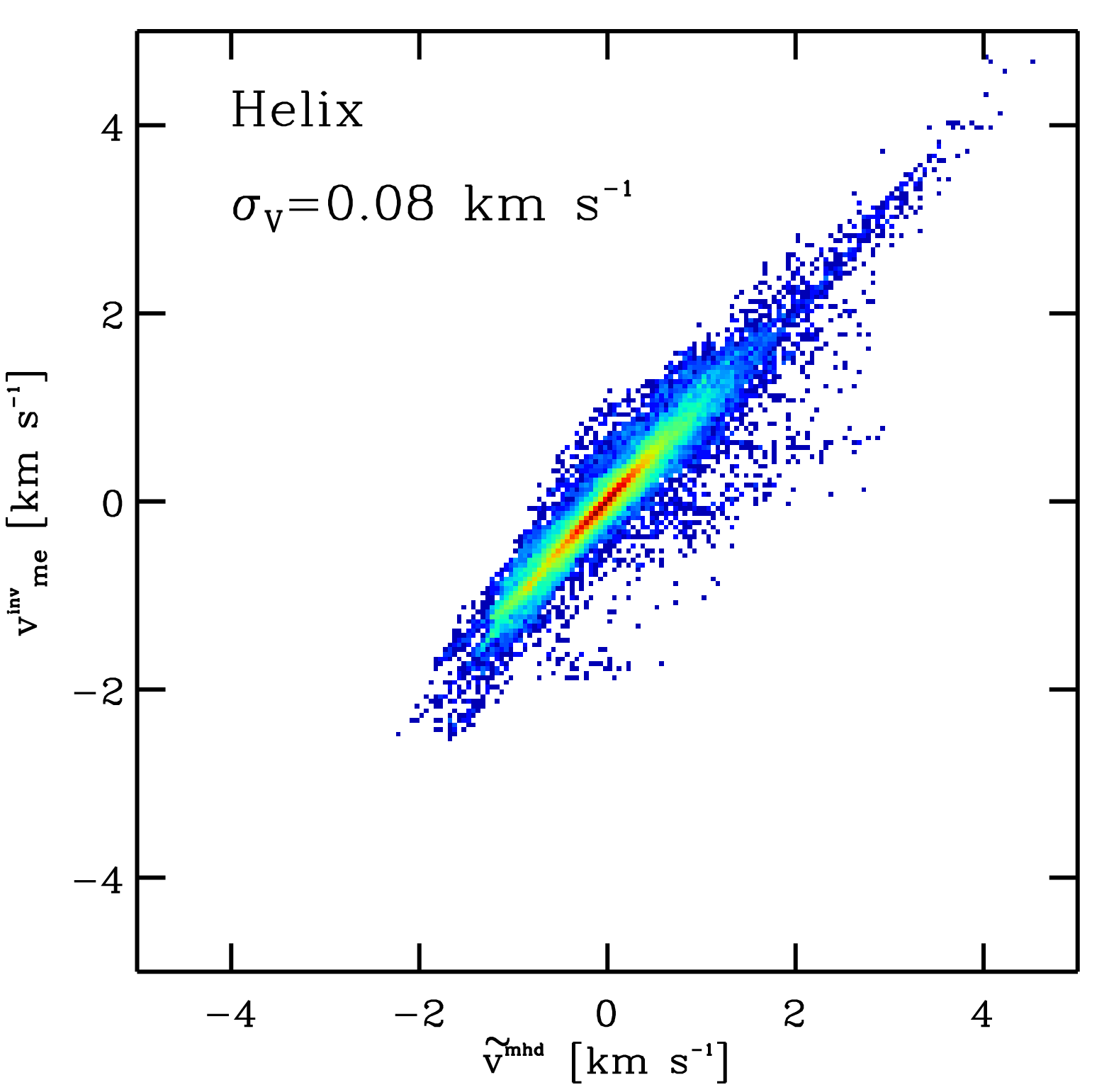} \\
\end{tabular}
\caption{Same as Figure~\ref{figure:ic_mhd_fixed} but comparing the results from the M-E inversions with the weighted averages of the MHD 
simulations $\widetilde{B}^{\rm mhd}$, $\widetilde{\gamma}^{\rm mhd}$, $\widetilde{\rm v}^{\rm mhd}_{\rm los}$ instead of the values at some 
fixed optical depth value $\tau_c^{*}$.}
\label{figure:ic_mhd_rf}
\end{center}
\end{figure*}

The results of the comparison employing response functions are presented in Figure~\ref{figure:ic_mhd_rf}. This figure is analogous to Figs.~\ref{figure:ic_ic}
and ~\ref{figure:ic_mhd_fixed}. When compared to Fig.~\ref{figure:ic_mhd_fixed}, we observe that employing proper kernels to determine the heights at which the spectral lines
are sensitive, significantly increases the agreement between MHD simulations and the M-E inversions: $\sigma_B < 90$ G, $\sigma_\gamma < 3\deg$, the line-of-sight 
component of the velocity $\sigma_{\rm v} < 90$\ms. These results strongly support the idea that {\it M-E inversions provide an average of the physical parameters
across the region in which the spectral lines are formed}. Although not the first time this fact is pointed out \citep[see e.g.,][Sect.~3.1]{westendorp1998}, our
work here certainly provides the most exhaustive demonstration to date, insofar as we have employed realistic numerical simulations of sunspots (Sect.~\ref{section:description_mhd}) 
to produce a large number of synthetic profiles (Sect.~\ref{section:description_synthesis}) that were subsequently analyzed using three different M-E inversion codes 
(Sect.~\ref{section:description_ic}). Moreover, the results presented in this section also ward off criticism about the lack of uniqueness in the inversion results,
as we have now demonstrated, these are very similar to the original values from the MHD simulations.\\

Last but not least, we have also followed the more rigorous approach, described in \citet{jorge1996}, to determine the weighting function $w_{\rm x}$ 
through the use of {\it generalized Response Functions}. The advantage of these functions is that they are positively defined, and therefore, 
there is no need to introduce the (somewhat artificial) absolute value in Eq.~\ref{equation:weight_function}. In particular, the term $\partial g/\partial I_i$
in Eq.~14 in \citet{jorge1996} should become negative whenever the response function is also negative, so that their product (see Eq.~8 in the cited paper)
remains positive (S\'anchez Almeida, {\it private communication}). Interestingly, after we tested this method using the stratification from the MHD simulations
we observed that this is often not the case. This forced us to, again, introduce an absolute value in this more rigorous formulation of the problem\footnote{Using
an absolute value in Eq.~15 in \citet{jorge1996} leads to the same results as using our Eq.~\ref{equation:xmhd_rf}}. We suspect that
the inconsistency arises from the large variations with optical depth present in the MHD simulations (see e.g., solid-black line in Fig.~\ref{figure:weight_function}), 
that can break down the assumption of linear perturbations implicit in our Eq.~\ref{equation:xmhd_rf} \citep[see also Eq.~10 in][]{jorge1996}. We tried to confirm 
this point by evaluating the physical parameters in the simulations in a grid with $\Delta\log\tau_c=10^{-3}$ instead of $\Delta\log\tau_c=10^{-2}$ (see 
Sect.~\ref{section:description_synthesis}) but to no avail. For these reasons we decided not to pursue this strategy further.\\

\section{Optical depths for the formation of the Fe \ion{I} line pair at 630 nm}
\label{section:hof}

\begin{figure*}
\begin{center}
\begin{tabular}{ccc}
\includegraphics[width=5.5cm]{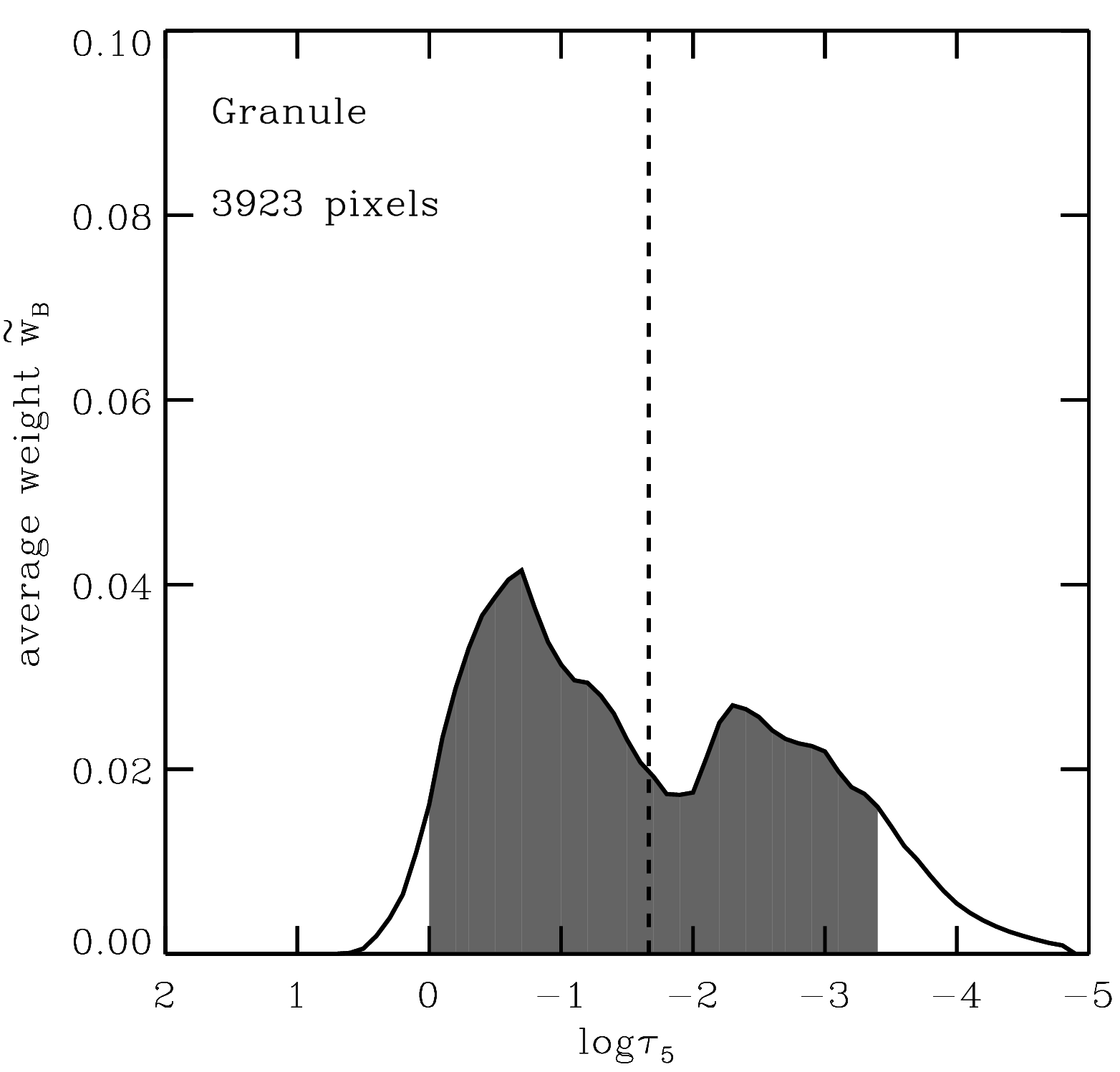} &
\includegraphics[width=5.5cm]{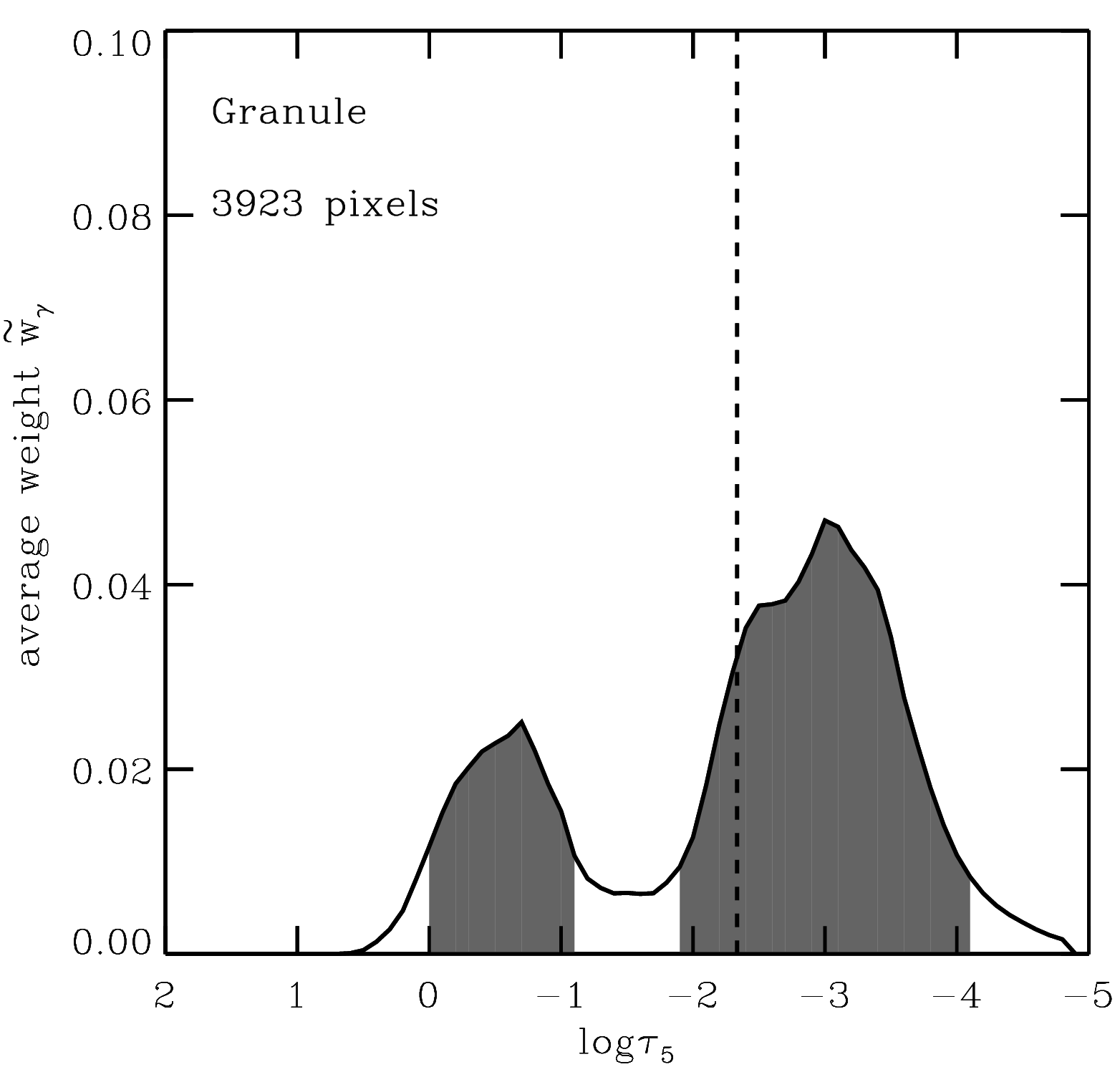} &
\includegraphics[width=5.5cm]{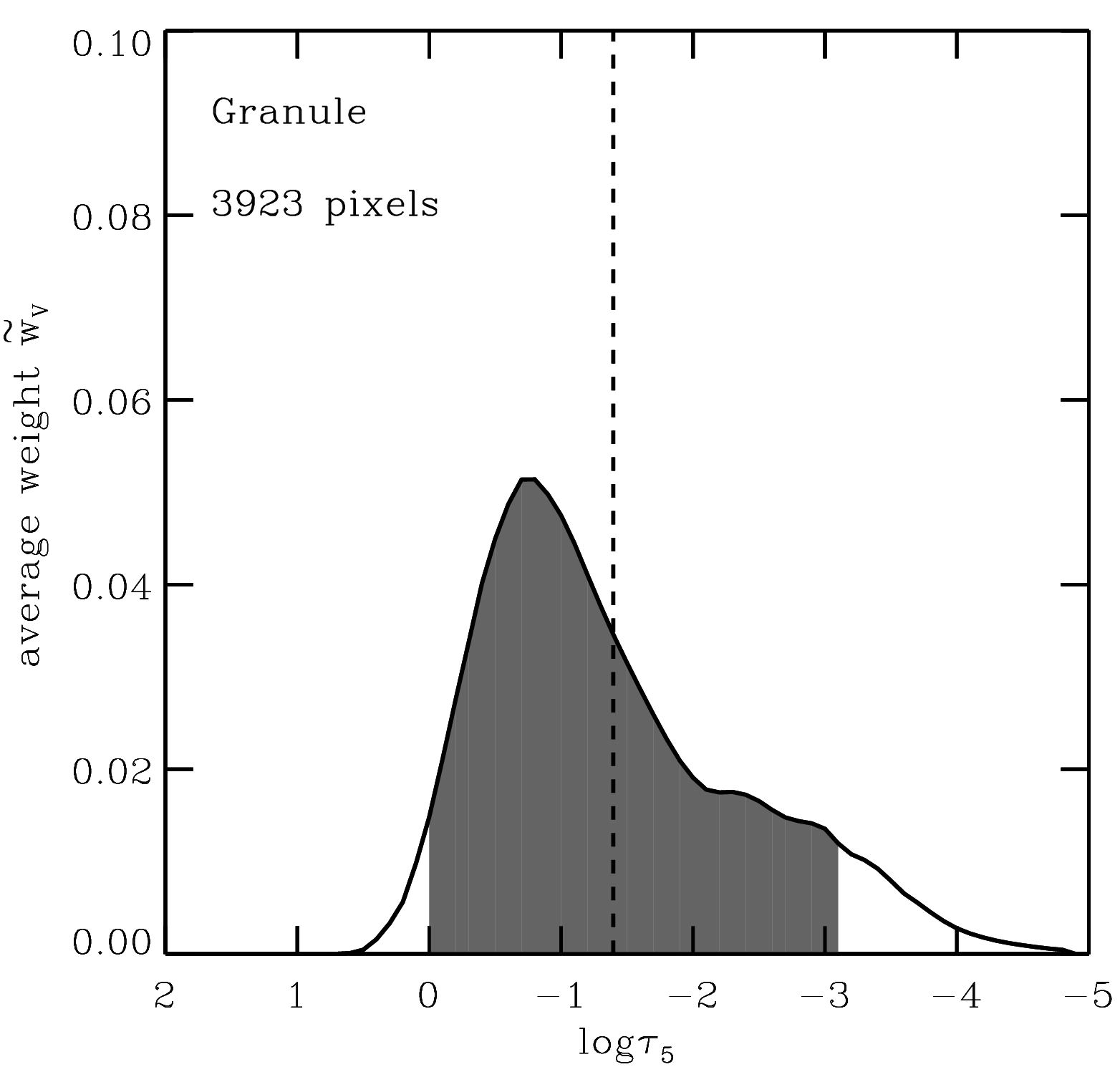} \\
\includegraphics[width=5.5cm]{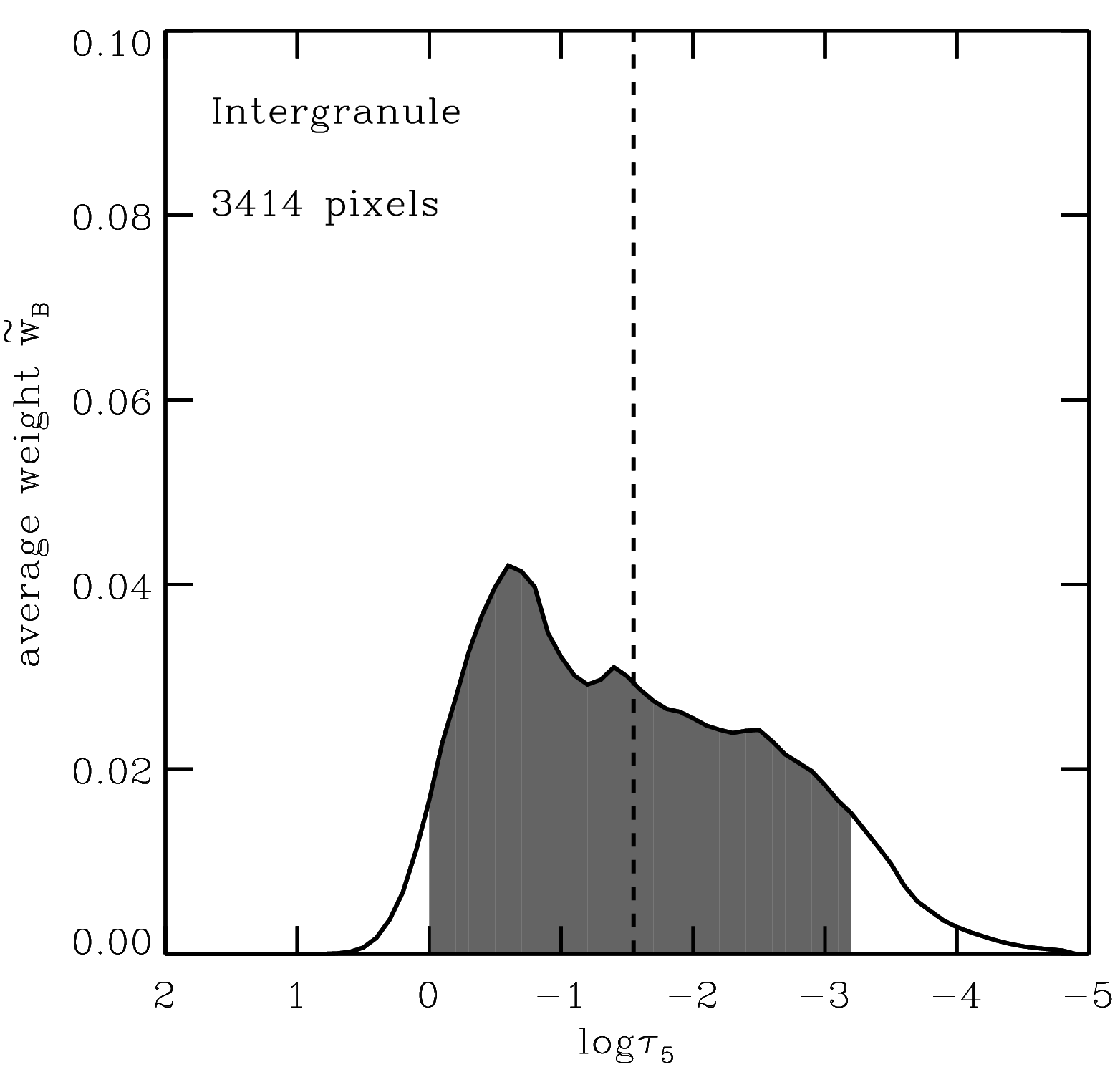} &
\includegraphics[width=5.5cm]{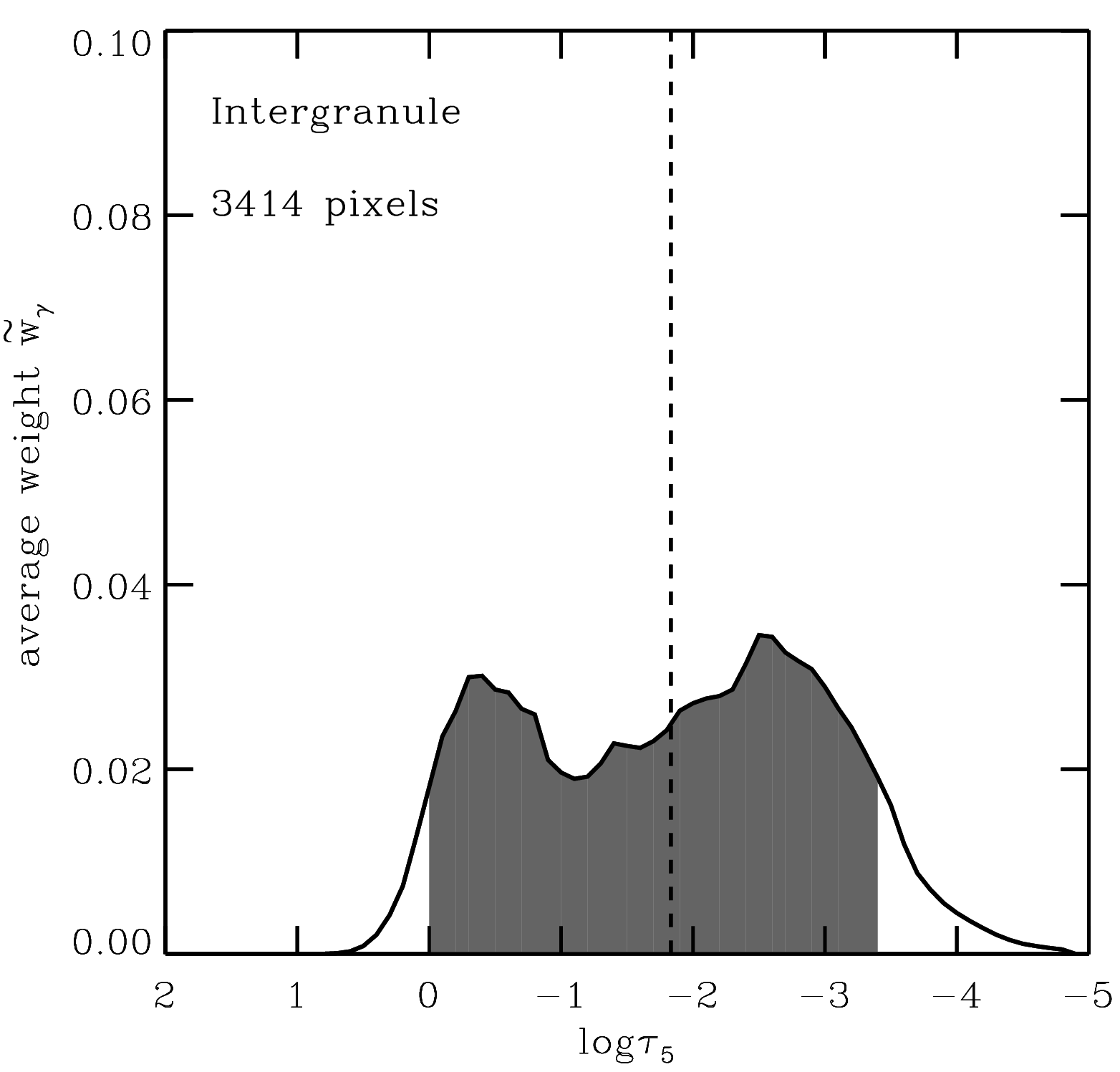} &
\includegraphics[width=5.5cm]{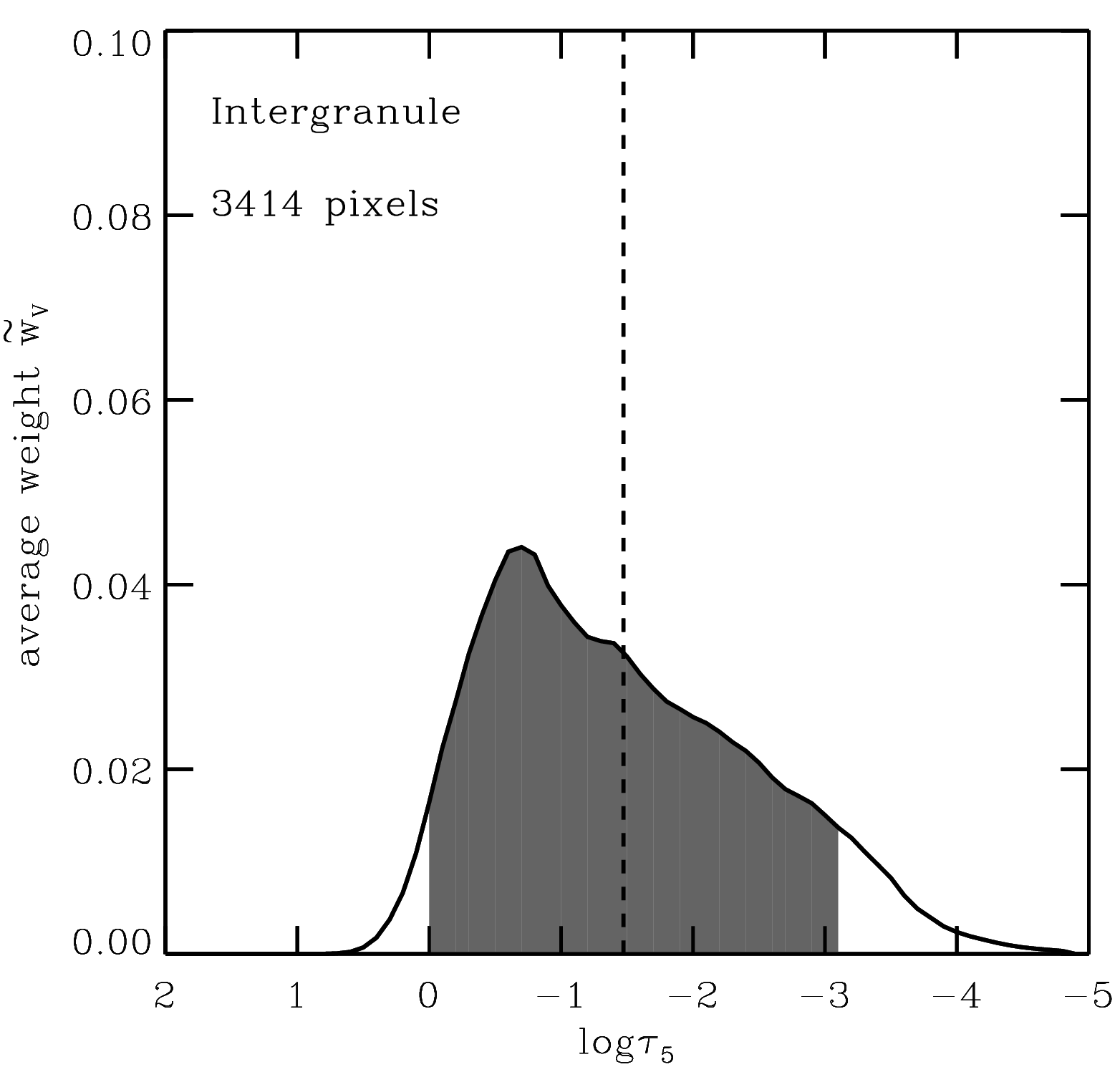} \\
\includegraphics[width=5.5cm]{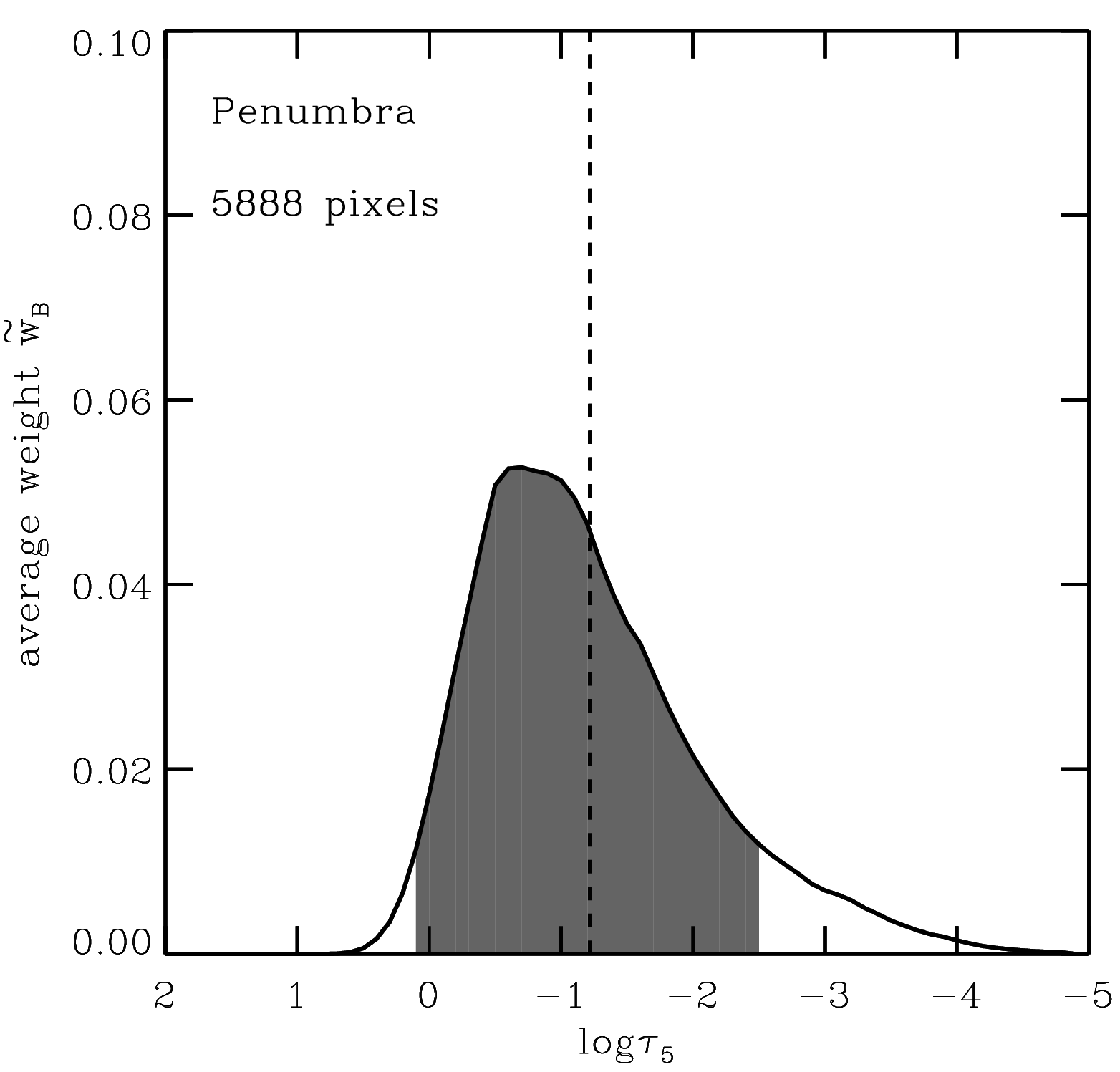} &
\includegraphics[width=5.5cm]{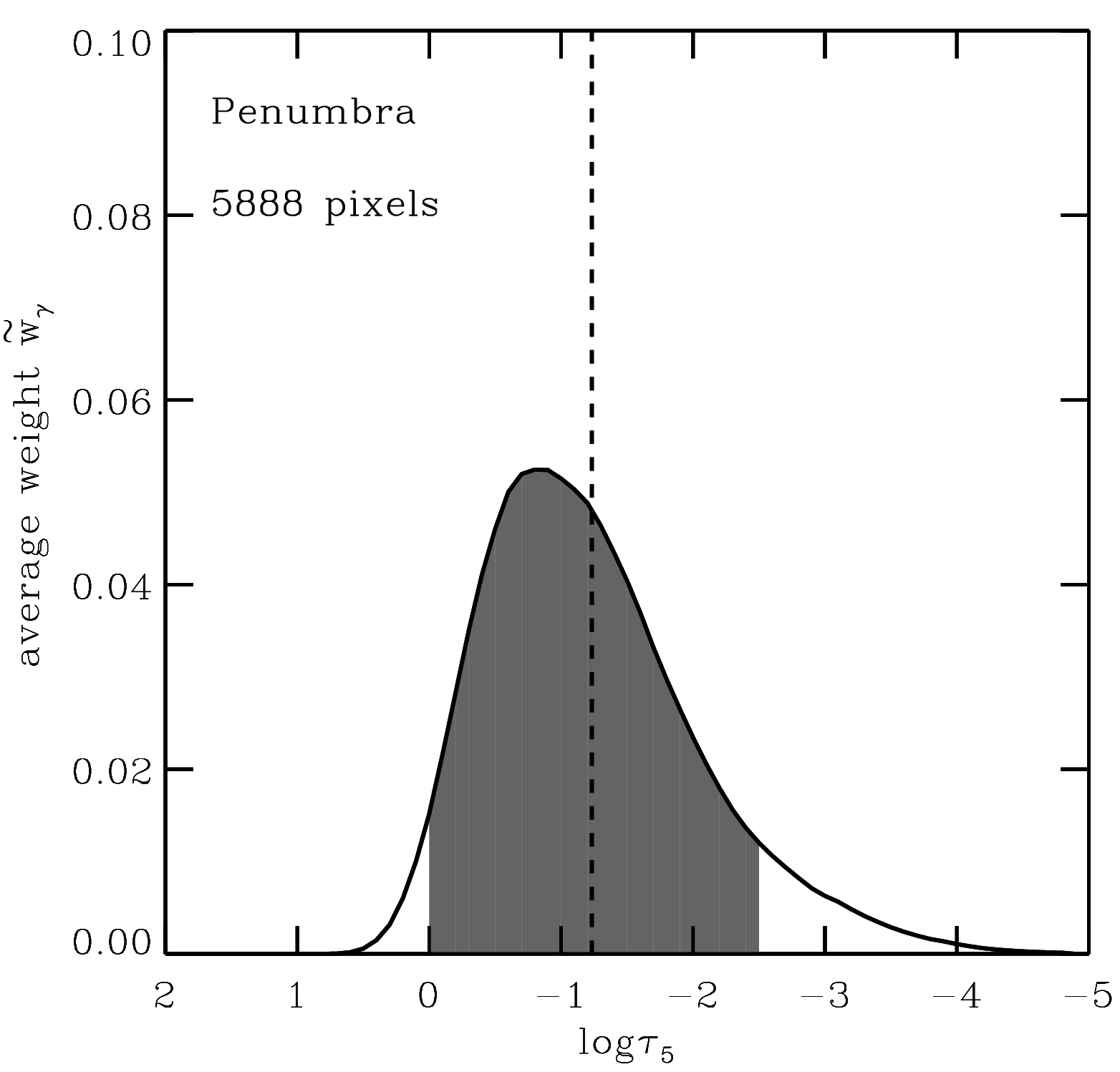} &
\includegraphics[width=5.5cm]{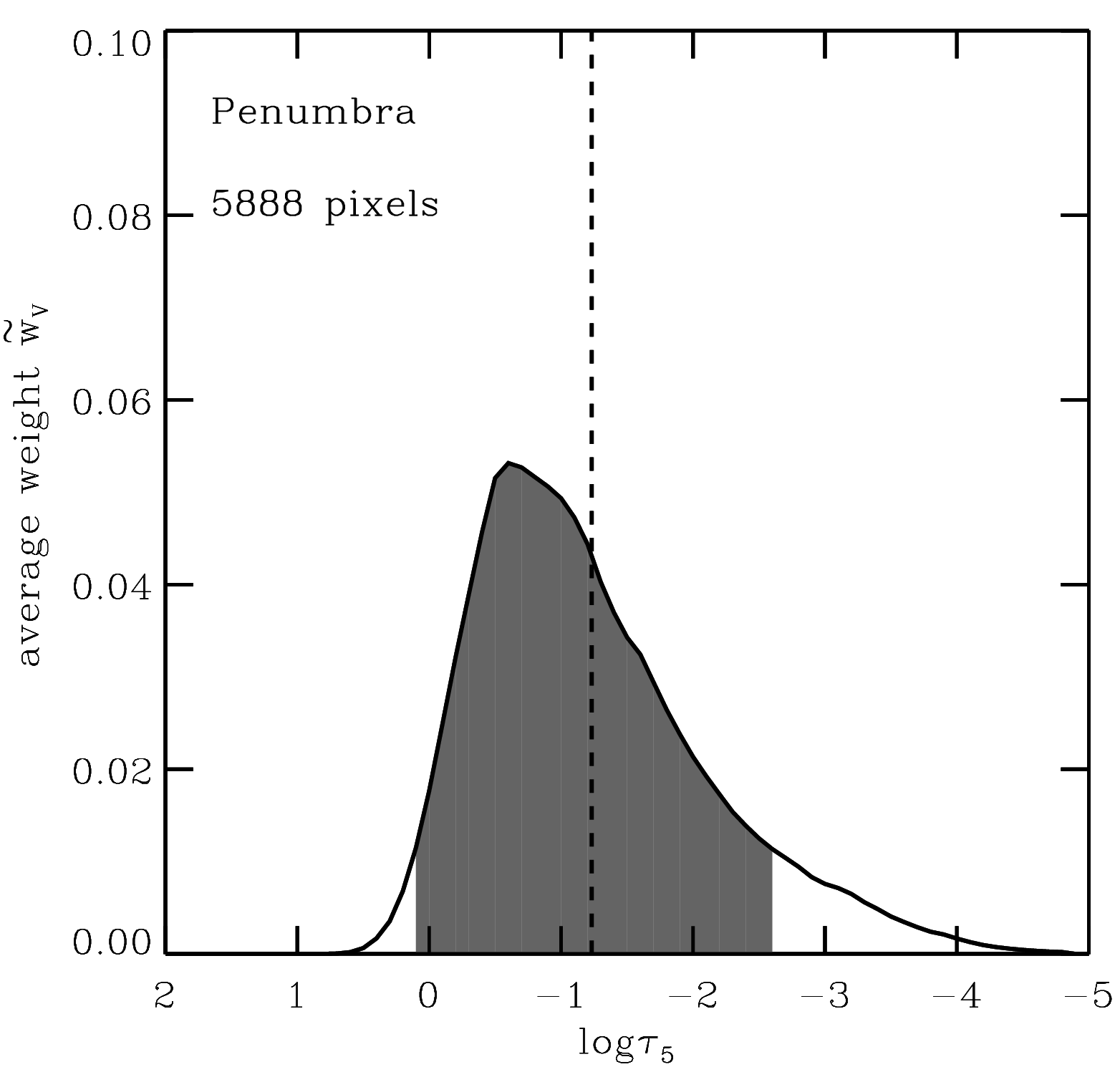} \\
\includegraphics[width=5.5cm]{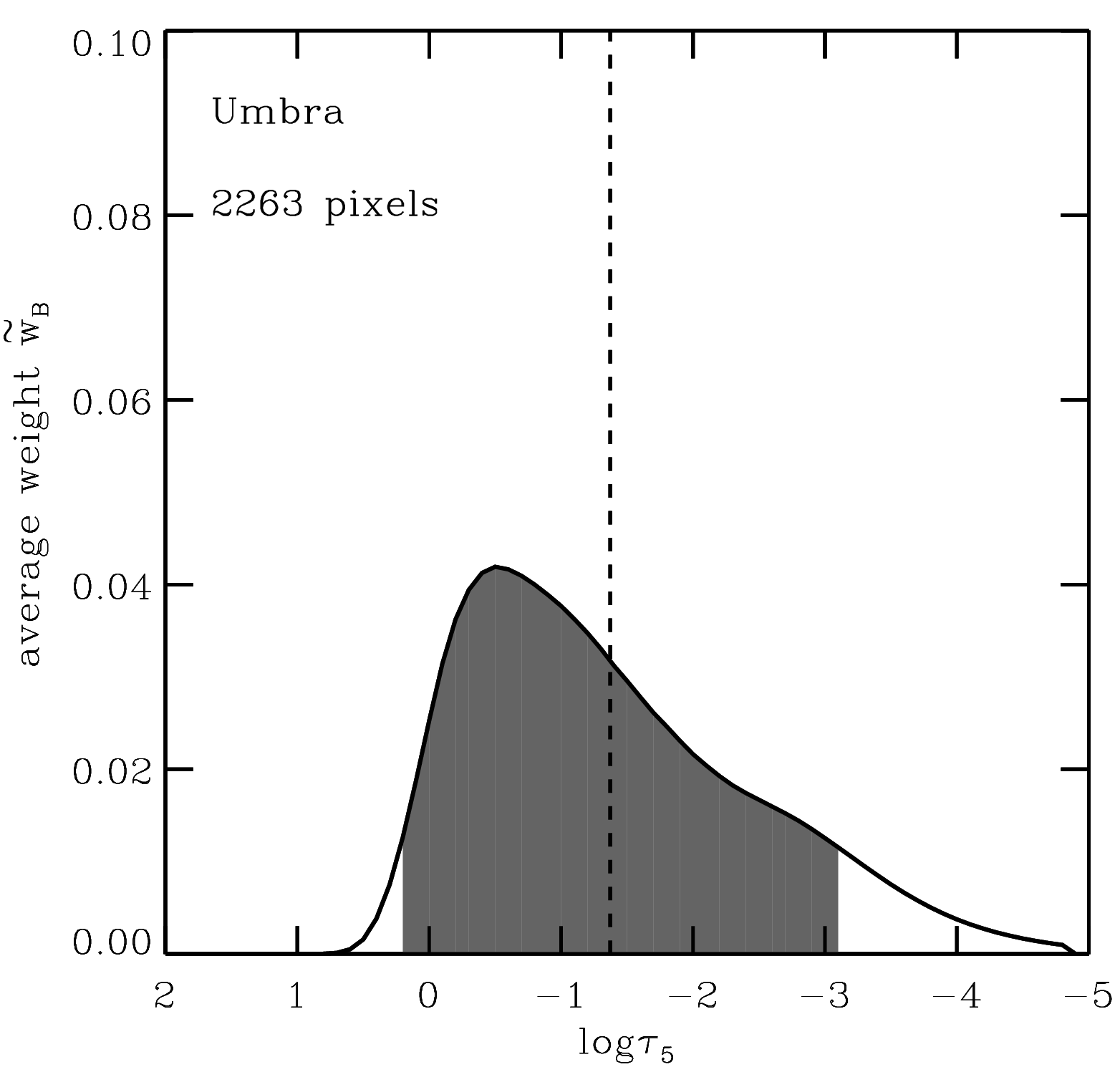} &
\includegraphics[width=5.5cm]{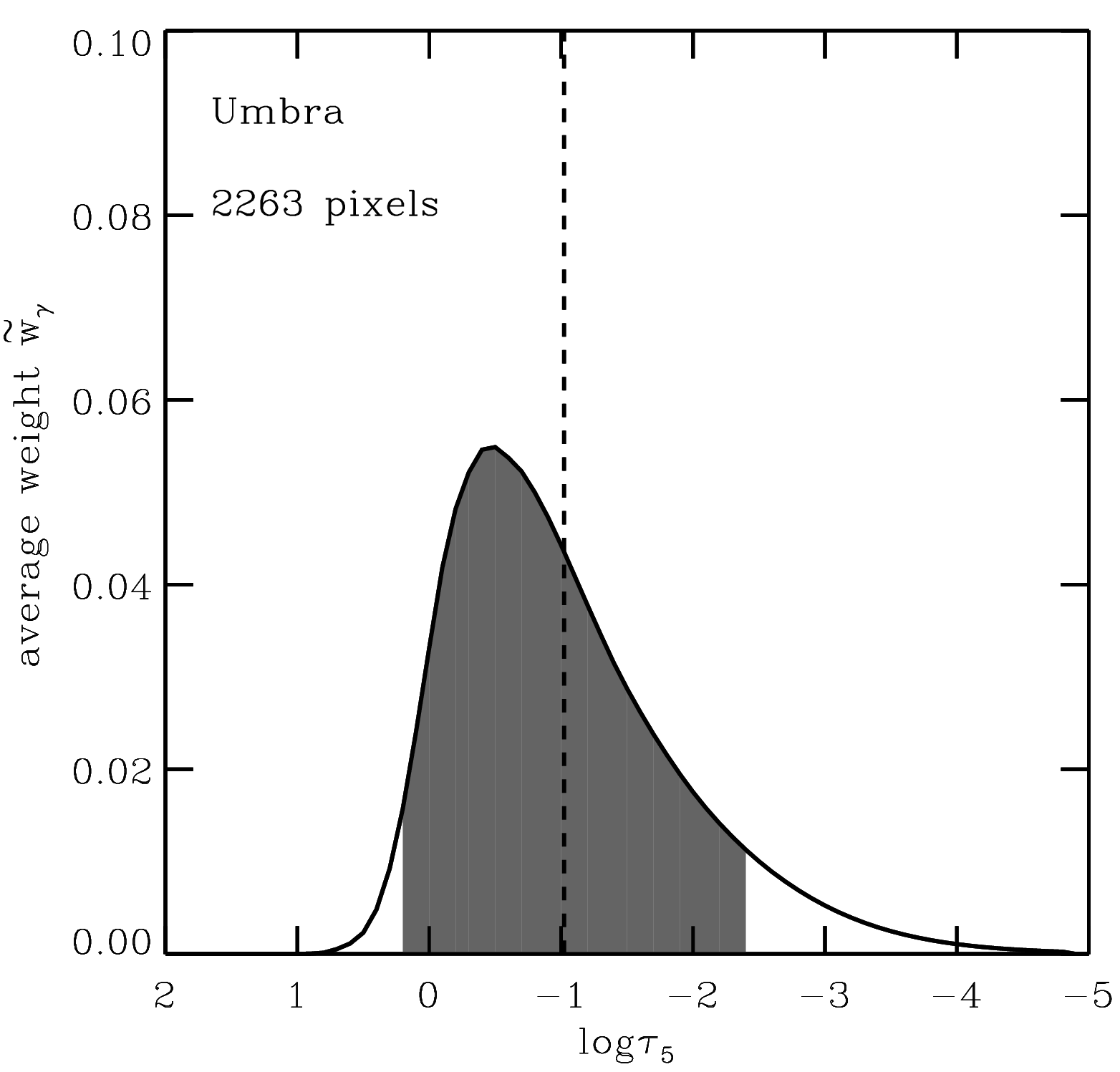} &
\includegraphics[width=5.5cm]{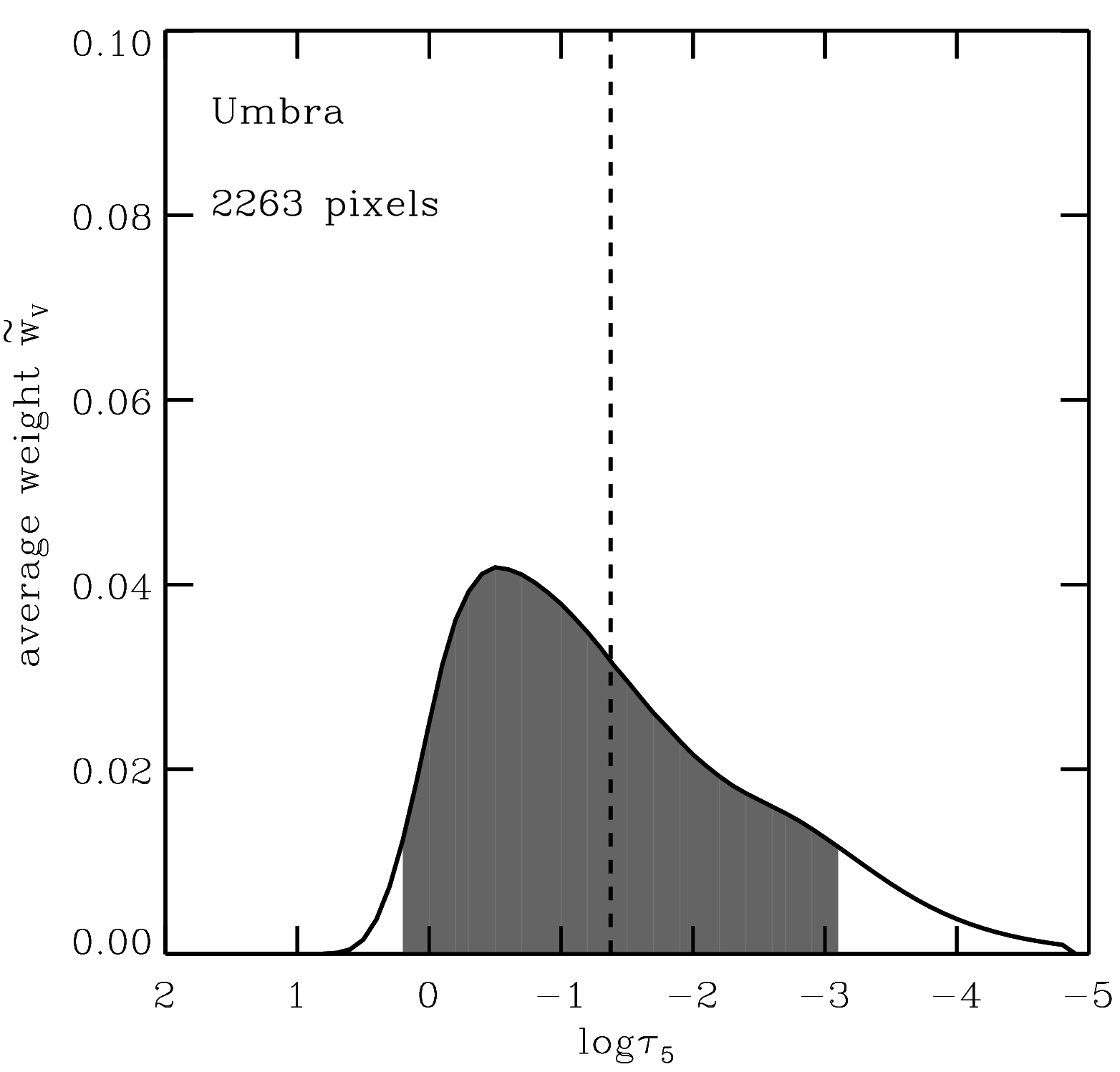} \\
\end{tabular}
\caption{Averaged weighting functions $\widetilde{w}_{\rm x}$ for different physical parameters and different solar structures. From left to right: $B$, $\gamma$,
and ${\rm v}_{\rm los}$. From top to bottom: granules, intergranules, penumbra, and umbra. Each panel indicates the number of points from Figure~\ref{figure:simul_highlight}
employed in the average. The vertical dashed lines correspond to the median of each distribution. Shaded areas encompass the regions of highest sensitivity (see text
for details).}
\label{figure:hof}
\end{center}
\end{figure*}

In Section~\ref{section:comparison_rf} we demonstrated that, after taking into account the atmospheric layers sampled by the spectral lines in Table~\ref{table:lines}
when measuring $B$, $\gamma$, and ${\rm v}_{\rm los}$, the agreement between the M-E inversions and MHD simulations improved greatly. For instance,
the standard deviations in Fig.~\ref{figure:ic_mhd_rf} are up to a factor of 2-3 smaller than those in Fig.~\ref{figure:ic_mhd_fixed}. This illustrates the need
for a proper account of the layers sensed by spectral lines when measuring different physical parameters. In an attempt to provide such account
we present, in Figure~\ref{figure:hof}, the average of the weighting functions $\widetilde{w}_{\rm x}(\tau_c)$ employed in the previous section for $B$ (left column)
, $\gamma$ (middle column), and ${\rm v}_{\rm los}$ (right column) in several solar structures: granules (first row), intergranules (second row), penumbra (third row)
and umbra (fourth row). These averages were determined by selecting points in the simulations (Fig.~\ref{figure:simul_highlight}) that identify these
structures, and then averaging the individual weighting functions from each of those pixels.\\

The dashed-vertical lines in Figure~\ref{figure:hof} represent the median of the distribution. This line indicates, by definition, that the spectral lines 
employed in this work are equally responsive to the layers above it than to the layers beneath it. The shaded areas in each panel include the layers where
the sum of the sorted (from highest to lowest) values of $\widetilde{w}_{x}$ adds up to 90 \% of the total area under $\widetilde{w}_{x}(\tau_c)$. These regions 
can be then considered as the regions where most of the sensitivity of the spectral lines comes from.\\

From this figure we infer that, on average, the response of the spectral lines to variations in $B$, $\gamma$, and ${\rm v}_{\rm los}$ is restricted to a narrower region 
($\log\tau_c \in [0,-2.5]$) in the penumbra than in granules and intergranules ($\log\tau_c \in [0,-3.5]$). The response in the umbra is spread mostly over a 
region between $\log\tau_c \in [0,-3.0]$. In some cases, in particular in the umbra and penumbra, $\widetilde{w}_{\rm x}(\tau_c)$ presents a strong narrow peak
that could be used to argue in favor of the idea that the spectral lines are formed at some particular height. As discussed in Section~\ref{section:comparison_rf}
and demonstrated here \citep[see also][]{jorge1996,jc1996} this interpretation is, however, misleading. Particularly interesting is the case of granules
and intergranules (top two rows in Fig.~\ref{figure:hof}), where layers spread over wide optical depth regions are equally sensitive to variations in the physical
parameters. An extreme example of this is the case of the granular response to variations in ${\rm v}_{\rm los}$ (top-middle panel in Fig.~\ref{figure:hof}), where
$\widetilde{w}_{\rm v}(\tau_c)$ presents a bimodal distribution. Here, we distinguish two regions of large sensitivity, $\log\tau_c \in [0,-1]$ and $\log\tau_c \in [-2,-4]$, but 
very little response to variations in ${\rm v}_{\rm los}$ in the layers located between these two regions: $\log\tau_c \in [-1,-2]$.\\

The averaged weighting functions $\widetilde{w}_{\rm x}(\tau_c)$ from Figure~\ref{figure:hof} can be useful to interpret the inferences from M-E inversion codes, 
as they can help assign (to first-order) \emph{where} in the Photosphere the measurement is done. Here we note that the response functions depend 
on the physical parameters of the atmosphere (see Sect.~\ref{section:comparison_rf}). For this reason we have tried to average $\widetilde{w}_{\rm x}(\tau_c)$ from 
individual pixels that correspond to similar solar structures so that the physical parameters, and thus also the response functions, are similar. However, 
this is never strictly the case (e.g., the physical conditions are not the same at the center of a granule as at the edge) and consequently the interpretation 
of $\widetilde{w}_{\rm x}(\tau_c)$ should be done with care. We emphasize, however, that employing the functions $\widetilde{w}_{\rm x}(\tau_c)$ from Figure~\ref{figure:hof} 
is clearly preferable to assigning a single optical depth to the measurement (see Sect.~\ref{section:comparison_fixed_tau} and references therein).\\

\section{Summary and conclusions}
\label{section:summary}

Milne-Eddington (M-E) inversion codes for the radiative transfer equation are the most widely used tools to infer the magnetic field vector from observations of 
the polarization signals in photospheric and chromospheric spectral lines. A comprehensive comparison between the different M-E codes available 
to the solar physics community is still missing, and so is a physical interpretation of their inferences. To address these questions we have carried out a comparison 
between three of those codes: VFISV, ASP/HAO, and \helixp. The three M-E codes have been used to invert synthetic Stokes profiles that were previously obtained 
from realistic non-grey 3D MHD simulations.\\

Our results indicate that the three tested M-E codes agree within approximately 30 Gauss in the determination of the magnetic field strength $B$, within 
1$\deg$ in the determination of the inclination of the magnetic field $\gamma$, and finally, within 100\ms in the determination of the line-of-sight component of
the velocity $v_{\rm los}$. Compared with the 3D MHD numerical simulations at a fixed optical depth, M-E codes retrieve the correct values within 130 Gauss, 5$\deg$, and 320\ms
in $B$, $\gamma$, and $v_{\rm los}$ respectively. We have argued, however, that comparisons at a fixed optical depth or geometrical height are misleading because they do not 
consider that the Stokes parameters convey information about a wide range of optical depths, and these ranges vary with the physical parameter being inferred. Moreover,
the atmosphere itself plays a role, and therefore the same physical parameter is measured in different atmospheric layers if we look, for instance, at granules and intergranules.\\

To properly account for all the aforementioned effects we have employed the response functions of the Stokes vector to the different physical parameters to determine the exact
range of optical depths that should be employed in the comparison between 3D MHD numerical simulations and the inferences made by M-E inversion codes. Once this is accounted for,
the agreement between the numerical simulations and M-E codes improves: 90 Gauss in $B$, 3$\deg$ in $\gamma$, and 90\ms in $v_{\rm los}$. Finally, we have provided the 
approximate optical depth regions that convey information, in the Fe I line pair at 630 nm, about the magnetic field strength, inclination and line-of-sight velocity in granules, 
intergranules, penumbral and umbral regions.\\

\begin{acknowledgements}
This research has been carried out in the frame of the two meetings held in February 2010 and December 2012 at the International 
Space Science Institute (ISSI) in Bern (Switzerland) as part of the International Working group {\it Extracting Information from spectropolarimetric 
observations: comparison of inversion codes}. We are particularly grateful to Dr. Maurizio Falanga, Andrea Fischer and Jennifer Zaugg for their
hospitality and help organizing the meetings. Discussions with Drs. Michiel van Noort, Arturo L\'opez, Andr\'es Asensio Ramos, H\'ector Socas-Navarro, 
and Nikola Vitas are also greatfully acknowledged. R.R. acknowledges financial support by DFG grant RE 3282/1-1. This work has made use of the NASA 
Astrophysical Data System. The National Center for Atmospheric Research is sponsored by the National Science Foundation. This investigation is based 
on work supported by the National Science Foundation under Grant numbers 0711134, 0933959, 1041709, and 1041710 and the University of Tennessee 
through the use of the Kraken computing resource at the National Institute for Computational Sciences (\url{http://www.nics.tennessee.edu}).
\end{acknowledgements}

\bibliographystyle{aa}
\bibliography{issi_me_aa}  
\end{document}